\documentclass[a4paper,12pt]{article}
\pdfoutput=1
\usepackage{epsfig}
\usepackage{amssymb}
\usepackage{amsfonts}
\usepackage{amsmath}
\usepackage{euscript}
\usepackage{verbatim}
\usepackage{latexsym}
\usepackage{graphicx}
\usepackage{caption}
\usepackage{float}
\usepackage{ytableau}
\usepackage{svg}
\usepackage{mathabx}
\usepackage{wrapfig}
	\usepackage[T1]{fontenc}
	\usepackage{tikz}
	\usetikzlibrary{decorations.pathmorphing}
\usepackage{tikz}
\usetikzlibrary{shapes.geometric, arrows,patterns,snakes}
\tikzstyle{ellip} = [ellipse, minimum width=3cm, minimum height=1cm,text centered, draw=black]

\newskip\humongous \humongous=0pt plus 1000pt minus 1000pt

\newif\ifdtup

\allowdisplaybreaks[1]

\catcode`\@=11

\@addtoreset{equation}{section}

\def\@normalsize{\@setsize\normalsize{15pt}\xiipt\@xiipt
\abovedisplayskip 14pt plus3pt minus3pt%
\belowdisplayskip \abovedisplayskip
\abovedisplayshortskip \z@ plus3pt%
\belowdisplayshortskip 7pt plus3.5pt minus0pt}

\def\small{\@setsize\small{13.6pt}\xipt\@xipt
\abovedisplayskip 13pt plus3pt minus3pt%
\belowdisplayskip \abovedisplayskip
\abovedisplayshortskip \z@ plus3pt%
\belowdisplayshortskip 7pt plus3.5pt minus0pt
\def\@listi{\parsep 4.5pt plus 2pt minus 1pt
     \itemsep \parsep
     \topsep 9pt plus 3pt minus 3pt}}

\relax

\catcode`@=12

\catcode`\@=11

\def\section{\@startsection{section}{1}{\z@}{3.5ex plus 1ex minus
   .2ex}{2.3ex plus .2ex}{\large\bf}}

\def\SymBoxes#1#2#3#4{\newdimen\un@t \un@t#3%
\raisebox{#1}{\rule{#2\un@t}{#4}\hskip-#2\un@t
\@tempdimb\un@t \advance\@tempdimb by-#4\@tempcntb#2\relax%
\@whilenum{\@tempcntb>0}\do{
\rule{#4}{\un@t}\hskip\@tempdimb \advance\@tempcntb by\m@ne}%
\hskip-#2\un@t \rule[\un@t]{#2\un@t}{#4}%
\rule[\un@t]{#4}{#4}\hskip-#4
\rule{#4}{\un@t}}\hskip-#4}                

\begin{document}

\newcommand{\beq}{\begin{equation}}
\newcommand{\eeq}{\end{equation}}
\newcommand{\bea}{\begin{eqnarray}}
\newcommand{\eea}{\end{eqnarray}}
\newcommand{\beas}{\begin{eqnarray*}}
\newcommand{\eeas}{\end{eqnarray*}}
\newcommand{\defi}{\stackrel{\rm def}{=}}
\newcommand{\non}{\nonumber}
\newcommand{\bquo}{\begin{quote}}
\newcommand{\enqu}{\end{quote}}
\renewcommand{\(}{\begin{equation}}
\renewcommand{\)}{\end{equation}}
\def \eqn#1#2{\begin{equation}#2\label{#1}\end{equation}}
\def\IZ{{\mathbb Z}}
\def\IR{{\mathbb R}}
\def\IC{{\mathbb C}}
\def\IQ{{\mathbb Q}}
\def\de{\partial}
\def\Tr{ \hbox{\rm Tr}}
\def\H{ \hbox{\rm H}}
\def\HE{ \hbox{$\rm H^{even}$}}
\def\HO{ \hbox{$\rm H^{odd}$}}
\def\K{ \hbox{\rm K}}
\def\Im{ \hbox{\rm Im}}
\def\Ker{ \hbox{\rm Ker}}
\def\const{\hbox {\rm const.}}
\def\o{\over}
\def\im{\hbox{\rm Im}}
\def\re{\hbox{\rm Re}}
\def\bra{\langle}\def\ket{\rangle}
\def\Arg{\hbox {\rm Arg}}
\def\Re{\hbox {\rm Re}}
\def\Im{\hbox {\rm Im}}
\def\exo{\hbox {\rm exp}}
\def\diag{\hbox{\rm diag}}
\def\longvert{{\rule[-2mm]{0.1mm}{7mm}}\,}
\def\a{\alpha}
\def\dag{{}^{\dagger}}
\def\tq{{\widetilde q}}
\def\p{{}^{\prime}}
\def\W{W}
\def\N{{\cal N}}
\def\hsp{,\hspace{.7cm}}

\def\br{\nonumber\\}
\def\IZ{{\mathbb Z}}
\def\IR{{\mathbb R}}
\def\IC{{\mathbb C}}
\def\IQ{{\mathbb Q}}
\def\IP{{\mathbb P}}
\def \eqn#1#2{\begin{equation}#2\label{#1}\end{equation}}

\newcommand{\sgm}[1]{\sigma_{#1}}
\newcommand{\idd}{\mathbf{1}}

\newcommand{\C}{\ensuremath{\mathbb C}}
\newcommand{\Z}{\ensuremath{\mathbb Z}}
\newcommand{\R}{\ensuremath{\mathbb R}}
\newcommand{\rp}{\ensuremath{\mathbb {RP}}}
\newcommand{\cp}{\ensuremath{\mathbb {CP}}}
\newcommand{\vac}{\ensuremath{|0\rangle}}
\newcommand{\vact}{\ensuremath{|00\rangle}                    }
\newcommand{\oc}{\ensuremath{\overline{c}}}
\begin{titlepage}
\begin{flushright}
CHEP XXXXX
\end{flushright}
\bigskip
\def\thefootnote{\fnsymbol{footnote}}

\begin{center}
{\Large
{\bf Complete Solution of a Gauged Tensor Model
}
}
\end{center}

\bigskip
\begin{center}
{\large  Chethan KRISHNAN$^a$\footnote{\texttt{chethan.krishnan@gmail.com}}, and K. V. Pavan KUMAR$^a$\footnote{\texttt{kumar.pavan56@gmail.com}} \vspace{0.15in} \\ }
\vspace{0.1in}

\end{center}

\renewcommand{\thefootnote}{\arabic{footnote}}

\begin{center}
$^a$ {Center for High Energy Physics,\\
Indian Institute of Science, Bangalore 560012, India}

\end{center}

\noindent
\begin{center} {\bf Abstract} \end{center}
Building on a strategy introduced in arXiv:1706.05364, we present exact analytic expressions for all the singlet eigenstates and eigenvalues of the simplest non-linear ($n=2, d=3$) gauged Gurau-Witten tensor model. This solves the theory completely. The ground state eigenvalue is $-2\sqrt{14}$ in suitable conventions. This matches the result obtained for the ground state energy in the ungauged model, via brute force diagonalization on a computer. We find that the leftover degeneracies in the gauged theory, are only partially accounted for by its known discrete symmetries, indicating the existence of previously unidentified ``hidden" global symmetries in the system. We discuss the spectral form factor, the beginnings of chaos, and the distinction between theories with $SO(n)$ and $O(n)$ gaugings. Our results provide the complete analytic solution of a non-linear gauge theory in 0+1 dimensions, albeit for a specific value of $N$. A summary of the main results in this paper were presented in the companion letter arXiv:1802.02502.


\vspace{1.6 cm}
\vfill

\end{titlepage}

\setcounter{footnote}{0}


\section{Introduction}

The purpose of this paper is to present the complete analytic solution of a gauged non-linear 0+1 dimensional quantum theory of strongly interacting fermions. More specifically, we will present the eigenvalues and eigenstates of the $n=2, d=3$ colored tensor model in the language of \cite{witten}. A letter version of this paper with only (some of) the results was presented in \cite{PRL}, here we will present the various technical details of both the method as well as the complete results. Before proceeding, we will present a few words of motivation as well as make some general comments. 

Gauged tensor models \cite{tensordump} have been around for a while, but they experienced a resurgence in interest after Witten argued \cite{witten} that certain classes of them can mimic the large-$N$ diagrammatics of the SYK model \cite{Polchinski, Maldacena, coredump} and therefore they might be of interest for holographic purposes (see related discussions in \cite{Murugan, Minwalla}). Unlike the SYK model, these theories are not disorder averaged and therefore are perfectly legitimate quantum mechanical theories. In this paper, we will view these tensor models as a class of strongly coupled gauge theories in 0+1 dimensions, with tunable parameters that capture $N$. Our goal will be to see how far we can proceed in exactly (non-pertubatively) solving them, and we will stick to relatively small $N$.  

The strategy we present here is in principle general enough to go through for all colored Gurau-Witten tensor models with gauge group $O(n)^4\times O(2)^2$, as well as possibly other related classes of tensor models. Of course, at larger $n$ the implementatioon of the approach is more complicated. What we have explicitly solved here is the $n=2$ case. Even though the gauge group is Abelian, because of the quartic self-interactions the theory is still highly non-linear. One of the advantages of 0+1 dimensions is that the gauge field is not dynamical and merely imposes a singlet constraint (see e.g., \cite{Milekhin}) for any $n$. We take systematic advantage of this, as well as of the crucial fact observed in \cite{finite-N} that the Hilbert space is a spinor and that its Clifford structure can be exploited profitably. We should emphasize that while the ungauged model is more or less straightforwardly diagonalized numerically on a computer, the gauged theory is not. This is because constructing singlets and singlet eigenstates in a useful way in these theories is difficult. In particular, it is unclear (to us) whether having the numerical eigenvalue spectrum of the ungauged theory is helpful in solving the gauged theory. This is what makes this problem interesting. In fact, even the counting of singlets is a non-trivial problem, see \cite{klebanov-counting}. The dimensionality of the gauged Hilbert space we find matches with the count in \cite{klebanov-counting}.

It is easy to convince oneself that the eigenvalues of the gauged model should form a subset of the ungauged model. Since the ungauged model can be diagonalized numerically, this offers us a non-trivial test of our solution. Indeed, we find that all the eigenvalues that we find (our eigenvalues are all square roots of integres as it turns out) match with (a subset of) the numerical  eigenvalues of \cite{bala} upto six decimal places. In particular, the ground state energy is $-2 \sqrt{14}$ in units where the coupling $J=1$\footnote{See next section for definition of $J$. The coupling is dimensionful and setting it to one corresponds merely to a choice of unit, and we will do so in most of the paper. We can always re-instate it by dimensional analysis. Note that the running coupling one expects at long distances in melonic theories should not be directly compared to this. What we are solving is the UV theory. To get IR correlators one should calculate correlators and look at their long time behavior. The effective running coupling will emerge then.}, and agrees with the ungauged ground state energy. We will offer a qualitative understanding of the origin of the irrational energies in a later section.

Despite the relative smallness of $N$ we find that the theory does lead to some rudimentary large-$N$ features like chaos. In particular, the Spectral Form Factor (SFF) is qualitatively identical to that of the ungauged model found in \cite{bala, uncolored-numerical} which in turn was related to the dip-ramp-plateau structure of SYK \cite{cotler}. This should be contrasted to the simplest uncolored model \cite{klebanov} whose gauged version was solved in \cite{loga, dario}, but it was found to be a rather trivial two-state system. The ungauged Hilbert space there was only 16 dimensional, and after gauging only two states were left. Here, we start with an ungauged Hilbert space that is 65536 dimensional and after gauging we end up with a Hilbert space that is 140 dimensional. There are 11 distinct eigenvalues in the spectrum.  We are able to explain many of the degeneracies in the final Hilbert space in terms of the known \cite{witten, bala} discrete global symmetries of the tensor model. However, there remain degeneracies which are unaccounted for by the known symmetries of the system, suggesting that there are (so far) unidentified hidden symmetries in the Gurau-Witten tensor model. It is clearly of interest to identify them.

Note that it is the {\em finding} of the solution that is the difficult part in the problem. Once we find the solution it is easy enough to {\em verify}: by explicitly acting with the Hamiltonian on the eigenstates. This means that we can have quite a bit of confidence that the solution is indeed right. Further tests of the solution include the fact that the eigenvalues match with previous numerical results of a (subset of) eigenvalues in the ungauged tensor model, as well as the match of the dimensionality of the Hilbert space with the indirect count of singlets in \cite{klebanov-counting}\footnote{Note that \cite{klebanov-counting} appeared after the Letter version of this paper \cite{PRL} on the arXiv, so this is {\em not} a retrodiction.}.

We will mostly discuss the $SO(n)$ version of the model in detail in this paper. It is also possible to consider the $O(n)$ model, which will remove many more of the states from the spectrum. For completeness, we present a discussion of that in section \ref{so(n) vs o(n)}.  But it should be kept in mind that in the rest of the paper we will not emphasize the distinction between the two, even though we always have the SO(n) case in mind.

\section{Gurau-Witten Model}

Gurau-Witten model is a quantum mechanical model in 0+1 dimensions. The model is constructed using fermionic tensors of the form $\psi _A^{i_1i_2\ldots i_d}$. The index $A$ corresponds to color index and take values from $0$ to $d$ and the tensor indices take values from 1 to $n$. That is, the total degrees of freedom is given by $N=(d+1)n^d$. 

For each pair of colors $(A,B)$, we assign a symmetry group $O(n)_{AB}$. As there are $(d+1)$ colors in the theory, the overall symmetry group of the theory is given by:
\begin{align}
G \sim O(n)^{d(d+1)/2}
\end{align}
Further, we demand that under any group $O(n)_{AB}$, fermions belonging to the colors $A$ and $B$  transform in the vector representation and the fermions of rest of the colors transform trivially. The interaction term of the theory is an invariant under the symmetry group $G$ and contains fermions belonging to all the $(d+1)$ colors. 

From now on, we work with $d=3$ i.e., we work with a theory that has 4 colors and has a quartic interaction term. The Lagrangian of $d=3$ Gurau-Witten model is given by:
\begin{align}
{\cal L}&=\frac{i}{2}\psi ^{ijk}_A\partial _t \psi ^{ijk}_A+\frac{J}{n^{3/2}}\sum \psi ^{ijk}_0\psi ^{ilm}_1\psi ^{njm}_2\psi ^{nlk}_3
\end{align}
where $(0,1,2,3)$ correspond to the color indices and each of the tensor indices takes values from 1 to $n$. $J$ is the dimensionful coupling and we set it to unity in the rest of the discussion. Quantizing this theory gives rise to the following anti-commutation relations:
\begin{align}
\{\psi_{A}^{ijk},\psi_{B}^{pqr}\}&=\delta_{AB}\delta^{ip}\delta^{jq}\delta^{kr}
\end{align}

The kinetic term of $d=3$ Gurau-Witten Lagrangian has $O(4n^3)$ symmetry which is broken down to $O(n)^6$ due to the presence of interaction term. That is, the symmetry group $G$ of the theory is given by:
\begin{align}
G\sim O(n)_{01}\times O(n)_{02} \times O(n)_{03}\times O(n)_{12} \times O(n)_{13} \times O(n)_{23}
\end{align}
. More specifically, each of the fermionic tensors transform under $G$ as follows:
\begin{align}
\psi ^{ijk}_0\rightarrow M_{01}^{ii'}~M_{02}^{jj'}~M_{03}^{kk'}~\psi ^{i'j'k'}_0 \nonumber \\
\psi ^{ijk}_1\rightarrow M_{01}^{ii'}~M_{13}^{jj'}~M_{12}^{kk'}~\psi ^{i'j'k'}_1 \nonumber \\
\psi ^{ijk}_2\rightarrow M_{23}^{ii'}~M_{02}^{jj'}~M_{12}^{kk'}~\psi ^{i'j'k'}_2 \nonumber \\
\psi ^{ijk}_3\rightarrow M_{23}^{ii'}~M_{13}^{jj'}~M_{03}^{kk'}~\psi ^{i'j'k'}_3 
\end{align}
where $M_{AB}$ are the matrices that correspond to the group $O(n)_{AB}$. Using this information, we can compute the Noether charges corresponding to the symmetry group $G$ as follows:
\begin{align}
\label{charge-1}
Q^{i_1i_2}_{01}&=i \left(\psi _0^{i_1jk}\psi _0^{i_2jk}+\psi _1^{i_1jk}\psi _1^{i_2jk}\right)\\
\label{charge-2}
Q^{i_1i_2}_{23}&=i \left(\psi _2^{i_1jk}\psi _2^{i_2jk}+\psi _3^{i_1jk}\psi _3^{i_2jk}\right)\\
\label{charge-3}
Q^{j_1j_2}_{02}&=i \left(\psi _0^{ij_1k}\psi _0^{ij_2k}+\psi _2^{ij_1k}\psi _2^{ij_2k}\right)\\
\label{charge-4}
Q^{j_1j_2}_{13}&=i \left(\psi _1^{ij_1k}\psi _1^{ij_2k}+\psi _3^{ij_1k}\psi _3^{ij_2k}\right)\\
\label{charge-5}
Q^{k_1k_2}_{03}&=i \left(\psi _0^{ijk_1}\psi _0^{ijk_2}+\psi _3^{ijk_1}\psi _3^{ijk_2}\right)\\
\label{charge-6}
Q^{k_1k_2}_{12}&=i \left(\psi _1^{ijk_1}\psi _1^{ijk_2}+\psi _2^{ijk_1}\psi _2^{ijk_2}\right)
\end{align}
where the subscripts are the color indices. Note that the upper indices on any of the charges should not be equal. 

\section{The Clifford Basis}

Before explaining our strategy to find the singlet spectrum, we define the basis that we work with. The construction of basis is based on the fact that the Hilbert space of our theory forms a spinor representation of $O(n)$. 

To start with, following and slightly generalizing \cite{dario, finite-N}, we define for even $n$ the ``colored'' creation and annihilation operators as: 
\begin{align}
\psi ^{ij{k^\pm}}_A&=\frac{1}{\sqrt{2}}\left(\psi ^{ijk}_A\pm i \psi ^{ij(k+1)}_A\right)
\end{align}
where $k$ takes only odd values and $k^{\pm}$ is given by the relation:
\begin{align}
k&=2k^{\pm}-1
\end{align}
Further, we can show that the $\psi ^{ijk^{\pm}}_A$ obey the following anti-commutation relations:
\begin{align}
\{\psi ^{ij{k^+}}_A,\psi ^{lm{n^+}}_B\}=0; \hspace{5 mm}\{\psi ^{ij{k^-}}_A,\psi ^{lm{n^-}}_B\}=0; \hspace{5 mm} \{\psi ^{ij{k^+}}_A,\psi ^{lm{n^-}}_B\}=\delta _{AB}\delta ^{il}\delta ^{jm}\delta ^{k^+,n^-}
\end{align}

We define the Clifford vacuum as the state that is annihilated by all $\psi ^-$'s i.e.,
\begin{align}
\psi _A^{ijk^-}|~\rangle =0
\end{align}
Acting with the creation operators, we can construct the entire Hilbert space. Since there are $2n^3$ (fermionic) creation operators, we can see that the dimensionality of Hilbert space is $2^{2n^3}$.

For forthcoming purposes, we define four level operators $L_A$, which counts the number of creation operators of each color in a state. These $L_A$ are defined as:
\begin{align}
\label{level operator of one color}
L_A&=\psi ^{ijk+}_A\psi ^{ijk-}_A
\end{align} 
Note that there is no summation over the color index $A$. The commutation relations with the fermionic tensors are given by:
\begin{align}
[L_A,\psi _B^{ijk^\pm}]=\pm \psi _B^{ijk^\pm} \delta _{AB} 
\end{align}
There is no summation over $B$ on the RHS.

The Hamiltonian can be written in terms of the creation and annihilation operators as follows:
\begin{align}
H=&\sum \psi ^{ij{k^+}}_0\psi ^{il{m^+}}_1\psi ^{nj{m^-}}_2\psi ^{nl{k^-}}_3+\psi ^{ij{k^+}}_0\psi ^{il{m^-}}_1\psi ^{nj{m^+}}_2\psi ^{nl{k^-}}_3 \nonumber \\
&+~~\psi ^{ij{k^-}}_0\psi ^{il{m^+}}_1\psi ^{nj{m^-}}_2\psi ^{nl{k^+}}_3+\psi ^{ij{k^-}}_0\psi ^{il{m^-}}_1\psi ^{nj{m^+}}_2\psi ^{nl{k^+}}_3
\end{align}
Note that each of the four terms in the Hamiltonian is manifestly invariant under $O(n)^4\times U(\frac{n}{2})^2$. From the explicit form of the Hamiltonian, it is clear that it does not commute with the level operators corresponding to individual colors. As we will explain later, this makes finding the eigenstates of the Hamiltonian more difficult in the case of $n=2$ as compared to that of identifying the singlets.

\section{The Singlet Spectrum}

In this section, we describe our strategy to find the singlet spectrum of the Gurau-Witten model. Our strategy here is a ``colored'' generalization of the one presented in \cite{dario} for uncolored models and in principle can be implemented to identify singlet spectrum of Gurau-Witten model with arbitrary $d$ and $n$. Note that one can gauge first and then solve the theory or solve the theory and then gauge it afterwards. The second approach leads to complications related to Young tableaux proliferation \cite{Young}, so we will stick to the first.

We start by noting that the singlet states by definition are the states that transform trivially under the symmetry group. This definition can be operationally implemented by demanding that the singlet states are annihilated by the Noether charges \eqref{charge-1}-\eqref{charge-6} corresponding to the symmetry group $G$ i.e.,
\begin{align}
\label{singlet condition}
Q_{AB}~|\text{singlet}\rangle =0
\end{align}
So, our goal is to find the linear combination of our basis states that are annihilated by all the Noether charges. Since the number of basis states are exponentially large, this seems a daunting task. But the following simplification mitigates the situation partially and  in particular makes the model of our interest ($n=2$ model) tractable. We emphasize that the last statement does \textit{not} mean that our method of finding singlet states using equation \eqref{singlet condition} is restricted to $O(2)^6$ GW model alone and indeed can be extended to $O(n)^6$ model. Further, the techniques and simplifications that we discuss in this paper for $O(2)^6$ model  will be applicable to the case of $O(n)^4\times O(2)^2$ model as well and work is in progress along this direction.      

The simplification is that all the gauge singlet states are present in the mid-Clifford level. It can be shown as follows. Taking $k_2=k_1+1$ in the last two Noether charges \eqref{charge-5} and \eqref{charge-6} and summing over all the odd $k_1$'s, we get:
\begin{align}
\label{mid-level condition}
\left(L_0+L_3-\frac{n^3}{2}\right)~|\text{singlet}\rangle =0 \nonumber \\
\left(L_1+L_2-\frac{n^3}{2}\right)~|\text{singlet}\rangle =0
\end{align}
These conditions imply that the singlets are at the mid-Clifford ($=n^3$) level in which $\frac{n^3}{2}$ creation operators belong to the colors $A=0,3$ and the other half of them belong to the colors $A=1,2$.

Even with this simplification, identifying all the singlets is still a 
non-trivial task and we currently\footnote{Finding all the singlets in uncolored model with arbitrary $n$ is comparatively simpler problem and work towards this is in progress \cite{avinash}.} do not have a solution for Gurau-Witten model with arbitrary $n$.  As we will explain later, $n=2$ case has some extra simplifications which helps us in identifying all the singlets. 

Once we have a strategy to identify the singlets, the next step is to find linear combinations of singlets such that they form eigenstates of the Hamiltonian. As the Noether charges commute with the Hamiltonian, acting on any singlet state with the Hamiltonian necessarily gives a combination of singlet states. Finding the energy eigenstates is in general a hard task and the judicious application of residual symmetries of the Hamiltonian helps us by reducing the number of computations that we need to do. We will elaborate on the residual symmetries in the next section.

\section{Discrete (Residual) Symmetries}

The two tasks we have at hand- identifying the singlets and constructing singlet eigenstates of the Hamiltonian are conceptually simple. But the computations involved are often tedious and can not be evaded. But, we can reduce their number by exploiting the residual symmetries\cite{witten} of the Gurau-Witten Hamiltonian. In this section, we identify these set of (discrete) residual symmetries. These symmetries are related to the permutation of colors and are not part of the $O(n)^6$ group that we will gauge later in the paper. Thus, these symmetries will be helpful in identifying the singlet eigenstates of the Hamiltonian and also in explaining some of the degeneracies that we find in the singlet spectrum. 

The first set of these symmetries are denoted as:
\begin{align}
S_{01;23}; ~S_{02;13}; ~S_{03;12}
\end{align}
The action of these symmetries is as follows. The operator $S_{AB;CD}$ acting on a state exchanges the colors $A\leftrightarrow B$ and $C\leftrightarrow D$ simultaneously. The action of these operators on the Noether charges is as follows:
\begin{align}
S_{01;23}~Q_{01}~S_{01;23}^{-1}&=Q_{01}; ~S_{01;23}~Q_{23}~S_{01;23}^{-1}=Q_{23}; ~S_{01;23}~Q_{02}~S_{01;23}^{-1}=Q_{13}; ~S_{01;23}~Q_{13}~S_{01;23}^{-1}=Q_{02}\nonumber \\
S_{02;13}~Q_{01}~S_{02;13}^{-1}&=Q_{23}; ~S_{02;13}~Q_{23}~S_{02;13}^{-1}=Q_{01}; ~S_{02;13}~Q_{02}~S_{02;13}^{-1}=Q_{13}; ~S_{02;13}~Q_{13}~S_{02;13}^{-1}=Q_{02}\nonumber \\
S_{03;12}~Q_{01}~S_{03;12}^{-1}&=Q_{23}; ~S_{03;12}~Q_{23}~S_{03;12}^{-1}=Q_{01}; ~S_{03;12}~Q_{02}~S_{03;12}^{-1}=Q_{13}; ~S_{03;12}~Q_{13}~S_{03;12}^{-1}=Q_{02}
\end{align} 
From these relations, it is easy to see that if $|a\rangle $ is a singlet state then $S_{AB;CD}|a\rangle $ is also a singlet state. Before moving ahead, we note that the operators $S_{01;23}$, $S_{02;13}$ and $ S_{03;12} $ commute with the Hamiltonian. 

The next set of operators are:
\begin{align}
S_{03}; S_{12}
\end{align}
The operator $S_{AB}$ exchanges the colors $A\leftrightarrow B$ along with the exchange of first two indices on each $\psi $. For instance,
\begin{align}
S_{01}\psi _0^{ij+}|~\rangle &=\psi _1^{ji+}|~\rangle
\end{align}
The action of these symmetries on the Noether charges is given as follows:
\begin{align}
S_{03}~Q_{01}~S_{03}^{-1}&=Q_{13}; ~S_{03}~Q_{23}~S_{03}^{-1}=Q_{02}; ~S_{03}~Q_{02}~S_{03}^{-1}=Q_{23}; ~S_{03}~Q_{13}~S_{03}^{-1}=Q_{01}\nonumber\\
S_{12}~Q_{01}~S_{12}^{-1}&=Q_{02}; ~S_{12}~Q_{23}~S_{12}^{-1}=Q_{13}; ~S_{12}~Q_{02}~S_{12}^{-1}=Q_{01}; ~S_{12}~Q_{13}~S_{12}^{-1}=Q_{23}
\end{align}
From the action of $S_{AB}$ on Noether charges, we can see that if $|a\rangle $ is a singlet then $S_{AB}|a\rangle $ is also a singlet state.

The next set of operators we define are quite non-trivial as they do not commute with the level operators $L_A$ defined in \eqref{level operator of one color}. The first such symmetry operator is $S_{23}$. $S_{23}$ exchanges the colors 2 and 3 along with exchanging the second and third indices on each fermion. For instance,
\begin{align}
S_{23}\psi _2^{ijk}S_{23}^{-1}&=\psi _3^{ikj}
\end{align}
As this operator exchanges second and third indices and since we constructed our basis by breaking the $O(n)^2$ symmetry corresponding to the third indices, $S_{23}$ can be thought of as an operator that relates our basis (the one obtained by breaking $O(n)$'s of the third indices) to another basis that needs breaking of $O(n)$'s corresponding to the second indices. As a result, the action of $S_{23}$ on our Clifford vacuum is quite non-trivial. To understand the action of $S_{23}$ on the Clifford vacuum, we need to know the operator $S_{23}$ explicitly. The construction of $S_{23}$ is straightforward\footnote{For more details, see \cite{uncolored-numerical}} and uses the following identities:
\begin{align}
\frac{1}{2}\left(\psi _A^{ijk}+\psi _B^{i'jk}\right)\psi _{A}^{ijk}\left(\psi _A^{ijk}+\psi _B^{i'j'k'}\right)&=\psi _B^{i'j'k'} \nonumber \\
\frac{1}{2}\left(\psi _A^{ijk}+\psi _B^{i'jk}\right)\psi _{B}^{i'j'k'}\left(\psi _A^{ijk}+\psi _B^{i'j'k'}\right)&=\psi _A^{ijk}
\end{align}   
Using these identities, we can write down the explicit form of $S_{23}$ operator as:
\begin{align}
S_{23}=&\psi _0^{111}\psi _0^{122}\psi _0^{211}\psi _0^{222}\left(\psi _0^{112}+\psi _0^{121}\right)\left(\psi _0^{212}+\psi _0^{221}\right)\psi _1^{111}\psi _1^{122}\psi _1^{211}\psi _1^{222}\left(\psi _1^{112}+\psi _1^{121}\right)\left(\psi _1^{212}+\psi _1^{221}\right)\nonumber \\
&\left(\psi _2^{111}+\psi _3^{111}\right)\left(\psi _2^{112}+\psi _3^{121}\right)\left(\psi _2^{121}+\psi _3^{112}\right)\left(\psi _2^{122}+\psi _3^{122}\right)\nonumber \\
&\left(\psi _2^{211}+\psi _3^{211}\right)\left(\psi _2^{212}+\psi _3^{221}\right)\left(\psi _2^{221}+\psi _3^{212}\right)\left(\psi _2^{222}+\psi _3^{222}\right)
\end{align}
The action of this operator on the Clifford vacuum is given by:
\begin{align}
&S_{23}|~\rangle = \left[\left(\psi _0^{11+}\psi _0^{21+}-\psi _0^{12+}\psi _0^{22+}\right)-i\left(\psi _0^{11+}\psi _0^{22+}+\psi _0^{12+}\psi _0^{21+}\right)\right]\left[0\rightarrow 1,2,3\right]|~\rangle 
\end{align}
The operator $S_{23}$ permutes the Noether charges \eqref{charge-1}-\eqref{charge-6} among themselves. More precisely, we have:
\begin{align}
S_{23}Q_{01}S_{23}^{-1}&=Q_{01}; ~~~~S_{23}Q_{23}S_{23}^{-1}=Q_{23} \nonumber \\
S_{23}Q_{12}S_{23}^{-1}&=Q_{13}; ~~~~S_{23}Q_{13}S_{23}^{-1}=Q_{12} \nonumber \\
S_{23}Q_{02}S_{23}^{-1}&=Q_{03}; ~~~~S_{23}Q_{03}S_{23}^{-1}=Q_{02}
\end{align} 
From these relations, it is clear that if $|a\rangle $ is a singlet, then $S_{23}|a\rangle $ is also a singlet.

Likewise, we can define an operator $S_{13}$ that exchanges the colors 1 and 3 along with exchanging the first and third indices. Similar to the case of $S_{23}$, we can write down the explicit form of $S_{13}$. The action of $S_{13}$ on the Clifford vacuum is given as:
\begin{align}
S_{13}|~\rangle =&\left[\left(\psi _0^{11+}\psi _0^{12+}-\psi _0^{21+}\psi _0^{22+}\right)-i\left(\psi _0^{11+}\psi _0^{22+}-\psi _0^{12+}\psi _0^{21+}\right)\right]\left[0\rightarrow 1,2,3\right]|~\rangle
\end{align}
Note that $S_{13}$ is not an independent operator and can be obtained from the operators we have already defined. For instance, it can be written as a combination of $S_{23}$ and $S_{12}$ as follows:
\begin{align}
S_{13}&=S_{12}~S_{23}~S_{12}^{-1}
\end{align}
We can also define operators $S_{01}$ and $S_{02}$ that have an action analogous to $S_{23}$ and $S_{13}$ respectively. These operators are also not independent and can be obtained from the operators we have defined already. Note that the operators of the form $S_{AB}$ anti-commute with the Hamiltonian.

Further, we can define operators of the form:
\begin{align}
S_A&=\prod _{i,j,k=1}^n \psi _A^{ijk}
\end{align}  
From the anti-commutation relations, we see that the action of this operator is as follows:
\begin{align}
S_A\psi _BS_A^{-1}&=(-1)^{n-1}\psi _B ~~~~ \text{if} ~~A=B \\
&=(-1)^n\psi _B ~~~~~~~ \text{if} ~~A\neq B
\end{align}
The operators $S_A$ commute with the Noether charges but anti-commute with the Hamiltonian.

\section{$n=2$}

From now on, we specialize to the case of $n=2$. The Clifford levels are $2n^3=16$ in number and thus the Hilbert space is $2^{16}$ dimensional.  From the mid-level condition \eqref{mid-level condition}, we find that all the singlets should be at 8th Clifford level. Out of the 8 creation operators, four of them should have $A=\{0,3\}$ color indices and the remaining four should have $A=\{1,2\}$ color indices. A generic candidate singlet state satisfying these constraints is of the form:
\begin{align}
\label{CSS-n=2}
\sum \alpha ^{0/3 \ldots ,0/3,1/2 \ldots 1/2}_{i_1j_1;i_2j_2;\ldots i_8j_8} ~\psi _{0/3}^{i_1j_11^+}\psi _{0/3}^{i_2j_21^+}\psi _{0/3}^{i_3j_31^+}\psi _{0/3} ^{i_4j_41^+}\psi _{1/2} ^{i_5j_51^+} \psi _{1/2}^{i_6j_61^+}\psi _{1/2}^{i_7j_71^+}\psi _{1/2}^{i_8j_81^+}|~\rangle
\end{align}
The total number of $\alpha $'s are given by ${8\choose 4}{8\choose 4}=4900$. These $\alpha $'s can be divided into 25 different groups based on the bi-partitions of 4 which are given by:
\begin{align}
4=4+0=3+1=2+2=1+3=0+4
\end{align}
We call these partitions as $p_1,\ldots p_5$. Each of the candidate singlet states\footnote{By candidate singlet state, we mean a state in the Clifford basis that satisfies the mid-level condition.} belong to a unique group that can be denoted by an ordered pair $(p_a,p_b)$ where $p_a$ denotes the partition of the colors $0$ and $3$ whereas $p_b$ denotes the partition corresponding to other two colors. For instance, the group of states denoted by $(p_1,p_2)$ has four $\psi ^+_0$'s; three $\psi ^+_1$'s and a  $\psi ^+_2$. The number of states in each of the groups is given in table \ref{states in groups}.

\begin{table}
\centering
\begin{tabular}{c|c|c}
$(p_a,p_b)$ & Number of states & Number of Singlets \\
\hline
$(p_1,p_1)$ & 1 & 1\\
\hline
$(p_1,p_2)$ & 16 & 0\\
\hline
$(p_1,p_3)$ & 36 & 4\\
\hline
$(p_1,p_4)$ & 16 & 0\\
\hline
$(p_1,p_5)$ & 1 & 1\\
\hline
$(p_2,p_1)$ & 16 & 0\\
\hline
$(p_2,p_2)$ & 256 & 16\\
\hline
$(p_2,p_3)$ & 576 &0\\
\hline
$(p_2,p_4)$ & 256 &16 \\
\hline
$(p_2,p_5)$ & 16 &0\\
\hline
$(p_3,p_1)$ & 36 &4\\
\hline
$(p_3,p_2)$ & 576 &0\\
\hline
$(p_3,p_3)$ & 1296 &16+8+8+8+8+4+4=56\\
\hline
$(p_3,p_4)$ & 576 &0\\
\hline
$(p_3,p_5)$ & 36 &4\\
\hline
$(p_4,p_1)$ & 16 &0\\
\hline
$(p_4,p_2)$ & 256 &16\\
\hline 
$(p_4,p_3)$ & 576 &0\\
\hline
$(p_4,p_4)$ & 256 &16\\
\hline
$(p_4,p_5)$ & 16 &0\\
\hline
$(p_5,p_1)$ & 1 &1\\
\hline
$(p_5,p_2)$ & 16 &0\\
\hline
$(p_5,p_3)$ & 36 &4\\
\hline
$(p_5,p_4)$ & 16 &0\\
\hline
$(p_5,p_5)$ &1 &1 \\
\hline
Total & 4900 & $1\times 4+4\times4+16\times4+56=140$
\end{tabular}
\caption{Number of states and singlets in various groups $(p_a,p_b)$}
\label{states in groups}
\end{table}

For the case of $n=2$, the mid-level condition exhausts the information with respect to the equations:
\begin{align}
\label{midlevel-n=2}
Q^{k_1k_2}_{03}~|\text{singlet}\rangle &=0 \nonumber \\
Q^{k_1k_2}_{12}~|\text{singlet}\rangle &=0
\end{align}
The remaining four independent charges in the language of creation and annihilation operators can be written as\footnote{In the rest of the paper we denote the last index as $\pm$ instead of $1^{\pm}$. }:
\begin{align}
\label{charges-n=2}
Q^{12}_{01}&=i \left(\psi _0^{1j+}\psi _0^{2j-}-\psi _0^{2j+}\psi _0^{1j-}+\psi _1^{1j+}\psi _1^{2j-}-\psi _1^{2j+}\psi _1^{1j-}\right)	\equiv R_0^{12}+R_1^{12}\nonumber \\
Q^{12}_{23}&=i \left(\psi _2^{1j+}\psi _2^{2j-}-\psi _2^{2j+}\psi _2^{1j-}+\psi _3^{1j+}\psi _3^{2j-}-\psi _3^{2j+}\psi _3^{1j-}\right)\equiv R_2^{12}+R_3^{12}\nonumber \\
Q^{12}_{02}&=i \left(\psi _0^{i1+}\psi _0^{i2-}-\psi _0^{i2+}\psi _0^{i1-}+\psi _2^{i1+}\psi _2^{i2-}-\psi _1^{i2+}\psi _1^{i1-}\right) \equiv S_0^{12}+S_2^{12}	\nonumber \\
Q^{12}_{13}&=i \left(\psi _1^{i1+}\psi _1^{i2-}-\psi _1^{i2+}\psi _1^{i1-}+\psi _3^{i1+}\psi _3^{i2-}-\psi _3^{i2+}\psi _3^{i1-}\right)\equiv S_1^{12}+S_3^{12}
\end{align}
where the charges $R$ and $S$ are the colored analogues of the charges\footnote{See appendix for more details.} $Q_1$ and $Q_2$ in $n=2$ uncolored model. As the level operators $L_A$ commute with the above charges, we can find the singlet states in each of the groups separately. Note that this simplification is unique to $n=2$ case. More generally, this simplification happens whenever we construct the Clifford basis by breaking a $O(2)$ group. For instance, this simplification also occurs in $O(n)^4\times O(2)^2$ Gurau-Witten model and hence the singlets in that model can be written down straightforwardly following our method-I to construct singlets. We leave further details to a future work.

\section{Singlets of $n=2$}\label{singlets of SO(n)}

In this section, we identify the singlets of $n=2$ Gurau-Witten model in all the groups using two different methods. In the first method, we will make use of the group-theoretic facts about the orthogonal group to list down the singlets. In the second method, we solve the equations $Q_{AB}|\text{singlets}\rangle =0$ explicitly to find the singlet states. We find that there are 140 singlets in $n=2$ Gurau-Witten model spreading over only 13 out of the total 25 groups. See table \ref{states in groups} for details. Note that this number of singlets matches exactly with that of \cite{klebanov-counting} where a systematic way to count the number of singlet states  in the Gurau-Witten and uncolored tensor models is presented.

\subsection{Method-I}

In this method, we take advantage of various facts about orthogonal groups. Firstly, we note that there are only two invariant tensors of $SO(n)$ that are given by:
\begin{itemize}
	\item Kronecker Delta- Due to the following property of the orthogonal group:
	\begin{align}
	M^TM=I
	\end{align}
	where $M\in SO(n)$ or $O(n)$. Note that Kronecker delta is the only  invariant tensor of $O(n)$.
	\item  Levi-Civita tensor - Because the determinant of $SO(n)$ matrices is equal to $+1$.
\end{itemize}
   
Secondly, note that the Clifford vacuum is invariant under $SO(2)^4\times U(1)^2$ by definition and thus is annihilated by the charges \eqref{charges-n=2}. Further, the Noether charges \eqref{charges-n=2} correspond to those four orthogonal groups. This implies that the quantity that appears before the Clifford vacuum of any singlet state should be an invariant of $SO(2)^4$. Combining all these observations with the mid-level condition, we can list down all the singlet states. More operationally, if we start with a generic singlet state of the form \eqref{CSS-n=2}, then the $\alpha $'s are made up of Kronecker deltas and Levi-Civita tensors. Concerning our future work \cite{avinash}, we mention that the singlets under the generalizations of the charges \eqref{charges-n=2} to arbitrary $n$ case can be found in a similar way. 

We have listed down all the 140 independent singlet states that can be constructed using this method in appendix \ref{method-I}. Note that not all possible contractions lead to different singlets.

\subsection{Method-II}

Our strategy here is to start with a generic linear combination of states in a particular\footnote{This can be done because the Noether charges in $n=2$ case commute with the level operators $L_A$ as explained in the last section.} group of the form \eqref{CSS-n=2} and then demand that the Noether charges \eqref{charges-n=2} annihilate this state to find the numerical coefficients $\alpha$ in \eqref{CSS-n=2}. This gives us the singlets. Instead of identifying all the singlets using this strategy, we can also use the discrete symmetry operators defined in the last section to identify some of the singlets. As an example, consider the following singlet state:
\begin{align}
&\left[(\psi _2^{11+}\psi _2^{12+}-\psi _2^{21+}\psi _2^{22+})(2\rightarrow 3)+(\psi _2^{11+}\psi _2^{22+}-\psi _2^{12+}\psi _2^{21+})(2\rightarrow 3)\right]\nonumber \\
&~(\psi _0^{11+}\psi _0^{12+}+\psi _0^{21+}\psi _0^{22+})(\psi _1^{11+}\psi _1^{12+}+\psi _1^{21+}\psi _1^{22+})
\end{align} 
From the last section, we know that $S_{02;13}$ acting on the above state is also a singlet i.e., the following state is also a singlet\footnote{This can be verified by doing an explicit computation.}:
\begin{align}
&\left[(\psi _0^{11+}\psi _0^{12+}-\psi _0^{21+}\psi _0^{22+})(0\rightarrow 1)+(\psi _0^{11+}\psi _0^{22+}-\psi _0^{12+}\psi _0^{21+})(0\rightarrow 1)\right]\nonumber \\
&~(\psi _2^{11+}\psi _2^{12+}+\psi _2^{21+}\psi _2^{22+})(\psi _3^{11+}\psi _3^{12+}+\psi _3^{21+}\psi _3^{22+})
\end{align}
Likewise, we can identify three more singlets in this particular case using other symmetry operators. We can also use these symmetries as a (rough) check that we have identified a complete set of singlets. We take any singlet state and act with these discrete symmetry operators and then verify whether the resultant state is also present in our singlet spectrum. The singlets we have listed in the next section are indeed closed under the action of these operators. Now, we move on to finding the singlets.

We start by noting that the charges $R_A$ and $S_A$ act non-trivially on the fermions belonging to color $A$ and acts trivially on the objects of other colors. Also, for a specific color $A$, there are four singlets\footnote{Just to avoid any confusions, we emphasize that singlets under $R_A$ and $S_A$ (for a particular $A$) are not necessarily singlets under the Noether charges \eqref{charges-n=2}.} with respect to $R_A$ and $S_A$ and are given by:
\begin{align}
\label{singlets of a color}
1&. f(\psi _B)|~\rangle \nonumber \\
2&. \left(\psi _A^{11+}\psi _A^{12+}+\psi _A^{21+}\psi _A^{22+}\right) f(\psi _B)|~\rangle \hspace{100 mm}\nonumber \\
3&. \left(\psi _A^{11+}\psi _A^{21+}+\psi _A^{12+}\psi _A^{22+}\right)f(\psi _B)|~\rangle \hspace{100 mm}\nonumber \\
4&. ~\psi _A^{11+}\psi _A^{12+}\psi _A^{21+}\psi _A^{22+}f(\psi _B)|~\rangle
\end{align}
where $f(\psi _B)$ denote functions of fermions that do not belong to the color $A$. Also, we note that the charges $R_B$ and $S_B$ act trivially on $\psi _A$'s when $A\neq B$. This information along with the mid-level conditions \eqref{midlevel-n=2} is sufficient to show that the single state present in each of the four groups $(p_{1,5},p_{1,5})$ is a singlet state. Also, it helps us to list down the first 16 singlet states in the group $(p_3,p_3)$ as we discuss later in the section.

Let us now consider the group $(p_1,p_2)$ which has 16 states. The states have four $\psi _0$'s, three $\psi _1$'s and one $\psi _2$. The generic form of a singlet state is given by:
\begin{align}
\psi _0^{11+}\psi _0^{12+}\psi _0^{21+}\psi _0^{22+}&\left[\psi _1^{11+}\psi _1^{12+}\psi _1^{21+}\left(\alpha _1\psi _2^{11+}+\alpha _2\psi _2^{12+}+\alpha _3\psi _2^{21+}+\alpha _4\psi _2^{22+} \right)\right. \nonumber \\
+&~\left.\psi _1^{11+}\psi _1^{12+}\psi _1^{22+}\left(\alpha _5\psi _2^{11+}+\alpha _6\psi _2^{12+}+\alpha _7\psi _2^{21+}+\alpha _8\psi _2^{22+} \right)\right. \nonumber \\
+&~\left.\psi _1^{11+}\psi _1^{21+}\psi _1^{22+}\left(\alpha _9\psi _2^{11+}+\alpha _{10}\psi _2^{12+}+\alpha _{11}\psi _2^{21+}+\alpha _{12}\psi _2^{22+} \right)\right. \nonumber \\
+&~\left.\psi _1^{11+}\psi _1^{12+}\psi _1^{21+}\left(\alpha _{13}\psi _2^{11+}+\alpha _{14}\psi _2^{12+}+\alpha _{15}\psi _2^{21+}+\alpha _{16}\psi _2^{22+} \right)\right] |~\rangle
\end{align}
Noting that $\psi _0^{11+}\psi _0^{12+}\psi _0^{21+}\psi _0^{22+}$ is a singlet under $S_0$, and imposing the condition that $Q_{02}=S_0+S_2$ should annihilate this state, we see that all $\alpha $'s need to be zero. That is, there are no singlets in this group. In a similar way, we can show that there are no singlets in $(p_1,p_{4})$, $(p_5,p_{2,4})$ and $(p_{2,4},p_{1,5})$ groups.

Let us now consider the states in the group $(p_2,p_2)$. A generic singlet state is of the form:
\begin{align}
\label{p2,p2 generic singlet}
&\psi _0^{12+}\psi _0^{21+}\psi _0^{22+}\psi _1^{11+}\psi _1^{21+}\psi _1^{22+}\left[\psi _2^{11+}\left(\alpha _1\psi _3^{11+}+\alpha _2\psi _3^{12+}+\alpha _3\psi _3^{21+}+\alpha _4\psi _3^{22+}\right)+\ldots \right]\nonumber \\
+&\psi _0^{11+}\psi _0^{12+}\psi _0^{22+}\psi _1^{11+}\psi _1^{12+}\psi _1^{21+}\left[\psi _2^{11+}\left(\alpha _{17}\psi _3^{11+}+\alpha _{18}\psi _3^{12+}+\alpha _{19}\psi _3^{21+}+\alpha _{20}\psi _3^{22+}\right)+\ldots \right]\nonumber \\
+&\ldots |~\rangle
\end{align}
Let us start by imposing the condition that $Q_{01}\equiv R_0+R_1$ annihilates the state. We note that a linear combination of $\psi _0\psi _0\psi _0$ (or $\psi _1\psi _1\psi _1$) can not form a singlet under $R_0$ (or $R_1$). Further, since the charge $Q_{01}$ does not affect the colors 2 \& 3 , the state can be annihilated by $Q_{01}$ only by some suitable combinations of $\psi _0$'s and $\psi _1$'s. The important point is that under the action of $Q_{01}$, the cancellations can only happen between the terms of the form $\psi _0\psi _0\psi _0\psi _1\psi _1\psi _1$ and $R_0\left(\psi _0\psi _0\psi _0\right) R_1\left(\psi _1\psi _1\psi _1\right)$. This can be shown as follows. The action of $Q_{01}$ on  $\psi _0\psi _0\psi _0\psi _1\psi _1\psi _1$ is given by:
\begin{align}
\label{Q01 on p2,p2}
Q_{01}\left(\psi _0\psi _0\psi _0\psi _1\psi _1\psi _1\right) &=R_0\left(\psi _0\psi _0\psi _0\right)\psi _1\psi _1\psi _1+\psi _0\psi _0\psi _0 R_1\left(\psi _1\psi _1\psi _1\right)
\end{align}
From the appendix, we see that acting with $R_0$ twice on $\psi _0\psi _0\psi _0 $ gives us the negative of the same state. Hence the action of $Q_{01}$ on  $R_0\left(\psi _0\psi _0\psi _0\right) R_1\left(\psi _1\psi _1\psi _1\right)$ is the only other possibility that can reproduce\footnote{The action of $Q_{01}$ on  $R_0\left(\psi _0\psi _0\psi _0\right) R_1\left(\psi _1\psi _1\psi _1\right)$ is  given as:
\begin{align*}
Q_{01}\left[R_0\left(\psi _0\psi _0\psi _0\right) R_1\left(\psi _1\psi _1\psi _1\right)\right]&=-\psi _0\psi _0\psi _0 R_1\left(\psi _1\psi _1\psi _1\right)-R_0\left(\psi _0\psi _0\psi _0\right)\psi _1\psi _1\psi _1
\end{align*}} the terms in the RHS of \eqref{Q01 on p2,p2}. The final message from this discussion is that a singlet under $Q_{01}$ is of the form:
\begin{align}
\left[\psi _0\psi _0\psi _0\psi _1\psi _1\psi _1+R_0\left(\psi _0\psi _0\psi _0\right) R_1\left(\psi _1\psi _1\psi _1\right)\right] f(\psi _2,\psi _3)
\end{align}
Similar arguments go through for the other three charges as well. Putting this all together, we can see that a singlet in the group $(p_2,p_2)$ is of the following form:
\begin{align}
\label{general p2,p2 singlet}
&\left[\psi _0\psi _0\psi _0\psi _1\psi _1\psi _1+R_0\left(\psi _0\psi _0\psi _0\right) R_1\left(\psi _1\psi _1\psi _1\right)\right] \left[\psi _2\psi _3+R_2(\psi _2) R_3(\psi _3)\right]\nonumber \\
+&\left[\psi _0\psi _0\psi _0S_1(\psi _1\psi _1\psi _1)+R_0\left(\psi _0\psi _0\psi _0\right) S_1R_1\left(\psi _1\psi _1\psi _1\right)\right] \left[\psi _2S_3(\psi _3)+R_2(\psi _2) S_3R_3(\psi _3)\right]\nonumber \\
+&\left[S_0(\psi _0\psi _0\psi _0)\psi _1\psi _1\psi _1+S_0R_0\left(\psi _0\psi _0\psi _0\right) R_1\left(\psi _1\psi _1\psi _1\right)\right] \left[S_2(\psi _2)\psi _3+S_2R_2(\psi _2) R_3(\psi _3)\right]\nonumber \\
+&\left[S_0(\psi _0\psi _0\psi _0)S_1(\psi _1\psi _1\psi _1)+S_0R_0\left(\psi _0\psi _0\psi _0\right) S_1R_1\left(\psi _1\psi _1\psi _1\right)\right] \left[S_2(\psi _2)S_3(\psi _3)+S_2R_2(\psi _2) S_3R_3(\psi _3)\right]
\end{align}
From the structure of the singlet, it is clear that starting from any one of the terms, one can construct the entire singlet uniquely. This observation suggests that there are 16 singlet states in the group $(p_2,p_2)$. In a similar way, we can show that there are 16 singlets in each of  $(p_2,p_4)$, $(p_4,p_2)$ and $(p_4,p_4)$. 

Now, we move on to the group $(p_1,p_3)$. The states in this group comprises of four $\psi _0$'s, two $\psi _1$'s and two $\psi _2$'s. Hence the singlets are of the form:
\begin{align}
\psi _0\psi _0\psi _0\psi _0 f(\psi _1\psi _1,\psi _2\psi _2)
\end{align}
Since $\psi _0\psi _0\psi _0\psi _0$ is a singlet under the charges \eqref{charges-n=2}, $f(\psi _1\psi _1,\psi _2\psi _2)$ should also be made up of singlets under those charges. From the list of singlets \eqref{singlets of a color}, we see that there are 2 singlets that can be constructed using two $\psi $'s  of same color.  So, there are totally four singlets in this group. Similar arguments show that there are four singlets in each of $(p_3,p_1)$, $(p_3,p_5)$ and $(p_5,p_3)$.

We now move on to the group $(p_2,p_3)$. The states have three $\psi _0$'s, one $\psi _3$ and two of each of $\psi _1$ and $\psi _2$. Hence the singlets are of the form:
\begin{align}
\psi _0\psi _0\psi _0 f_1(\psi _1\psi _1,\psi _2\psi _2,\psi _3)+\psi _0\psi _0\psi _0 f_2(\psi _1\psi _1,\psi _2\psi _2,\psi _3)+\ldots
\end{align}
Since a linear combination of $\psi _0\psi _0\psi _0$ can not form a singlet under $R_0$ or $S_0$, the function $f_i(\psi _1\psi _1,\psi _2\psi _2,\psi _3)$ can not include singlets of $\{1,2\}$ colors under $R_{1,2}$ or $S_{1,2}$. This suggests that the singlets are of the form:
\begin{align}
&\psi _0\psi _0\psi _0\left[\alpha _1 (\psi ^{11+}_1\psi ^{21+}_1-\psi ^{12+}_1\psi ^{22+}_1)+\alpha _2 (\psi ^{11+}_1\psi ^{12+}_1-\psi ^{21+}_1\psi ^{22+}_1)\right.\nonumber \\
&~~~~~~~~~+\left.\alpha _3 (\psi ^{11+}_1\psi ^{22+}_1-\psi ^{12+}_1\psi ^{21+}_1)+\alpha _4 (\psi ^{11+}_1\psi ^{22+}_1+\psi ^{12+}_1\psi ^{21+}_1)\right]\nonumber \\
+R_0&(\psi _0\psi _0\psi _0)\left[\alpha _5 (\psi ^{11+}_1\psi ^{21+}_1-\psi ^{12+}_1\psi ^{22+}_1)+ \ldots \right]|~\rangle +\ldots
\end{align}    
Demanding that $Q_{01}$ annihilates the above state sets all the $\alpha $'s to be zero and hence there are no singlets in this group. Similarly, we can show that there are no singlets in $(p_4,p_3)$, $(p_3,p_2)$ and $(p_3,p_4)$. 

We now consider the last group: $(p_3,p_3)$. The states in this group include two fermions of each color. From the appendix, we have the following observations:
\begin{itemize}
\item Under the charge $R_A$, the following four states are singlets:
\begin{align}
&\psi _A^{11+}\psi _A^{21+}f(\psi _B); ~~(\psi _A^{11+}\psi _A^{12+}+\psi _A^{21+}\psi _A^{22+})f(\psi _B) \nonumber \\
&\psi _A^{12+}\psi _A^{22+}f(\psi _B); ~~(\psi _A^{11+}\psi _A^{22+}+\psi _A^{12+}\psi _A^{21+})f(\psi _B) 
\end{align}
where $f(\psi _B)$ denotes some function of fermions of all the colors except the ones that belong to color $A$. Note that there is no summation over $A$ in the above states. 
\item Under the charge $S_A$, the following four states are singlets:
\begin{align}
&\psi _A^{11+}\psi _A^{12+}f(\psi _B); ~~(\psi _A^{11+}\psi _A^{21+}+\psi _A^{12+}\psi _A^{22+})f(\psi _B) \nonumber \\
&\psi _A^{21+}\psi _A^{22+}f(\psi _B); ~~(\psi _A^{11+}\psi _A^{22+}-\psi _A^{12+}\psi _A^{21+})f(\psi _B) 
\end{align}
\item Under both the charges $R_A$ and $S_A$, the following two states are singlets:
\begin{align}
\label{2-level singlet of a color}
(\psi _A^{11+}\psi _A^{21+}+\psi _A^{12+}\psi _A^{22+})f(\psi _B); ~~ (\psi _A^{11+}\psi _A^{12+}+\psi _A^{21+}\psi _A^{22+})f(\psi _B)
\end{align}
\end{itemize}

There are different types of singlets (with respect to the charges \eqref{charges-n=2}) in this group. The first type of singlets is obtained by taking a product of singlets under both $R_A$ and $S_A$ for each of the colors $A=\{0,1,2,3\}$. For example, the following singlet belongs to this type:
\begin{align}
&(\psi _0^{11+}\psi _0^{21+}+\psi _0^{12+}\psi _0^{22+})(\psi _1^{11+}\psi _1^{21+}+\psi _1^{12+}\psi _1^{22+})\times \nonumber \\
&\hspace{55 mm}\times(\psi _2^{11+}\psi _2^{21+}+\psi _2^{12+}\psi _2^{22+})(\psi _3^{11+}\psi _3^{21+}+\psi _3^{12+}\psi _3^{22+})
\end{align} 
There are 16 singlets of this type. Singlets of this type range from 21 to 36 in the list given in the next section.

Let us now take three of the colors to have singlets of the form \eqref{2-level singlet of a color} i.e., we consider states of the form:
\begin{align}
(u_A/v_A)(u_B/v_B)(u_C/v_C)&~\left[\alpha _1(\psi _D^{11+}\psi _D^{12+}-\psi _D^{21+}\psi _D^{22+})+\alpha _2(\psi _D^{11+}\psi _D^{21+}-\psi _D^{12+}\psi _D^{22+})\right. \nonumber \\
&+\left.\alpha _3(\psi _D^{11+}\psi _D^{22+}-\psi _D^{12+}\psi _D^{21+})+\alpha _4(\psi _D^{11+}\psi _D^{22+}+\psi _D^{12+}\psi _D^{21+})\right]
\end{align} 
where $u_A\equiv (\psi _A^{11+}\psi _A^{21+}+\psi _A^{12+}\psi _A^{22+})$ and $v_A\equiv (\psi _A^{11+}\psi _A^{12+}+\psi _A^{21+}\psi _A^{22+})$. Demanding that the Noether charges \eqref{charges-n=2} annihilates this state sets all the $\alpha $'s to be zero and hence there are no singlets of this form. 

Now, we consider states  of the form:
\begin{align}
\label{2 singlets, 2 non-singlets}
(u_A/v_A)(u_B/v_B)&~\left[(\psi _C^{11+}\psi _C^{12+}-\psi _C^{21+}\psi _C^{22+})\left\{\alpha _1(\psi _D^{11+}\psi _D^{12+}-\psi _D^{21+}\psi _D^{22+})+\alpha _2(\psi _D^{11+}\psi _D^{21+}-\psi _D^{12+}\psi _D^{22+})\right.\right. \nonumber \\
&+\left.\left.\alpha _3(\psi _D^{11+}\psi _D^{22+}-\psi _D^{12+}\psi _D^{21+})+\alpha _4(\psi _D^{11+}\psi _D^{22+}+\psi _D^{12+}\psi _D^{21+})\right\}\right. \nonumber \\
&+\left. (\psi _C^{11+}\psi _C^{21+}-\psi _C^{12+}\psi _C^{22+})\left\{\alpha _5(\psi _D^{11+}\psi _D^{12+}-\psi _D^{21+}\psi _D^{22+})+\alpha _6(\psi _D^{11+}\psi _D^{21+}-\psi _D^{12+}\psi _D^{22+})\right.\right. \nonumber \\
&+\left.\left.\alpha _7(\psi _D^{11+}\psi _D^{22+}-\psi _D^{12+}\psi _D^{21+})+\alpha _8(\psi _D^{11+}\psi _D^{22+}+\psi _D^{12+}\psi _D^{21+})\right\}\right. \nonumber \\
&+\left. (\psi _C^{11+}\psi _C^{22+}-\psi _C^{12+}\psi _C^{21+})\left\{\alpha _9(\psi _D^{11+}\psi _D^{12+}-\psi _D^{21+}\psi _D^{22+})+\alpha _{10}(\psi _D^{11+}\psi _D^{21+}-\psi _D^{12+}\psi _D^{22+})\right.\right. \nonumber \\
&+\left.\left.\alpha _{11}(\psi _D^{11+}\psi _D^{22+}-\psi _D^{12+}\psi _D^{21+})+\alpha _{12}(\psi _D^{11+}\psi _D^{22+}+\psi _D^{12+}\psi _D^{21+})\right\}\right. \nonumber \\ 
&+\left. (\psi _C^{11+}\psi _C^{22+}+\psi _C^{12+}\psi _C^{21+})\left\{\alpha _{13}(\psi _D^{11+}\psi _D^{12+}-\psi _D^{21+}\psi _D^{22+})+\alpha _{14}(\psi _D^{11+}\psi _D^{21+}-\psi _D^{12+}\psi _D^{22+})\right.\right. \nonumber \\
&+\left.\left.\alpha _{15}(\psi _D^{11+}\psi _D^{22+}-\psi _D^{12+}\psi _D^{21+})+\alpha _{16}(\psi _D^{11+}\psi _D^{22+}+\psi _D^{12+}\psi _D^{21+})\right\}\right.
\end{align}
Depending on the different choices of the colors $\{A,B,C,D\}$, we obtain different values for $\alpha $'s such that the above state is a singlet. We will outline the calculation for determining $\alpha $'s for one such choice and we just mention the singlets directly for the rest of the choices. Let us take $$A=0; B=1; C=2; D=3$$. The above state is trivially a singlet under $Q_{01}$. For it to be a singlet under $Q_{23}$, we need:
\begin{align}
\alpha _2&=\alpha _4=\alpha _5=\alpha _7=\alpha _{10}=\alpha _{12}=\alpha _{13}=\alpha _{15}=0 \nonumber \\
\alpha _1&=\alpha _{11}; ~\alpha _3=-\alpha _9
\end{align} 
Demanding that this state is annihilated by $Q_{02}$ will give us:	
\begin{align}
\alpha _6=\alpha _8=\alpha _{14}=\alpha _{16}=0
\end{align}
The charge $Q_{13}$ does not give any new conditions. Hence we have two different\footnote{One of them is obtained if we take $\alpha _1=\alpha _{11}=0$ and the other one for the choice $\alpha _3=-\alpha _9=0$ } types of independent singlet states in this case:
\begin{align}
&(u_0/v_0)(u_1/v_1)\left[(\psi _2^{11+}\psi _2^{12+}-\psi _2^{21+}\psi _2^{22+})(2\rightarrow 3)+(\psi _2^{11+}\psi _2^{22+}-\psi _2^{12+}\psi _2^{21+})(2\rightarrow 3)\right]\\
&(u_0/v_0)(u_1/v_1)\left[(\psi _2^{11+}\psi _2^{12+}-\psi _2^{21+}\psi _2^{22+})(\psi _3^{11+}\psi _3^{22+}-\psi _3^{12+}\psi _3^{21+})-(2\leftrightarrow 3)\right]
\end{align}
We can follow a similar strategy and obtain the singlets for other choices of colors. See (37)-(68) in the list of singlets in the next section for explicit expressions of these type of singlets. 

Now consider the states of the form:
\begin{align}
(u_A/v_A) f(\psi _B,\psi _C,\psi _D)
\end{align}
where $f(\psi _B,\psi _C,\psi _D)$ does not contain either $u_{B,C,D}$ or $v_{B,C,D}$. We can show that there are no singlets that can be constructed from these states.

Lastly, we consider the states that does not include either $u_{A,B,C,D}$ or $v_{A,B,C,D}$. Following a similar analysis as presented for the states of the form \eqref{2 singlets, 2 non-singlets}, we find that there are eight singlets in this final set. These singlets are listed from (69) to (76) in the next section.

We should have ideally done this entire analysis for the $(p_3,p_3)$ group starting from the most general state that can be written down for this group. Instead we have identified singlets by considering different type of states in this group. This is consistent because the Noether charges indeed do not mix these different type of states and hence our results are not affected.   

\section{$SO(n)$ vs $O(n)$} \label{so(n) vs o(n)}

In the last section, we found the singlets by demanding that the Noether charges \eqref{charges-n=2} annihilate the singlet states. The  Noether charges we have computed takes into account only the continuous part of the group i.e., the part of the group that is continuously connected to the identity. This means that we have obtained the singlets of $SO(n)$ in the last section. In this section, we give a strategy on how to obtain the $O(n)$ singlets starting from the $SO(n)$ singlets.

To begin with, note that the $O(n)$ group contains an extra parity transformation i.e.,
\begin{align}
	O(n)\sim SO(n)\times Z_2
\end{align}
We denote the parity transformation by $P^{i}_{AB}$ where $A/B$ are color indices and its action is given by:
\begin{align}
P^i_{AB}\psi_{A/B}^{i'j'k'}{P^i_{AB}}^{-1}&=- \psi_{A/B}^{i'j'k'} \hspace{7 mm} \text{if $i=i'$}\\
&=+\psi_{A/B}^{i'j'k'} \hspace{7 mm} \text{if $i\neq i'$}
\end{align}
That is, the action of parity operator $P^i_{AB}$ changes the sign of fermions $\psi_{A/B}^i$ and leaves the rest of them unchanged. Also, the product of two such parity transformations within the same orthogonal group $O(n)$ corresponds to a $SO(n)$ rotation.  Hence, we need to consider only one such parity transformation to obtain the singlets of $O(n)$.  There are five more such parity operators, corresponding to the ${}^4C_2$ different pairs of colors.

On $SO(n)$ singlets we obtained in the last section, we need to impose extra constraint that the $O(n)$ singlets are invariant under the parity transformations. This constraint is easier to implement in the method-I of construction of singlets.   Note that under the parity transformations, the $SO(n)$ singlets constructed using Kronecker deltas are invariant whereas the  ones constructed using Levi-Civita tensor change sign. So, the $O(n)$ singlets are the ones that are constructed using only\footnote{ This is consistent with the fact that  Kronecker delta is the only invariant tensor of $O(n)$.} the Kronecker deltas. This strategy works for the groups $O(n)_{01}$, $O(n)_{23}$, $O(n)_{02}$ and $O(n)_{13}$.

For the groups $O(n)_{03}$ and $O(n)_{12}$, things are bit involved as the Clifford vacuum that we are working with transforms non-trivially under these groups. To obtain the action of parity operators corresponding to these two groups on a generic state, we need to know their action on the Clifford vacuum. To that end, we construct explicit forms of the parity operators using their definitions. For $n=2$, the parity operators are given by:
\begin{align}
P_{03}^{k=1}=2^4~\psi _0^{111}\psi _0^{121}\psi _0^{211}\psi _0^{221}\psi _3^{111}\psi _3^{121}\psi _3^{211}\psi _3^{221}\\
P_{12}^{k=1}=2^4~\psi _1^{111}\psi _1^{121}\psi _1^{211}\psi _1^{221}\psi _2^{111}\psi _2^{121}\psi _2^{211}\psi _2^{221}
\end{align} 
From the explicit forms, it is easy to verify that they satisfy the following as expected:
\begin{align}
P_{03}\psi_{0/3} ^{ij1}P_{03}^{-1}&=-\psi_{0/3} ^{ij1};~~ P_{03}\psi_{0/3} ^{ij2}P_{03}^{-1}=+\psi_{0/3} ^{ij2} \\
P_{12}\psi_{1/2} ^{ij1}P_{12}^{-1}&=-\psi_{1/2} ^{ij1};~~ P_{12}\psi_{1/2} ^{ij2}P_{12}^{-1}=+\psi_{1/2} ^{ij2}
\end{align}
From these relations, we can find the action of these parity operators on the creation and annihilation operators as:
\begin{align}
P_{03}\psi_{0/3} ^{ij\pm}P_{03}^{-1}&=-\psi_{0/3} ^{ij\mp};~~ P_{12}\psi_{1/2} ^{ij\pm}P_{12}^{-1}=-\psi_{1/2} ^{ij\mp}
\end{align}
Further, we can find the action of these parity operators on the Clifford vacuum as:
\begin{align}
P_{03}|~\rangle &=\psi_0^{11+}\psi_0^{12+}\psi_0^{21+}\psi_0^{22+}\psi_3^{11+}\psi_3^{12+}\psi_3^{21+}\psi_3^{22+}|~\rangle \\
P_{12}|~\rangle &=\psi_1^{11+}\psi_1^{12+}\psi_1^{21+}\psi_1^{22+}\psi_2^{11+}\psi_2^{12+}\psi_2^{21+}\psi_2^{22+}|~\rangle
\end{align}

Now, we have all the information needed to identify the singlets of $O(2)_{03}$ and $O(n)_{12}$. Among the $SO(2)_{03}$ (and $SO(2)_{12}$) singlets that we already have, the $O(2)_{03}$ (and $O(2)_{12}$) singlets can be identified  as the ones that are left invariant under the respective parity operators.

Out of the 140 $SO(2)^6$ singlets,  only six of them are invariant under the parity transformations corresponding to all the $O(2)^6$ groups. More explicitly, following are the $O(2)^6$ singlets:
\begin{align}
|1\rangle +|2\rangle +|3\rangle +|4\rangle ; ~~|21\rangle ; ~~|36\rangle ; ~~|69\rangle ; ~~|73\rangle ; ~~|77\rangle +|93\rangle +|109\rangle +|125\rangle
\end{align} 
where $|i\rangle $ correspond to the $i^{\text{th}}$ singlet in the $SO(2)^6$ singlet list in the appendix. It is interesting to note that all the $O(2)^6$ singlets are from\footnote{See the next section for more details on the independent sets.} the first independent set. Further, the $O(2)^6$ invariant eigenstates of the Hamiltonian are given as follows:
\begin{align}
&|36\rangle -|21\rangle ; ~~ |73\rangle -4|21\rangle ; ~~ |69\rangle -4|36\rangle ; ~~ |36\rangle -\left(|1\rangle +|2\rangle +|3\rangle +|4\rangle\right) \nonumber \\
4&\left(|1\rangle +|2\rangle +|3\rangle +|4\rangle\right)+|36\rangle +|21\rangle +|69\rangle +|73\rangle \pm \sqrt{\frac{7}{2}}\left(|77\rangle +|93\rangle +|109\rangle +|125\rangle\right)
\end{align} 
The eigenstates in the first line have zero energy and the ones in the next line have eigenvalues of $\pm 2\sqrt{14}$. Note that the latter eigenvalues correspond to the highest energy state and the lowest energy state (ground state).

\section{Singlet Eigenstates of the Hamiltonian}

In the last section, we have identified all the singlets in the theory. Now, we want to identify the energy eigenstates among those singlets. Before doing that, we note that the Hamiltonian is a singlet of $O(2)^6$ and hence commutes with the Noether charges \eqref{charges-n=2}. As a result, we have:
\begin{align}
H|\text{singlet}\rangle =\sum _a \beta _a |\text{singlet}\rangle _a 
\end{align}  
where $a$ denotes the singlets in the theory and runs from 1 to 140 and $\beta _a$ are some numerical coefficients. The point we emphasize here is that the singlets are a closed set under the action of the Hamiltonian.

We now proceed to identify the eigenstates. First of all,  let us write down the Hamiltonian for the $n=2$ case explicitly:
\begin{align}
H&= \psi ^{ij{+}}_0\psi ^{il{+}}_1\psi ^{nj{-}}_2\psi ^{nl{-}}_3+\psi ^{ij{+}}_0\psi ^{il{-}}_1\psi ^{nj{+}}_2\psi ^{nl{-}}_3 +\psi ^{ij{-}}_0\psi ^{il{+}}_1\psi ^{nj{-}}_2\psi ^{nl{+}}_3+\psi ^{ij{-}}_0\psi ^{il{-}}_1\psi ^{nj{+}}_2\psi ^{nl{+}}_3 \nonumber \\
&=\psi ^{11{+}}_0\psi ^{11{+}}_1\psi ^{11{-}}_2\psi ^{11{-}}_3+\psi ^{11{+}}_0\psi ^{11{+}}_1\psi ^{21{-}}_2\psi ^{21{-}}_3+\psi ^{11{+}}_0\psi ^{12{+}}_1\psi ^{11{-}}_2\psi ^{12{-}}_3+\psi ^{11{+}}_0\psi ^{12{+}}_1\psi ^{21{-}}_2\psi ^{22{-}}_3 \nonumber \\
&+\psi ^{12{+}}_0\psi ^{11{+}}_1\psi ^{12{-}}_2\psi ^{11{-}}_3+\psi ^{12{+}}_0\psi ^{11{+}}_1\psi ^{22{-}}_2\psi ^{21{-}}_3+\psi ^{12{+}}_0\psi ^{12{+}}_1\psi ^{12{-}}_2\psi ^{12{-}}_3+\psi ^{12{+}}_0\psi ^{12{+}}_1\psi ^{22{-}}_2\psi ^{22{-}}_3 \nonumber \\
&+\psi ^{21{+}}_0\psi ^{21{+}}_1\psi ^{11{-}}_2\psi ^{11{-}}_3+\psi ^{21{+}}_0\psi ^{21{+}}_1\psi ^{21{-}}_2\psi ^{21{-}}_3+\psi ^{21{+}}_0\psi ^{22{+}}_1\psi ^{11{-}}_2\psi ^{12{-}}_3+\psi ^{21{+}}_0\psi ^{22{+}}_1\psi ^{21{-}}_2\psi ^{22{-}}_3 \nonumber \\
&+\psi ^{22{+}}_0\psi ^{21{+}}_1\psi ^{12{-}}_2\psi ^{11{-}}_3+\psi ^{22{+}}_0\psi ^{21{+}}_1\psi ^{22{-}}_2\psi ^{21{-}}_3+\psi ^{22{+}}_0\psi ^{22{+}}_1\psi ^{12{-}}_2\psi ^{12{-}}_3+\psi ^{22{+}}_0\psi ^{22{+}}_1\psi ^{22{-}}_2\psi ^{22{-}}_3 \nonumber \\
&=\psi ^{11{+}}_0\psi ^{11{-}}_1\psi ^{11{+}}_2\psi ^{11{-}}_3+\psi ^{11{+}}_0\psi ^{11{-}}_1\psi ^{21{+}}_2\psi ^{21{-}}_3+\psi ^{11{+}}_0\psi ^{12{-}}_1\psi ^{11{+}}_2\psi ^{12{-}}_3+\psi ^{11{+}}_0\psi ^{12{-}}_1\psi ^{21{+}}_2\psi ^{22{-}}_3 \nonumber \\
&+\psi ^{12{+}}_0\psi ^{11{-}}_1\psi ^{12{+}}_2\psi ^{11{-}}_3+\psi ^{12{+}}_0\psi ^{11{-}}_1\psi ^{22{+}}_2\psi ^{21{-}}_3+\psi ^{12{+}}_0\psi ^{12{-}}_1\psi ^{12{+}}_2\psi ^{12{-}}_3+\psi ^{12{+}}_0\psi ^{12{-}}_1\psi ^{22{+}}_2\psi ^{22{-}}_3 \nonumber \\
&+\psi ^{21{+}}_0\psi ^{21{-}}_1\psi ^{11{+}}_2\psi ^{11{-}}_3+\psi ^{21{+}}_0\psi ^{21{-}}_1\psi ^{21{+}}_2\psi ^{21{-}}_3+\psi ^{21{+}}_0\psi ^{22{-}}_1\psi ^{11{+}}_2\psi ^{12{-}}_3+\psi ^{21{+}}_0\psi ^{22{-}}_1\psi ^{21{+}}_2\psi ^{22{-}}_3 \nonumber \\
&+\psi ^{22{+}}_0\psi ^{21{-}}_1\psi ^{12{+}}_2\psi ^{11{-}}_3+\psi ^{22{+}}_0\psi ^{21{-}}_1\psi ^{22{+}}_2\psi ^{21{-}}_3+\psi ^{22{+}}_0\psi ^{22{-}}_1\psi ^{12{+}}_2\psi ^{12{-}}_3+\psi ^{22{+}}_0\psi ^{22{-}}_1\psi ^{22{+}}_2\psi ^{22{-}}_3 \nonumber \\
&=\psi ^{11{-}}_0\psi ^{11{+}}_1\psi ^{11{-}}_2\psi ^{11{+}}_3+\psi ^{11{-}}_0\psi ^{11{+}}_1\psi ^{21{-}}_2\psi ^{21{+}}_3+\psi ^{11{-}}_0\psi ^{12{+}}_1\psi ^{11{-}}_2\psi ^{12{+}}_3+\psi ^{11{-}}_0\psi ^{12{+}}_1\psi ^{21{-}}_2\psi ^{22{+}}_3 \nonumber \\
&+\psi ^{12{-}}_0\psi ^{11{+}}_1\psi ^{12{-}}_2\psi ^{11{+}}_3+\psi ^{12{-}}_0\psi ^{11{+}}_1\psi ^{22{-}}_2\psi ^{21{+}}_3+\psi ^{12{-}}_0\psi ^{12{+}}_1\psi ^{12{-}}_2\psi ^{12{+}}_3+\psi ^{12{-}}_0\psi ^{12{+}}_1\psi ^{22{-}}_2\psi ^{22{+}}_3 \nonumber \\
&+\psi ^{21{-}}_0\psi ^{21{+}}_1\psi ^{11{-}}_2\psi ^{11{+}}_3+\psi ^{21{-}}_0\psi ^{21{+}}_1\psi ^{21{-}}_2\psi ^{21{+}}_3+\psi ^{21{-}}_0\psi ^{22{+}}_1\psi ^{11{-}}_2\psi ^{12{+}}_3+\psi ^{21{-}}_0\psi ^{22{+}}_1\psi ^{21{-}}_2\psi ^{22{+}}_3 \nonumber \\
&+\psi ^{22{-}}_0\psi ^{21{+}}_1\psi ^{12{-}}_2\psi ^{11{+}}_3+\psi ^{22{-}}_0\psi ^{21{+}}_1\psi ^{22{-}}_2\psi ^{21{+}}_3+\psi ^{22{-}}_0\psi ^{22{+}}_1\psi ^{12{-}}_2\psi ^{12{+}}_3+\psi ^{22{-}}_0\psi ^{22{+}}_1\psi ^{22{-}}_2\psi ^{22{+}}_3 \nonumber \\
&=\psi ^{11{-}}_0\psi ^{11{-}}_1\psi ^{11{+}}_2\psi ^{11{+}}_3+\psi ^{11{-}}_0\psi ^{11{-}}_1\psi ^{21{+}}_2\psi ^{21{+}}_3+\psi ^{11{-}}_0\psi ^{12{-}}_1\psi ^{11{+}}_2\psi ^{12{+}}_3+\psi ^{11{-}}_0\psi ^{12{-}}_1\psi ^{21{+}}_2\psi ^{22{+}}_3 \nonumber \\
&+\psi ^{12{-}}_0\psi ^{11{-}}_1\psi ^{12{+}}_2\psi ^{11{+}}_3+\psi ^{12{-}}_0\psi ^{11{-}}_1\psi ^{22{+}}_2\psi ^{21{+}}_3+\psi ^{12{-}}_0\psi ^{12{-}}_1\psi ^{12{+}}_2\psi ^{12{+}}_3+\psi ^{12{-}}_0\psi ^{12{-}}_1\psi ^{22{+}}_2\psi ^{22{+}}_3 \nonumber \\
&+\psi ^{21{-}}_0\psi ^{21{-}}_1\psi ^{11{+}}_2\psi ^{11{+}}_3+\psi ^{21{-}}_0\psi ^{21{-}}_1\psi ^{21{+}}_2\psi ^{21{+}}_3+\psi ^{21{-}}_0\psi ^{22{-}}_1\psi ^{11{+}}_2\psi ^{12{+}}_3+\psi ^{21{-}}_0\psi ^{22{-}}_1\psi ^{21{+}}_2\psi ^{22{+}}_3 \nonumber \\
&+\psi ^{22{-}}_0\psi ^{21{-}}_1\psi ^{12{+}}_2\psi ^{11{+}}_3+\psi ^{22{-}}_0\psi ^{21{-}}_1\psi ^{22{+}}_2\psi ^{21{+}}_3+\psi ^{22{-}}_0\psi ^{22{-}}_1\psi ^{12{+}}_2\psi ^{12{+}}_3+\psi ^{22{-}}_0\psi ^{22{-}}_1\psi ^{22{+}}_2\psi ^{22{+}}_3
\end{align}
We need to act with this Hamiltonian on each of the singlet states and then identify appropriate linear combinations such that:
\begin{align}
H\sum _a\alpha _a|\text{singlet}\rangle _a &=\lambda \sum _a\alpha _a|\text{singlet}\rangle _a
\end{align}
Even though this is conceptually straightforward, the calculations are tedious. Additionally, the Hamiltonian does not\footnote{Note that the Hamiltonian commutes with the overall level operator defined by the sum of level operators of individual colors.} commute with the level operators \eqref{level operator of one color} of specific colors. As a result, the Hamiltonian mixes the states from different groups and hence we lost the simplification that happened in the case of identifying singlets. 

The discrete symmetry operators we have defined earlier make things a bit easier. To begin with, we note that the Hamiltonian commutes with the operators $S_{01;23}$, $S_{02;13}$ and $S_{03;12}$ whereas it anti-commutes with the operators $S_{AB}$ and $S_{A}$. We will describe the usefulness of these symmetry operators via an example. Consider the action of the Hamiltonian on the following singlet state: 
\begin{align}
H&~\psi _0^{11+}\psi _0^{12+}\psi _0^{21+}\psi _0^{22+}\psi _1^{11+}\psi _1^{12+}\psi _1^{21+}\psi _1^{22+} \equiv H|a\rangle \nonumber \\
=&\left(\psi _2^{11+}\psi _3^{11+}+\psi _2^{21+}\psi _3^{21+}\right)\left(\psi _0^{12+}\psi _0^{21+}\psi _0^{22+}\psi _1^{12+}\psi _1^{21+}\psi _1^{22+}+\psi _0^{11+}\psi _0^{12+}\psi _0^{22+}\psi _1^{11+}\psi _1^{12+}\psi _1^{22+}\right)\nonumber \\
-&\left(\psi _2^{11+}\psi _3^{12+}+\psi _2^{21+}\psi _3^{22+}\right)\left(\psi _0^{12+}\psi _0^{21+}\psi _0^{22+}\psi _1^{11+}\psi _1^{21+}\psi _1^{22+}+\psi _0^{11+}\psi _0^{12+}\psi _0^{22+}\psi _1^{11+}\psi _1^{12+}\psi _1^{21+}\right)\nonumber \\
-&\left(\psi _2^{12+}\psi _3^{11+}+\psi _2^{22+}\psi _3^{21+}\right)\left(\psi _0^{11+}\psi _0^{21+}\psi _0^{22+}\psi _1^{12+}\psi _1^{21+}\psi _1^{22+}+\psi _0^{11+}\psi _0^{12+}\psi _0^{21+}\psi _1^{11+}\psi _1^{12+}\psi _1^{22+}\right)\nonumber \\
+&\left(\psi _2^{12+}\psi _3^{12+}+\psi _2^{22+}\psi _3^{22+}\right)\left(\psi _0^{11+}\psi _0^{21+}\psi _0^{22+}\psi _1^{11+}\psi _1^{21+}\psi _1^{22+}+\psi _0^{11+}\psi _0^{12+}\psi _0^{21+}\psi _1^{11+}\psi _1^{12+}\psi _1^{21+}\right) \nonumber \\
\equiv &|(12,21,22),(12,21,22),(11),(11)\rangle
\end{align}
Suppose we now want to find the action of Hamiltonian on $$|b\rangle\equiv \psi _2^{11+}\psi _2^{12+}\psi _2^{21+}\psi _2^{22+}\psi _3^{11+}\psi _3^{12+}\psi _3^{21+}\psi _3^{22+}$$. Since $|b\rangle =S_{02;13}|a\rangle $, we see that $H|b\rangle =S_{02;13}\left(H|a\rangle \right)$. More explicitly, we find that:
\begin{align}
H&~\psi _2^{11+}\psi _2^{12+}\psi _2^{21+}\psi _2^{22+}\psi _3^{11+}\psi _3^{12+}\psi _3^{21+}\psi _3^{22+} \equiv H|b\rangle \nonumber \\
=&\left(\psi _0^{11+}\psi _1^{11+}+\psi _0^{21+}\psi _1^{21+}\right)\left(\psi _2^{12+}\psi _2^{21+}\psi _2^{22+}\psi _3^{12+}\psi _3^{21+}\psi _3^{22+}+\psi _2^{11+}\psi _2^{12+}\psi _2^{22+}\psi _3^{11+}\psi _3^{12+}\psi _3^{22+}\right)\nonumber \\
-&\left(\psi _0^{11+}\psi _1^{12+}+\psi _0^{21+}\psi _1^{22+}\right)\left(\psi _2^{12+}\psi _2^{21+}\psi _2^{22+}\psi _3^{11+}\psi _3^{21+}\psi _3^{22+}+\psi _2^{11+}\psi _2^{12+}\psi _2^{22+}\psi _3^{11+}\psi _3^{12+}\psi _3^{21+}\right)\nonumber \\
-&\left(\psi _0^{12+}\psi _1^{11+}+\psi _0^{22+}\psi _1^{21+}\right)\left(\psi _2^{11+}\psi _2^{21+}\psi _2^{22+}\psi _3^{12+}\psi _3^{21+}\psi _3^{22+}+\psi _2^{11+}\psi _2^{12+}\psi _2^{21+}\psi _3^{11+}\psi _3^{12+}\psi _3^{22+}\right)\nonumber \\
+&\left(\psi _0^{12+}\psi _1^{12+}+\psi _0^{22+}\psi _1^{22+}\right)\left(\psi _2^{11+}\psi _2^{21+}\psi _2^{22+}\psi _3^{11+}\psi _3^{21+}\psi _3^{22+}+\psi _2^{11+}\psi _2^{12+}\psi _2^{21+}\psi _3^{11+}\psi _3^{12+}\psi _3^{21+}\right)\nonumber \\
\equiv & |(11),(11),(12,21,22),(12,21,22)\rangle
\end{align}
By using the other symmetry operators, we can find the action of Hamiltonian on three more states in a similar way. Since many of the singlet states are related by these operators, the number of calculations we need to do are considerably reduced.

The symmetry operators are further helpful in identifying the eigenstates. Let $|E\rangle $ be an eigenstate of the Hamiltonian with eigenvalue $E$. Then the states $S_{01;23}|E\rangle $, $S_{02;13}|E\rangle $ and $S_{03;12}|E\rangle $ are also eigenstates with the same eigenvalue whereas the states $S_{AB}|E\rangle $ and $S_{A}|E\rangle $ are eigenstates of the Hamiltonian with eigenvalue $-E$.

We found that the 140 singlet eigenstates we have fall into 16 independent sets. By independent sets, we mean that the action of Hamiltonian on any of the states in an independent set produces the states in that set. From the appendix, it is clear that each of these independent sets has only one singlet from the $(p_2,p_2)$ group. This fact is useful to organize our calculations. We go about identifying the independent sets of eigenstates following the steps listed below:
\begin{itemize}
\item Start with any one of the singlet states of the $(p_2,p_2)$ group. Let us denote it by $|a\rangle $.
\item Act with the Hamiltonian on $|a\rangle $ and organize the result in terms of singlets i.e.,
\begin{align}
H|a\rangle =\sum _i\beta _i|b_i\rangle 
\end{align} 
where $|b_i\rangle $ are some singlets that depend on our choice of singlet state $|a\rangle $ and $\beta _i$ are numerical coefficients.  
\item Now act with the Hamiltonian on $|b_i\rangle $'s and organize the result in terms of singlets.
\item Repeat this until we have a set of singlet states that closes under the action of the Hamiltonian. We call this set of states as an independent set.
\item Take appropriate linear combination of states in the independent sets to form eigenstates.  
\end{itemize}    
We demonstrate these steps using an example in the following subsection. We will choose the example such that the ground state is a part of it. By ground state we mean the lowest energy state of the entire theory not just the gauged sector of it. The ground state being a part of our singlet spectrum is expected since we know that it is unique as found numerically in \cite{bala}. This is also consistent with the discussions in \cite{dario}.   

Before we proceed further, we emphasize that the reader should not be surprised by the fact that we find some of the eigenvalues to be irrational. We present here a simple case where the eigenvalues can be irrational. This example mirrors the situation that arises while finding the eigenvalues. Consider an operator $K$ whose action on two states $|p\rangle $ and $|q\rangle $ is given as following:
\begin{align}
K|p\rangle =\eta |q\rangle ; ~~~ K|q\rangle =\zeta |p\rangle
\end{align}
for some positive integers $\eta $ and $\zeta $. Two of the eigenstates of $K$ can then be constructed as:
\begin{align}
	K\left(|p\rangle \pm \sqrt{\frac{\eta}{\zeta}}|q\rangle\right)&=\pm \sqrt{\eta \zeta}	\left(|p\rangle \pm \sqrt{\frac{\eta}{\zeta}}|q\rangle\right)
\end{align} 
As we can see from this simple example, even though $\eta $ and $\zeta $ are integers, the eigenvalues $\pm\sqrt{\eta \zeta}$ can be irrational. This is similar to how we get some of the eigenvalues of the Gurau-Witten Hamiltonian to be irrational and it will become clearer after the following example.

\subsection{An example}

Here, we describe our strategy to find independent sets by  choosing the singlet state from the group $(p_2,p_2)$ to be:
\begin{align}
\label{example singlet of p2,p2}
 &|(12,21,22),(12,21,22),(11),(11)\rangle \nonumber \\
=&\left(\psi _2^{11+}\psi _3^{11+}+\psi _2^{21+}\psi _3^{21+}\right)\left(\psi _0^{12+}\psi _0^{21+}\psi _0^{22+}\psi _1^{12+}\psi _1^{21+}\psi _1^{22+}+\psi _0^{11+}\psi _0^{12+}\psi _0^{22+}\psi _1^{11+}\psi _1^{12+}\psi _1^{22+}\right)\nonumber \\
-&\left(\psi _2^{11+}\psi _3^{12+}+\psi _2^{21+}\psi _3^{22+}\right)\left(\psi _0^{12+}\psi _0^{21+}\psi _0^{22+}\psi _1^{11+}\psi _1^{21+}\psi _1^{22+}+\psi _0^{11+}\psi _0^{12+}\psi _0^{22+}\psi _1^{11+}\psi _1^{12+}\psi _1^{21+}\right)\nonumber \\
-&\left(\psi _2^{12+}\psi _3^{11+}+\psi _2^{22+}\psi _3^{21+}\right)\left(\psi _0^{11+}\psi _0^{21+}\psi _0^{22+}\psi _1^{12+}\psi _1^{21+}\psi _1^{22+}+\psi _0^{11+}\psi _0^{12+}\psi _0^{21+}\psi _1^{11+}\psi _1^{12+}\psi _1^{22+}\right)\nonumber \\
+&\left(\psi _2^{12+}\psi _3^{12+}+\psi _2^{22+}\psi _3^{22+}\right)\left(\psi _0^{11+}\psi _0^{21+}\psi _0^{22+}\psi _1^{11+}\psi _1^{21+}\psi _1^{22+}+\psi _0^{11+}\psi _0^{12+}\psi _0^{21+}\psi _1^{11+}\psi _1^{12+}\psi _1^{21+}\right) 
\end{align} 
As can be seen from the appendix, this choice corresponds to the independent set-I. Before finding the action of the Hamiltonian on this state, let us define the following for convenience:
\begin{align}
|a_1\rangle =&\left(\psi _0^{11+}\psi _0^{21+}+\psi _0^{12+}\psi _0^{22+}\right)\left(0\rightarrow 1,2,3 \right)|~\rangle \\
|a_2\rangle =&\left(\psi _0^{11+}\psi _0^{12+}+\psi _0^{21+}\psi _0^{22+}\right)\left(0\rightarrow 1,2,3 \right)|~\rangle\\
|a_3\rangle =&\left[\left(\psi _0^{11+}\psi _0^{22+}+\psi _0^{12+}\psi _0^{21+}\right)\left(0\leftrightarrow 2\right)+\left(\psi _0^{11+}\psi _0^{21+}-\psi _0^{12+}\psi _0^{22+}\right)\left(0\leftrightarrow 2 \right)\right]\times \nonumber \\
&\hspace{41 mm}\times \left(\psi _1^{11+}\psi _1^{21+}+\psi _1^{12+}\psi _1^{22+}\right)\left(\psi _3^{11+}\psi _3^{21+}+\psi _3^{12+}\psi _3^{22+}\right)|~\rangle\\
|a_4\rangle =&\left[\left(\psi _1^{11+}\psi _1^{22+}+\psi _1^{12+}\psi _1^{21+}\right)\left(1\leftrightarrow 3 \right)+\left(\psi _1^{11+}\psi _1^{21+}-\psi _1^{12+}\psi _1^{22+}\right)\left(1\leftrightarrow 3 \right)\right]\times \nonumber \\
&\hspace{41 mm}\times\left(\psi _0^{11+}\psi _0^{21+}+\psi _0^{12+}\psi _0^{22+}\right)\left(\psi _2^{11+}\psi _2^{21+}+\psi _2^{12+}\psi _2^{22+}\right)|~\rangle\\
|a_5\rangle =&\left[\left(\psi _0^{11+}\psi _0^{22+}-\psi _0^{12+}\psi _0^{21+}\right)\left(0\leftrightarrow 1 \right)+\left(\psi _0^{11+}\psi _0^{12+}-\psi _0^{21+}\psi _0^{22+}\right)\left(0\leftrightarrow 1 \right)\right]\times\nonumber \\
&\hspace{41 mm}\times\left(\psi _2^{11+}\psi _2^{12+}+\psi _2^{21+}\psi _2^{22+}\right)\left(\psi _3^{11+}\psi _3^{12+}+\psi _3^{21+}\psi _3^{22+}\right)|~\rangle\\
|a_6\rangle =&\left[\left(\psi _2^{11+}\psi _2^{22+}-\psi _2^{12+}\psi _2^{21+}\right)\left(2\leftrightarrow 3 \right)+\left(\psi _2^{11+}\psi _2^{12+}-\psi _2^{21+}\psi _2^{22+}\right)\left(2\leftrightarrow 3 \right)\right]\times\nonumber \\
&\hspace{41 mm}\times\left(\psi _0^{11+}\psi _0^{12+}+\psi _0^{21+}\psi _0^{22+}\right)\left(\psi _1^{11+}\psi _1^{12+}+\psi _1^{21+}\psi _1^{22+}\right)|~\rangle\\
|a_7\rangle =&\left[\left(\psi _0^{11+}\psi _0^{22+}+\psi _0^{12+}\psi _0^{21+}\right)\left(0\leftrightarrow 2 \right)+\left(\psi _0^{11+}\psi _0^{21+}-\psi _0^{12+}\psi _0^{22+}\right)\left(0 \leftrightarrow 2 \right)\right]\times\nonumber \\
&\times\left[\left(\psi _1^{11+}\psi _1^{22+}+\psi _1^{12+}\psi _1^{21+}\right)\left(1\leftrightarrow 3 \right)+\left(\psi _1^{11+}\psi _1^{21+}-\psi _1^{12+}\psi _1^{22+}\right)\left(1\leftrightarrow 3 \right)\right]|~\rangle\\
|a_8\rangle =&\left[\left(\psi _0^{11+}\psi _0^{22+}-\psi _0^{12+}\psi _0^{21+}\right)\left(0\leftrightarrow 1 \right)+\left(\psi _0^{11+}\psi _0^{12+}-\psi _0^{21+}\psi _0^{22+}\right)\left(0\leftrightarrow 1 \right)\right]\times\nonumber \\
&\times\left[\left(\psi _2^{11+}\psi _2^{22+}-\psi _2^{12+}\psi _2^{21+}\right)\left(2\leftrightarrow 3 \right)+\left(\psi _2^{11+}\psi _2^{12+}-\psi _2^{21+}\psi _2^{22+}\right)\left(2\leftrightarrow 3 \right)\right]|~\rangle\\
|a_9\rangle =&|(12,21,22),(12,21,22),(11),(11)\rangle \\
|a_{10}\rangle =&|(12,21,22),(11),(12,21,22),(11)\rangle \\
|a_{11}\rangle =&|(11),(12,21,22),(11),(12,21,22)\rangle \\
|a_{12}\rangle =&|(11),(11),(12,21,22),(12,21,22)\rangle \\
|a_{13}\rangle =& \psi _0^{11+}\psi _0^{12+}\psi _0^{21+}\psi _0^{22+}\psi _1^{11+}\psi _1^{12+}\psi _1^{21+}\psi _1^{22+}\\
|a_{14}\rangle =&  \psi _0^{11+}\psi _0^{12+}\psi _0^{21+}\psi _0^{22+}\psi _2^{11+}\psi _2^{12+}\psi _2^{21+}\psi _2^{22+}\\
|a_{15}\rangle =& \psi _1^{11+}\psi _1^{12+}\psi _1^{21+}\psi _1^{22+}\psi _3^{11+}\psi _3^{12+}\psi _3^{21+}\psi _3^{22+}\\
|a_{16}\rangle =& \psi _2^{11+}\psi _2^{12+}\psi _2^{21+}\psi _2^{22+}\psi _3^{11+}\psi _3^{12+}\psi _3^{21+}\psi _3^{22+}
\end{align}

The action of the Hamiltonian on the state \eqref{example singlet of p2,p2} gives:
\begin{align}
H&|a_9\rangle =16 |a_{13}\rangle +|a_1\rangle +|a_2\rangle -|a_3\rangle -|a_4\rangle +|a_5\rangle +|a_6\rangle +|a_7\rangle +|a_8\rangle 
\end{align}
Now, we need to act with the Hamiltonian on the singlets on RHS. This would give us:
\begin{align}
4H|a_1\rangle &= 4H|a_2\rangle = H|a_7\rangle =H|a_8\rangle =+4|a_9\rangle +4|a_{10}\rangle +4|a_{11}\rangle +4|a_{12}\rangle \\
H|a_3\rangle &= H|a_4\rangle = -2|a_9\rangle +2|a_{10}\rangle +2|a_{11}\rangle -2|a_{12}\rangle \\
H|a_5\rangle &= H|a_6\rangle = +2|a_9\rangle -2|a_{10}\rangle -2|a_{11}\rangle +2|a_{12}\rangle \\
H|a_{13}\rangle &= |a_9\rangle 
\end{align} 
Acting with the Hamiltonian on the extra singlets that appeared here leads to:
\begin{align}
H|a_{10}\rangle &=16 |a_{14}\rangle +|a_1\rangle +|a_2\rangle +|a_3\rangle +|a_4\rangle -|a_5\rangle -|a_6\rangle +|a_7\rangle +|a_8\rangle \\
H|a_{11}\rangle &=16 |a_{15}\rangle +|a_1\rangle +|a_2\rangle +|a_3\rangle +|a_4\rangle -|a_5\rangle -|a_6\rangle +|a_7\rangle +|a_8\rangle \\
H|a_{12}\rangle &=16 |a_{16}\rangle +|a_1\rangle +|a_2\rangle -|a_3\rangle -|a_4\rangle +|a_5\rangle +|a_6\rangle +|a_7\rangle +|a_8\rangle
\end{align}
Lastly, the action of the Hamiltonian on $|a_{14}\rangle $, $|a_{15}\rangle $ and $|a_{16}\rangle $ is given as follows:
\begin{align}
 H|a_{14}\rangle = |a_{10}\rangle ; ~H|a_{15}\rangle =|a_{11}\rangle ; ~ H|a_{16}\rangle =|a_{12}\rangle
\end{align}  
As can be seen from the explicit expressions, these singlet states $|a_1\rangle $ to $|a_{16}\rangle $ are closed under the action of Hamiltonian. 

From all the information at hand here, it is easy to construct the eigenstates and are given by:
\begin{itemize}
\item Zero energy eigenstates: (8)
\begin{align}
&|a_1\rangle -|a_2\rangle ; ~ |a_5\rangle -|a_6\rangle ; ~ |a_7\rangle -|a_8\rangle ; ~|a_3\rangle -|a_4\rangle ; |a_7\rangle -4|a_1\rangle ; |a_3\rangle +|a_5\rangle \nonumber \\
& |a_1\rangle -|a_{13}\rangle -|a_{14}\rangle -|a_{15}\rangle -|a_{16}\rangle ; ~ |a_3\rangle +2|a_{13}\rangle -2|a_{14}\rangle -2|a_{15}\rangle +2|a_{16}\rangle 
\end{align}

\item Eigenvalue of $\pm 4$: ($2\times 2$)
\begin{align}
4(|a_{13}\rangle -|a_{16}\rangle )&\pm \left(|a_9\rangle -|a_{12}\rangle \right)\\
4(|a_{14}\rangle -|a_{15}\rangle )&\pm \left(|a_{10}\rangle -|a_{11}\rangle \right)
\end{align}

\item Eigenvalue of $\pm 4\sqrt{3}$: ($1\times 2$)
\begin{align}
&4\left(|a_{13}\rangle -|a_{14}\rangle -|a_{15}\rangle +|a_{16}\rangle \right)-|a_{3}\rangle -|a_{4}\rangle +|a_{5}\rangle +|a_{6}\rangle\nonumber \\
&\hspace{55 mm}\pm \sqrt{3}\left(|a_9\rangle -|a_{10}\rangle -|a_{11}\rangle +|a_{12}\rangle \right)
\end{align}

\item Eigenvalue of $\pm 2\sqrt{14}$: ($1\times 2$)
\begin{align}
&4\left(|a_{13}\rangle +|a_{14}\rangle +|a_{15}\rangle +|a_{16}\rangle \right)+|a_{1}\rangle +|a_{2}\rangle +|a_{7}\rangle +|a_{8}\rangle \nonumber \\
&\hspace{55 mm}\pm \sqrt{\frac{7}{2}}\left(|a_9\rangle +|a_{10}\rangle +|a_{11}\rangle +|a_{12}\rangle \right)
\end{align}
\end{itemize}
The ground state corresponds to the eigenvalue of $-2\sqrt{14}$ and this value exactly matches with the one that is obtained via numerical diagonalization. In the numerical diagonallization, it was found that the ground state is unique. So, we verified that the ground state we obtained here is unique with respect to all the discrete symmetries that we have defined. This is a non-trivial test as some of the discrete symmetry operators act quite non-trivially on the singlet states.  

In the appendix, we give a list of all the eigenstates along with their eigenvalues. The eigenvalues and their degeneracies are summarized in the table \ref{evalues}. 

\begin{table}
\centering
\begin{tabular}{c|c|c|c|c|c|c|c|c|c|c|c|c}
Eigenvalue & $-2\sqrt{14}$ & $-4\sqrt{3}$ & $-2\sqrt{6}$ & $-4$ & $-2\sqrt{2}$ &  $0$ &  $2\sqrt{2}$  & $4$ & $2\sqrt{6}$ & $4\sqrt{3}$ & $2\sqrt{14}$\\
\hline 
Degeneracy &1 &3 & 4 & 6 &  31  &50  &31 &6 &4 &3 &1
\end{tabular}
\caption{Eigenvalues and corresponding degeneracy of the singlet eigenstates}
\label{evalues}
\end{table}

\section{Uniqueness of the ground state and the degeneracies}

In this section, we verify that the ground state is unique i.e., we show that it remains unchanged under the action of all the discrete symmetry operators we have defined. Also, we will explain the degeneracy of $+4\sqrt{3}$ energy eigenvalue using those operators. We conclude this section by commenting on the other symmetry operators that might exist in the theory.

To start with, let us write down the ground state explicitly:
\begin{align}
|g\rangle \equiv&\left(\psi _0^{11+}\psi _0^{21+}+\psi _0^{12+}\psi _0^{22+}\right)\left(0\rightarrow 1,2,3 \right)+\left(\psi _0^{11+}\psi _0^{12+}+\psi _0^{21+}\psi _0^{22+}\right)\left(0\rightarrow 1,2,3 \right) \nonumber \\
+&4\left(\psi _0^{11+}\psi _0^{12+}\psi _0^{21+}\psi _0^{22+}+\psi _3^{11+}\psi _3^{12+}\psi _3^{21+}\psi _3^{22+}\right)\left(\psi _1^{11+}\psi _1^{12+}\psi _1^{21+}\psi _1^{22+}+\psi _2^{11+}\psi _2^{12+}\psi _2^{21+}\psi _2^{22+}\right) \nonumber	\\
+&\left[\left(\psi _1^{11+}\psi _1^{22+}+\psi _1^{12+}\psi _1^{21+} \right)\left(1\leftrightarrow 3\right)+\left(\psi _1^{11+}\psi _1^{21+}-\psi _1^{12+}\psi _1^{22+} \right)\left(1\leftrightarrow 3\right)\right]\nonumber \\
&\left[\left(\psi _0^{11+}\psi _0^{22+}+\psi _0^{12+}\psi _0^{21+} \right)\left(0\leftrightarrow 2\right)+\left(\psi _0^{11+}\psi _0^{21+}-\psi _0^{12+}\psi _0^{22+} \right)\left(0\leftrightarrow 2\right)\right]\nonumber \\
+&\left[\left(\psi _0^{11+}\psi _0^{22+}-\psi _0^{12+}\psi _0^{21+}\right)\left(0\leftrightarrow 1 \right)+\left(\psi _0^{11+}\psi _0^{12+}-\psi _0^{21+}\psi _0^{22+}\right)\left(0\leftrightarrow 1 \right)\right]\nonumber \\
&\left[\left(\psi _2^{11+}\psi _2^{22+}-\psi _2^{12+}\psi _2^{21+}\right)\left(2\leftrightarrow 3 \right)+\left(\psi _2^{11+}\psi _2^{12+}-\psi _2^{21+}\psi _2^{22+}\right)\left(2\leftrightarrow 3 \right)\right]\nonumber \\
\pm &\sqrt{\frac{7}{2}}\left(|(12,21,22),(12,21,22),(11),(11)\rangle +|(12,21,22),(11),(12,21,22),(11)\rangle\right) \nonumber \\
\pm &\sqrt{\frac{7}{2}}\left(|(11),(12,21,22),(11),(12,21,22)\rangle +|(11),(11),(12,21,22),(12,21,22)\rangle\right) 
\end{align}
We want to show that $S|g\rangle =|g\rangle $ where $S$ is one of the discrete symmetry operators. It is easy to verify that under the action of the operators $S_{AB;CD}$, the ground state transforms into itself. Since the operators $S_{AB}$ and $S_A$ anti-commute with the Hamiltonian, we take a product of two such operators to construct operators ($S'$) that commute with the Hamiltonian. As long as we consider the operators $S'$ that include $S_{AB}$ with $(A,B)=(0,3)$ or $(A,B)=(1,2)$, it is straightforward to show that the ground state remains unchanged under its action.

The action of $S'$ becomes non-trivial if it includes the operators $S_{AB}$ with $(A,B)=(2,3)$ or $(A,B)=(0,1)$ or $(A,B)=(1,3)$ or $(A,B)=(0,2)$. This is because these operators have a non-trivial action on the Clifford vacuum. They do not commute with the level operators \eqref{level operator of one color} and hence their action on any singlet (in general) mixes singlets of various groups. 

For concreteness, let us consider the operator $S_{23}$. Now, we will consider the action of $S_{23}$ on each of the singlets present in the ground state. Let us start with the singlets of $(p_{1,5},p_{1,5})$. Before considering the entire singlet, let us consider the following:
\begin{align}
S_{23}(\psi _0^{11+}\psi _0^{12+}&\psi _0^{21+}\psi _0^{22+})|~\rangle \nonumber \\
=&-\left[(\psi _0^{11+}\psi _0^{21+}-\psi _0^{12+}\psi _0^{22+})+i(\psi _0^{11+}\psi _0^{22+}+\psi _0^{12+}\psi _0^{21+})\right]\times\nonumber \\
&\times\left[(\psi _1^{11+}\psi _1^{21+}-\psi _1^{12+}\psi _1^{22+})-i(\psi _1^{11+}\psi _1^{22+}+\psi _1^{12+}\psi _1^{21+})\right]\left[1\rightarrow 2,3\right]|~\rangle
\end{align}
Using this, we can show that: 
\begin{align}
\frac{1}{4}S_{23}&\left(\psi _0^{11+}\psi _0^{12+}\psi _0^{21+}\psi _0^{22+}+\psi _3^{11+}\psi _3^{12+}\psi _3^{21+}\psi _3^{22+}\right)\left(\psi _1^{11+}\psi _1^{12+}\psi _1^{21+}\psi _1^{22+}+\psi _2^{11+}\psi _2^{12+}\psi _2^{21+}\psi _2^{22+}\right) |~\rangle \nonumber \\
&=\left[(\psi _0^{11+}\psi _0^{22+}+\psi _0^{12+}\psi _0^{21+})(0\rightarrow 2)+(\psi _0^{11+}\psi _0^{21+}-\psi _0^{12+}\psi _0^{22+})(0\rightarrow 2)\right]\times\nonumber \\
&~\times\left[(\psi _1^{11+}\psi _1^{22+}+\psi _2^{12+}\psi _2^{21+})(1\rightarrow 3)+(\psi _1^{11+}\psi _1^{21+}-\psi _1^{12+}\psi _1^{22+})(2\rightarrow 3)\right]|~\rangle
\end{align}

Next, we move on to the singlets of the group $(p_3,p_3)$ in the ground state. To compute the action of $S_{23}$ on these singlets,  the following relations are useful:
\begin{align}
S_{23}(\psi _0^{11+}\psi _0^{22+}+\psi _0^{12+}\psi _0^{21+})|~\rangle &=-2i(\psi _0^{11+}\psi _0^{12+}\psi _0^{21+}\psi _0^{22+}+1)~f'(\psi _{1,2,3})|~\rangle \nonumber \\
S_{23}(\psi _0^{11+}\psi _0^{21+}-\psi _0^{12+}\psi _0^{22+})|~\rangle &=+2(\psi _0^{11+}\psi _0^{12+}\psi _0^{21+}\psi _0^{22+}-1)~f'(\psi _{1,2,3})|~\rangle\nonumber \\
S_{23}(\psi _0^{11+}\psi _0^{22+}-\psi _0^{12+}\psi _0^{21+})|~\rangle &=-2(\psi _0^{11+}\psi _0^{12+}-\psi _0^{21+}\psi _0^{22+})~f'(\psi _{1,2,3})|~\rangle\nonumber \\
S_{23}(\psi _0^{11+}\psi _0^{12+}-\psi _0^{21+}\psi _0^{22+})|~\rangle &=+2(\psi _0^{11+}\psi _0^{22+}-\psi _0^{12+}\psi _0^{21+}) ~f'(\psi _{1,2,3})|~\rangle\nonumber \\
S_{23}(\psi _0^{11+}\psi _0^{12+}+\psi _0^{21+}\psi _0^{22+})|~\rangle &=+2i(\psi _0^{11+}\psi _0^{21+}+\psi _0^{12+}\psi _0^{22+})~f'(\psi _{1,2,3})|~\rangle\nonumber \\
S_{23}(\psi _0^{11+}\psi _0^{21+}+\psi _0^{12+}\psi _0^{22+})|~\rangle &=+2i(\psi _0^{11+}\psi _0^{12+}+\psi _0^{21+}\psi _0^{22+})~f'(\psi _{1,2,3})|~\rangle
\end{align}
where we have defined the function $$f'(\psi _{1,2,3})=\left[(\psi _1^{11+}\psi _1^{21+}-\psi _1^{12+}\psi _1^{22+})-i(\psi _1^{11+}\psi _1^{22+}+\psi _1^{12+}\psi _1^{21+})\right]\left[1\rightarrow 2,3\right]$$
Using the above relations, we can show that:
\begin{align}
\frac{1}{16}S_{23}&\left[(\psi _0^{11+}\psi _0^{22+}-\psi _0^{12+}\psi _0^{21+})(0\rightarrow 1)+(\psi _0^{11+}\psi _0^{12+}-\psi _0^{21+}\psi _0^{22+})(0\rightarrow 1)\right]\times \nonumber \\
&\times\left.(\psi _2^{11+}\psi _2^{22+}-\psi _2^{12+}\psi _2^{21+})(2\rightarrow 3)+(\psi _2^{11+}\psi _2^{12+}-\psi _2^{21+}\psi _2^{22+})(2\rightarrow 3)\right]|~\rangle \nonumber \\
=&\left[(\psi _0^{11+}\psi _0^{22+}+\psi _0^{12+}\psi _0^{21+})(0\rightarrow 1)+(\psi _0^{11+}\psi _0^{12+}-\psi _0^{21+}\psi _0^{22+})(0\rightarrow 1)\right]\times \nonumber \\
&\times\left[(\psi _2^{11+}\psi _2^{22+}+\psi _2^{12+}\psi _2^{21+})(2\rightarrow 3)+(\psi _2^{11+}\psi _2^{12+}-\psi _2^{21+}\psi _2^{22+})(2\rightarrow 3)\right]|~\rangle \\
\frac{1}{16}S_{23}&\left[(\psi _0^{11+}\psi _0^{22+}+\psi _0^{12+}\psi _0^{21+})(0\rightarrow 2)+(\psi _0^{11+}\psi _0^{21+}-\psi _0^{12+}\psi _0^{22+})(0\rightarrow 2)\right]\times \nonumber \\
\times &\left[(\psi _1^{11+}\psi _1^{22+}+\psi _2^{12+}\psi _2^{21+})(1\rightarrow 3)+(\psi _1^{11+}\psi _1^{21+}-\psi _1^{12+}\psi _1^{22+})(2\rightarrow 3)\right]|~\rangle \nonumber \\
=4&\left(\psi _0^{11+}\psi _0^{12+}\psi _0^{21+}\psi _0^{22+}+0\leftrightarrow 3\right)\left(\psi _1^{11+}\psi _1^{12+}\psi _1^{21+}\psi _1^{22+}+1\leftrightarrow 2\right) |~\rangle \\
\frac{1}{16}S_{23}&\left[(\psi _0^{11+}\psi _0^{21+}+\psi _0^{12+}\psi _0^{22+})(0\rightarrow 1,2,3)\right]=\left[(\psi _0^{11+}\psi _0^{12+}+\psi _0^{21+}\psi _0^{22+})(0\rightarrow 1,2,3)\right] \\
\frac{1}{16}S_{23}&\left[(\psi _0^{11+}\psi _0^{12+}+\psi _0^{21+}\psi _0^{22+})(0\rightarrow 1,2,3)\right]=\left[(\psi _0^{11+}\psi _0^{21+}+\psi _0^{12+}\psi _0^{22+})(0\rightarrow 1,2,3)\right]
\end{align}

Lastly, we move to the singlets belonging to the groups $(p_{2,4},p_{2,4})$. Before finding the action of $S_{23}$ on these singlets, we need the following:
\begin{align}
S_{23}\psi _0^{11+}|~\rangle &=-\left(\psi _0^{11+}\psi _0^{12+}+i\right)(i\psi _0^{21+}+\psi _0^{22+})~f'(\psi _{1,2,3})|~\rangle \nonumber \\
S_{23}\psi _0^{12+}|~\rangle &=-\left(\psi _0^{11+}\psi _0^{12+}-i\right)(\psi _0^{21+}-i\psi _0^{22+}) ~f'(\psi _{1,2,3})|~\rangle\nonumber \\
S_{23}\psi _0^{21+}|~\rangle &=+\left(i\psi _0^{11+}+\psi _0^{12+}\right)(\psi _0^{21+}\psi _0^{22+}+i) ~f'(\psi _{1,2,3})|~\rangle\nonumber \\
S_{23}\psi _0^{22+}|~\rangle &=+\left(\psi _0^{11+}-i\psi _0^{12+}\right)(\psi _0^{21+}\psi _0^{22+}-i) ~f'(\psi _{1,2,3})|~\rangle\nonumber \\
S_{23}\psi _0^{12+}\psi _0^{21+}\psi _0^{22+}|~\rangle &=-\left(\psi _0^{11+}\psi _0^{12+}-i\right)(i\psi _0^{21+}-\psi _0^{22+}) ~f'(\psi _{1,2,3})|~\rangle\nonumber \\ 
S_{23}\psi _0^{11+}\psi _0^{21+}\psi _0^{22+}|~\rangle &=+\left(\psi _0^{11+}\psi _0^{12+}+i\right)(\psi _0^{21+}+i\psi _0^{22+}) ~f'(\psi _{1,2,3})|~\rangle\nonumber \\
S_{23}\psi _0^{11+}\psi _0^{12+}\psi _0^{22+}|~\rangle &=+\left(i\psi _0^{11+}-\psi _0^{12+}\right)(\psi _0^{21+}\psi _0^{22+}-i) ~f'(\psi _{1,2,3})|~\rangle\nonumber \\
S_{23}\psi _0^{11+}\psi _0^{12+}\psi _0^{21+}|~\rangle &=-\left(\psi _0^{11+}+i\psi _0^{12+}\right)(\psi _0^{21+}\psi _0^{22+}+i) ~f'(\psi _{1,2,3})|~\rangle
\end{align}
where we have defined the function $f'(\psi _{1,2,3})$ above. Computing the action of $S_{23}$ on the singlets of $(p_{2,4},p_{2,4})$ is now straightforward. The explicit expressions are as follows:
\begin{align}
-&\frac{1}{4} S_{23}|(12,21,22),(12,21,22),(11),(11)\rangle  \nonumber \\
=&|(12,21,22),(12,21,22),(11),(11)\rangle +|(12,21,22),(11),(12,21,22),(11)\rangle \nonumber \\
+&|(11),(12,21,22),(11),(12,21,22)\rangle +|(11),(11),(12,21,22),(12,21,22)\rangle \nonumber \\
-&|(11),(12),(11,21,22),(12,21,22)\rangle -|(12,21,22),(11,21,22),(12),(11)\rangle \nonumber \\
-&|(12,21,22),(12),(11,21,22),(11)\rangle -|(12),(12,21,22),(11),(11,21,22)\rangle \nonumber \\
-i&|(11,21,22),(12,21,22),(11),(11)\rangle -i|(11,21,22),(11),(12,21,22),(11)\rangle \nonumber \\
+i&|(12),(12,21,22),(11),(12,21,22)\rangle +i|(12),(11),(12,21,22),(12,21,22)\rangle \nonumber \\
-i&|(12,21,22),(11,21,22),(11),(11)\rangle +i|(12,21,22),(12),(12,21,22),(11)\rangle \nonumber \\
-i&|(11),(11,21,22),(11),(12,21,22)\rangle +i|(11),(11),(12,21,22),(11,21,22)\rangle \\
&\hspace{48 mm} \diamondsuit \hspace{10 mm}\diamondsuit \hspace{10 mm}\diamondsuit \nonumber \\
-&\frac{1}{4}S_{23}|(12,21,22),(11),(12,21,22),(11)\rangle  \nonumber \\
=&|(12,21,22),(12,21,22),(11),(11)\rangle +|(12,21,22),(11),(12,21,22),(11)\rangle \nonumber \\
+&|(11),(12,21,22),(11),(12,21,22)\rangle +|(11),(11),(12,21,22),(12,21,22)\rangle \nonumber \\
+&|(11),(12),(11,21,22),(12,21,22)\rangle +|(12,21,22),(11,21,22),(12),(11)\rangle \nonumber \\
+&|(12,21,22),(12),(11,21,22),(11)\rangle +|(12),(12,21,22),(11),(11,21,22)\rangle \nonumber \\
-i&|(11,21,22),(12,21,22),(11),(11)\rangle -i|(11,21,22),(11),(12,21,22),(11)\rangle \nonumber \\
+i&|(12),(12,21,22),(11),(12,21,22)\rangle +i|(12),(11),(12,21,22),(12,21,22)\rangle \nonumber \\
+i&|(12,21,22),(11,21,22),(11),(11)\rangle -i|(12,21,22),(12),(12,21,22),(11)\rangle \nonumber \\
+i&|(11),(11,21,22),(11),(12,21,22)\rangle -i|(11),(11),(12,21,22),(11,21,22)\rangle \\
&\hspace{48 mm} \diamondsuit \hspace{10 mm}\diamondsuit \hspace{10 mm}\diamondsuit \nonumber \\
-&\frac{1}{4}S_{23}|(11),(11),(12,21,22),(12,21,22)\rangle  \nonumber \\
=&|(12,21,22),(12,21,22),(11),(11)\rangle +|(12,21,22),(11),(12,21,22),(11)\rangle \nonumber \\
+&|(11),(12,21,22),(11),(12,21,22)\rangle +|(11),(11),(12,21,22),(12,21,22)\rangle \nonumber \\
-&|(11),(12),(11,21,22),(12,21,22)\rangle -|(12,21,22),(11,21,22),(12),(11)\rangle \nonumber \\
-&|(12,21,22),(12),(11,21,22),(11)\rangle -|(12),(12,21,22),(11),(11,21,22)\rangle \nonumber \\
+i&|(11,21,22),(12,21,22),(11),(11)\rangle +i|(11,21,22),(11),(12,21,22),(11)\rangle \nonumber \\
-i&|(12),(12,21,22),(11),(12,21,22)\rangle -i|(12),(11),(12,21,22),(12,21,22)\rangle \nonumber \\
+i&|(12,21,22),(11,21,22),(11),(11)\rangle -i|(12,21,22),(12),(12,21,22),(11)\rangle \nonumber \\
+i&|(11),(11,21,22),(11),(12,21,22)\rangle -i|(11),(11),(12,21,22),(11,21,22)\rangle \\
&\hspace{48 mm}\diamondsuit \hspace{10 mm}\diamondsuit \hspace{10 mm}\diamondsuit \nonumber \\
-&\frac{1}{4}S_{23}|(11),(12,21,22),(11),(12,21,22)\rangle  \nonumber \\
=&|(12,21,22),(12,21,22),(11),(11)\rangle +|(12,21,22),(11),(12,21,22),(11)\rangle \nonumber \\
+&|(11),(12,21,22),(11),(12,21,22)\rangle +|(11),(11),(12,21,22),(12,21,22)\rangle \nonumber \\
+&|(11),(12),(11,21,22),(12,21,22)\rangle +|(12,21,22),(11,21,22),(12),(11)\rangle \nonumber \\
+&|(12,21,22),(12),(11,21,22),(11)\rangle +|(12),(12,21,22),(11),(11,21,22)\rangle \nonumber \\
+i&|(11,21,22),(12,21,22),(11),(11)\rangle +i|(11,21,22),(11),(12,21,22),(11)\rangle \nonumber \\
-i&|(12),(12,21,22),(11),(12,21,22)\rangle -i|(12),(11),(12,21,22),(12,21,22)\rangle \nonumber \\
-i&|(12,21,22),(11,21,22),(11),(11)\rangle +i|(12,21,22),(12),(12,21,22),(11)\rangle \nonumber \\
-i&|(11),(11,21,22),(11),(12,21,22)\rangle +i|(11),(11),(12,21,22),(11,21,22)\rangle
\end{align}
Using these expressions, it is straightforward to show that under the action of $S_{23}S_A$, $S_{23}S_{03}$ or $S_{23}S_{12}$, the ground state transforms into itself. 

The rest of the discrete symmetry operators are not independent and can be constructed using the operators we have considered so far. So, the information we have is sufficient to show that the ground state is unique under all the symmetries we have identified.   

We now explain the degeneracy of the eigenvalue $+4\sqrt{3}$. This eigenvalue appears in the sets 1,15 and 16. The $+4\sqrt{3}$ eigenstates in the 15 and 16 sets transform into each other under the action of the operator $S_{12}S_A$. We can further show that under the action of $S_{23}S_A$, the $+4\sqrt{3}$ eigenstates of set 1 and set 16 transform into each other. The action of other operators does not lead to any other new states. That is, we find that the degeneracy of the eigenvalue $+4\sqrt{3}$ is three.

In a similar way, we can explain the degeneracies of all the eigenvalues except $0$ and $\pm 2\sqrt{2}$. For these exceptions, using our symmetry operators, we can explain the degeneracies partially. By that we mean that there are states having same\footnote{As an example, consider the first states of sets 1, 2 and 11. All of them have zero eigenvalue and are not related by symmetries. It is good to keep this example in mind for the rest of the discussion.} eigenvalue which are not related via any of the symmetries that we have identified. Thus, we need to find some other symmetries to explain all the degeneracies. One of the features of our symmetry operators is that the singlets of $(p_{2,4},p_{2,4})$ do not mix with the singlets of the other groups under any of our symmetry operators. Also, the singlets (21)-(36) of $(p_3,p_3)$ group transform among themselves under the action of our symmetries. So, the extra symmetries we need to identify should go beyond these features in order to  explain the observed degeneracies. The known symmetry operators treat the fermions of all the colors on an equal footing. We believe that the new symmetry operator(s) that can explain all the degeneracies would mix the various colors. One more point we note is that the discrete symmetry coming from the extra $Z_2$ in the $(O(2)$ vs. $SO(2)$ case is also not sufficient to explain the complete set of degeneracies.

\section{Chaos in the gauged $n=2$ Gurau-Witten model}

In the previous sections, we have identified the gauge spectrum of $n=2$ Gurau-Witten model explicitly. Now, we investigate whether there are any signs of chaos in the gauged sector. Even though the number of distinct eigenvalues is small, we find that the spectral form factor has a dip-ramp-plateau structure indicating the signs of chaos.

\begin{figure}
\centering
\includegraphics[scale=1]{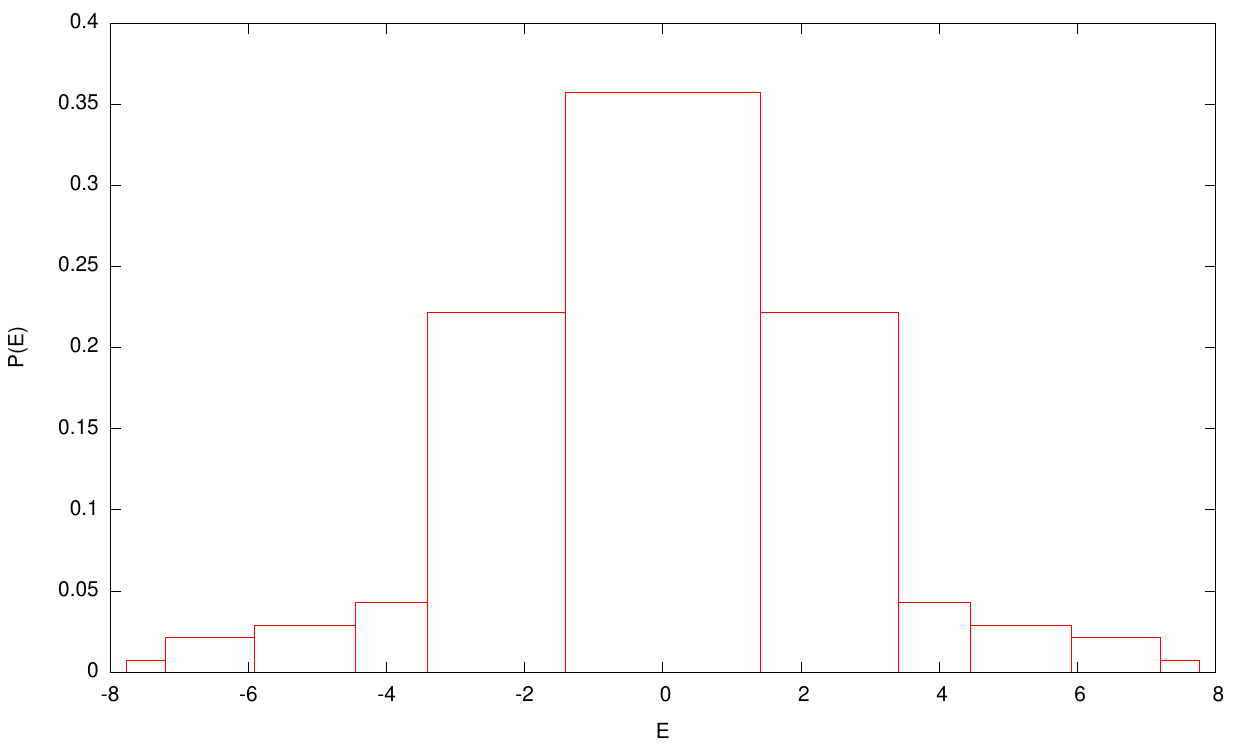}
\caption{Density of states for the singlet spectrum of $n=2$ Gurau-Witten model}
\label{DOS}
\end{figure}

Before investigating chaos, let us first understand the eigenvalue spectrum of the gauged model. There are 11 distinct eigenvalues in the spectrum. There is a large degeneracy at zero energy and the spectrum has spectral mirror symmetry as is obvious from the plot of density of eigenvalues in figure \ref{DOS}. Note that all the eigenvalues in the table \ref{evalues} are present in the numerical diagonalization as well and this provides a non-trivial check of our results. 

\begin{figure}
\centering
\includegraphics[scale=1]{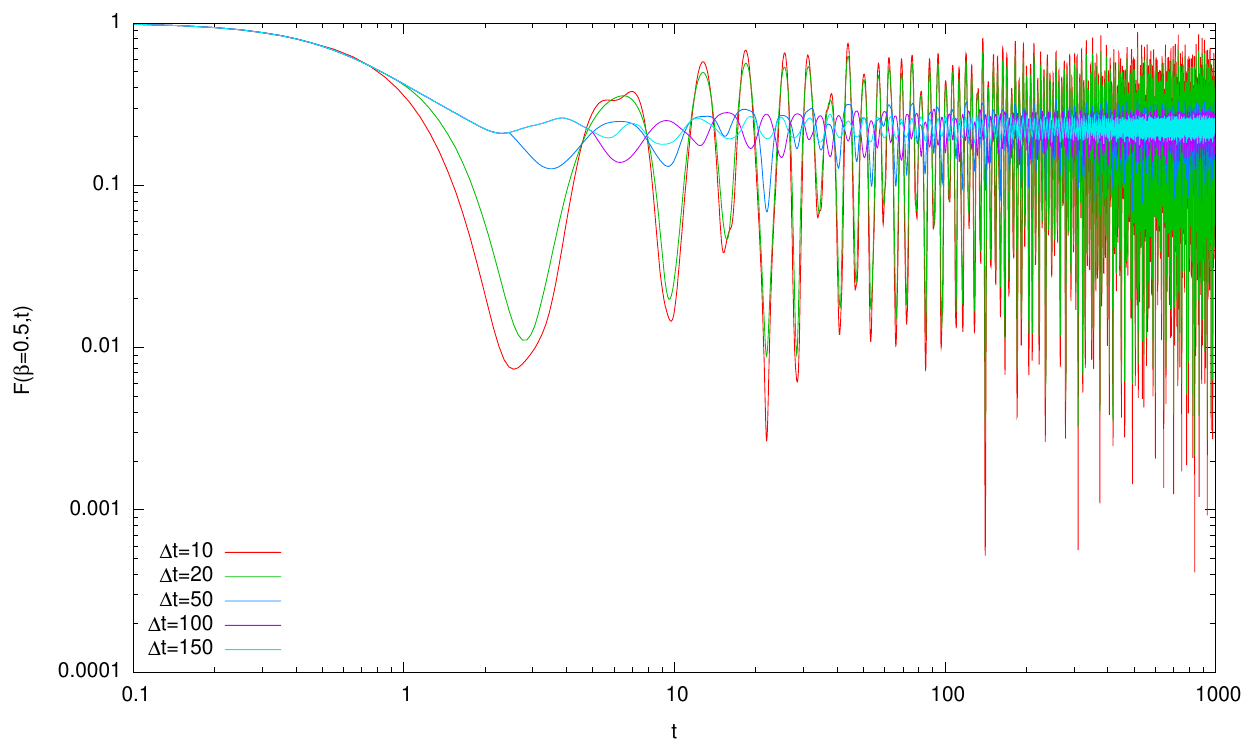}
\caption{SFF for the singlet spectrum of $n=2$ Gurau-Witten model for $\beta =0.5$}
\label{SFF}
\end{figure}

The tool we use to investigate chaos is the spectral form factor (SFF). It is defined as:
\begin{align}
F(\beta ,t)=\left|\frac{Z(\beta ,t)}{Z(\beta ,0)}\right|^2; ~~ Z(\beta ,t)=\text{Tr}\left(e^{-(\beta +it)H}\right)
\end{align}
For chaotic systems, SFF initially decays up to a certain time called the dip-time $(t_d)$.  After that, it starts raising until the plateau time $(t_p)$ and then finally stabilizes to a value called the plateau height. That is, the SFF of chaotic systems have a dip-ramp-plateau structure. We compute the SFF for the singlet spectrum  and report it in the figure \ref{SFF} after a sliding time average with different sliding intervals $\Delta t$. Even though we have only 11 distinct eigenvalues, the SFF qualitatively has a dip-ramp-plateau structure which can be understood as a primitive sign of chaos.

Note here however two caveats: one is that since the number of eigenvalues is small in each sector, these statements are preliminary and need further study at higher $N$ to be conclusive. Secondly, the structure of the SFF here is suggestive of random matrix like behavior, but the precise connection to chaos in the sense of OTOCs needs further investigation. Note that these two ideas are logically not identical. The random matrix features, via late time behavior is what is best captured by the SFF \cite{dario, dario2}. This is further strengthened by the fact that, we see signs of level repulsion in the ungauged model at low $N$  \cite{bala, uncolored-numerical}.

Perhaps a ``conservative'' guess is that non-linear systems with hints of randomness (like ours) are chaotic in the sense of OTOCs as well\footnote{In linear (ie., quadratic Hamiltonian) systems, clearly randomness in the parameters of the Hamiltonian cannot create chaos.}. But since our system provides a rare example of a non-linear but solvable system, this is perhaps not an automatic conclusion. Our system provides the possibility of studying this interplay between randomness and OTOCs, in a fairly unique setting. See \cite{bala, uncolored-numerical}, for evidence in the ungauged (and therefore with larger number of eigenvalues) model for random matrix behavior in the form of level repulsion. It will therefore be interesting to see the connection between solvability, chaos and randomness in these systems for higher $N$.

Exact solutions of strongly coupled (fermionic) gauge theories should be useful in extending the understanding of eigenstate thermalization \cite{Sonner}, entanglement/entanglement entropy \cite{Trivedi} and evolution of complexity \cite{Rifath} in the context of holography.

\section*{Acknowledgments}

We thank Avinash Raju, Dario Rosa and P. N. Bala Subramanian for discussions and related collaborations.

\appendix

\section{List of Singlets-Method-I}
\label{method-I}

In this appendix, we list down all the independent singlets that we obtain using method-I described in section \ref{singlets of SO(n)}. Various numerical coefficients are added so that $a^{\text{th}}$  singlet (for any $a\in \{1,140\}$) in both the methods match.  

\subsection{Groups $(p_1,p_1)$, $(p_1,p_5)$, $(p_5,p_1)$ and $(p_5,p_5)$}

\begin{align*}
1&. ~\frac{1}{64}\delta _{i_1i_2}\delta _{i_3i_4}\delta _{i_5i_6}\delta _{i_7i_8}\epsilon _{j_1j_2}\epsilon _{j_3j_4}\epsilon _{j_5j_6}\epsilon _{j_7j_8}\psi _0^{i_1j_1+}\psi _0^{i_2j_2+}\psi _0^{i_3j_3+}\psi _0^{i_4j_4+}\psi _1^{i_5j_5+}\psi _1^{i_6j_6+}\psi _1^{i_7j_7+}\psi _1^{i_8j_8+}\hspace{10 mm}\\
2&. ~\frac{1}{64}\delta _{i_1i_2}\delta _{i_3i_4}\delta _{i_5i_6}\delta _{i_7i_8}\epsilon _{j_1j_2}\epsilon _{j_3j_4}\epsilon _{j_5j_6}\epsilon _{j_7j_8}\psi _0^{i_1j_1+}\psi _0^{i_2j_2+}\psi _0^{i_3j_3+}\psi _0^{i_4j_4+}\psi _2^{i_5j_5+}\psi _2^{i_6j_6+}\psi _2^{i_7j_7+}\psi _2^{i_8j_8+}\hspace{10 mm}\\
3&. ~\frac{1}{64}\delta _{i_1i_2}\delta _{i_3i_4}\delta _{i_5i_6}\delta _{i_7i_8}\epsilon _{j_1j_2}\epsilon _{j_3j_4}\epsilon _{j_5j_6}\epsilon _{j_7j_8}\psi _3^{i_1j_1+}\psi _3^{i_2j_2+}\psi _3^{i_3j_3+}\psi _3^{i_4j_4+}\psi _1^{i_5j_5+}\psi _1^{i_6j_6+}\psi _1^{i_7j_7+}\psi _1^{i_8j_8+}\hspace{10 mm}\\
4&. ~\frac{1}{64}\delta _{i_1i_2}\delta _{i_3i_4}\delta _{i_5i_6}\delta _{i_7i_8}\epsilon _{j_1j_2}\epsilon _{j_3j_4}\epsilon _{j_5j_6}\epsilon _{j_7j_8}\psi _3^{i_1j_1+}\psi _3^{i_2j_2+}\psi _3^{i_3j_3+}\psi _3^{i_4j_4+}\psi _2^{i_5j_5+}\psi _2^{i_6j_6+}\psi _2^{i_7j_7+}\psi _2^{i_8j_8+}\hspace{10 mm}\\
\end{align*}

\subsection{Groups $(p_1,p_3)$, $(p_5,p_3)$, $(p_3,p_1)$ and $(p_3,p_5)$}

\begin{align*}
5&. ~\frac{1}{32}\delta _{i_1i_2}\delta _{i_3i_4}\delta _{i_5i_6}\delta _{i_7i_8}\epsilon _{j_1j_2}\epsilon _{j_3j_4}\epsilon _{j_5j_6}\epsilon _{j_7j_8}\psi _0^{i_1j_1+}\psi _0^{i_2j_2+}\psi _0^{i_3j_3+}\psi _0^{i_4j_4+}\psi _1^{i_5j_5+}\psi _1^{i_6j_6+}\psi _2^{i_7j_7+}\psi _2^{i_8j_8+}\hspace{5 mm}\\
6&. ~\frac{1}{32}\delta _{i_1i_2}\delta _{i_3i_4}\delta _{i_5i_6}\epsilon _{i_7i_8}\epsilon _{j_1j_2}\epsilon _{j_3j_4}\epsilon _{j_5j_6}\delta _{j_7j_8}\psi _0^{i_1j_1+}\psi _0^{i_2j_2+}\psi _0^{i_3j_3+}\psi _0^{i_4j_4+}\psi _1^{i_5j_5+}\psi _1^{i_6j_6+}\psi _2^{i_7j_7+}\psi _2^{i_8j_8+}\hspace{5 mm}\\
7&. ~\frac{1}{32}\delta _{i_1i_2}\delta _{i_3i_4}\epsilon _{i_5i_6}\delta _{i_7i_8}\epsilon _{j_1j_2}\epsilon _{j_3j_4}\delta _{j_5j_6}\epsilon _{j_7j_8}\psi _0^{i_1j_1+}\psi _0^{i_2j_2+}\psi _0^{i_3j_3+}\psi _0^{i_4j_4+}\psi _1^{i_5j_5+}\psi _1^{i_6j_6+}\psi _2^{i_7j_7+}\psi _2^{i_8j_8+}\hspace{5 mm}\\
8&. ~\frac{1}{32}\delta _{i_1i_2}\delta _{i_3i_4}\epsilon _{i_5i_6}\epsilon _{i_7i_8}\epsilon _{j_1j_2}\epsilon _{j_3j_4}\delta _{j_5j_6}\delta _{j_7j_8}\psi _0^{i_1j_1+}\psi _0^{i_2j_2+}\psi _0^{i_3j_3+}\psi _0^{i_4j_4+}\psi _1^{i_5j_5+}\psi _1^{i_6j_6+}\psi _2^{i_7j_7+}\psi _2^{i_8j_8+}\hspace{5 mm}\\
&\hspace{48 mm} \diamondsuit \hspace{10 mm}\diamondsuit \hspace{10 mm}\diamondsuit \nonumber \\
9&. ~\frac{1}{32}\delta _{i_1i_2}\delta _{i_3i_4}\delta _{i_5i_6}\delta _{i_7i_8}\epsilon _{j_1j_2}\epsilon _{j_3j_4}\epsilon _{j_5j_6}\epsilon _{j_7j_8}\psi _3^{i_1j_1+}\psi _3^{i_2j_2+}\psi _3^{i_3j_3+}\psi _3^{i_4j_4+}\psi _1^{i_5j_5+}\psi _1^{i_6j_6+}\psi _2^{i_7j_7+}\psi _2^{i_8j_8+}\hspace{5 mm}\\
10&. ~\frac{1}{32}\delta _{i_1i_2}\delta _{i_3i_4}\delta _{i_5i_6}\epsilon _{i_7i_8}\epsilon _{j_1j_2}\epsilon _{j_3j_4}\epsilon _{j_5j_6}\delta _{j_7j_8}\psi _3^{i_1j_1+}\psi _3^{i_2j_2+}\psi _3^{i_3j_3+}\psi _3^{i_4j_4+}\psi _1^{i_5j_5+}\psi _1^{i_6j_6+}\psi _2^{i_7j_7+}\psi _2^{i_8j_8+}\hspace{5 mm}\\
11&. ~\frac{1}{32}\delta _{i_1i_2}\delta _{i_3i_4}\epsilon _{i_5i_6}\delta _{i_7i_8}\epsilon _{j_1j_2}\epsilon _{j_3j_4}\delta _{j_5j_6}\epsilon _{j_7j_8}\psi _3^{i_1j_1+}\psi _3^{i_2j_2+}\psi _3^{i_3j_3+}\psi _3^{i_4j_4+}\psi _1^{i_5j_5+}\psi _1^{i_6j_6+}\psi _2^{i_7j_7+}\psi _2^{i_8j_8+}\hspace{5 mm}\\
12&. ~\frac{1}{32}\delta _{i_1i_2}\delta _{i_3i_4}\epsilon _{i_5i_6}\epsilon _{i_7i_8}\epsilon _{j_1j_2}\epsilon _{j_3j_4}\delta _{j_5j_6}\delta _{j_7j_8}\psi _3^{i_1j_1+}\psi _3^{i_2j_2+}\psi _3^{i_3j_3+}\psi _3^{i_4j_4+}\psi _1^{i_5j_5+}\psi _1^{i_6j_6+}\psi _2^{i_7j_7+}\psi _2^{i_8j_8+}\hspace{5 mm}\\
&\hspace{48 mm} \diamondsuit \hspace{10 mm}\diamondsuit \hspace{10 mm}\diamondsuit \nonumber \\
13&. ~\frac{1}{32}\delta _{i_1i_2}\delta _{i_3i_4}\delta _{i_5i_6}\delta _{i_7i_8}\epsilon _{j_1j_2}\epsilon _{j_3j_4}\epsilon _{j_5j_6}\epsilon _{j_7j_8}\psi _1^{i_1j_1+}\psi _1^{i_2j_2+}\psi _1^{i_3j_3+}\psi _1^{i_4j_4+}\psi _0^{i_5j_5+}\psi _0^{i_6j_6+}\psi _3^{i_7j_7+}\psi _3^{i_8j_8+}\hspace{5 mm}\\
14&. ~\frac{1}{32}\delta _{i_1i_2}\delta _{i_3i_4}\delta _{i_5i_6}\epsilon _{i_7i_8}\epsilon _{j_1j_2}\epsilon _{j_3j_4}\epsilon _{j_5j_6}\delta _{j_7j_8}\psi _1^{i_1j_1+}\psi _1^{i_2j_2+}\psi _1^{i_3j_3+}\psi _1^{i_4j_4+}\psi _0^{i_5j_5+}\psi _0^{i_6j_6+}\psi _3^{i_7j_7+}\psi _3^{i_8j_8+}\hspace{5 mm}\\
15&. ~\frac{1}{32}\delta _{i_1i_2}\delta _{i_3i_4}\epsilon _{i_5i_6}\delta _{i_7i_8}\epsilon _{j_1j_2}\epsilon _{j_3j_4}\delta _{j_5j_6}\epsilon _{j_7j_8}\psi _1^{i_1j_1+}\psi _1^{i_2j_2+}\psi _1^{i_3j_3+}\psi _1^{i_4j_4+}\psi _0^{i_5j_5+}\psi _0^{i_6j_6+}\psi _3^{i_7j_7+}\psi _3^{i_8j_8+}\hspace{5 mm}\\
16&. ~\frac{1}{32}\delta _{i_1i_2}\delta _{i_3i_4}\epsilon _{i_5i_6}\epsilon _{i_7i_8}\epsilon _{j_1j_2}\epsilon _{j_3j_4}\delta _{j_5j_6}\delta _{j_7j_8}\psi _1^{i_1j_1+}\psi _1^{i_2j_2+}\psi _1^{i_3j_3+}\psi _1^{i_4j_4+}\psi _0^{i_5j_5+}\psi _0^{i_6j_6+}\psi _3^{i_7j_7+}\psi _3^{i_8j_8+}\hspace{5 mm}\\
&\hspace{48 mm} \diamondsuit \hspace{10 mm}\diamondsuit \hspace{10 mm}\diamondsuit \nonumber \\
17&. ~\frac{1}{32}\delta _{i_1i_2}\delta _{i_3i_4}\delta _{i_5i_6}\delta _{i_7i_8}\epsilon _{j_1j_2}\epsilon _{j_3j_4}\epsilon _{j_5j_6}\epsilon _{j_7j_8}\psi _2^{i_1j_1+}\psi _2^{i_2j_2+}\psi _2^{i_3j_3+}\psi _2^{i_4j_4+}\psi _0^{i_5j_5+}\psi _0^{i_6j_6+}\psi _3^{i_7j_7+}\psi _3^{i_8j_8+}\hspace{5 mm}\\
18&. ~\frac{1}{32}\delta _{i_1i_2}\delta _{i_3i_4}\delta _{i_5i_6}\epsilon _{i_7i_8}\epsilon _{j_1j_2}\epsilon _{j_3j_4}\epsilon _{j_5j_6}\delta _{j_7j_8}\psi _2^{i_1j_1+}\psi _2^{i_2j_2+}\psi _2^{i_3j_3+}\psi _2^{i_4j_4+}\psi _0^{i_5j_5+}\psi _0^{i_6j_6+}\psi _3^{i_7j_7+}\psi _3^{i_8j_8+}\hspace{5 mm}\\
19&. ~\frac{1}{32}\delta _{i_1i_2}\delta _{i_3i_4}\epsilon _{i_5i_6}\delta _{i_7i_8}\epsilon _{j_1j_2}\epsilon _{j_3j_4}\delta _{j_5j_6}\epsilon _{j_7j_8}\psi _2^{i_1j_1+}\psi _2^{i_2j_2+}\psi _2^{i_3j_3+}\psi _2^{i_4j_4+}\psi _0^{i_5j_5+}\psi _0^{i_6j_6+}\psi _3^{i_7j_7+}\psi _3^{i_8j_8+}\hspace{5 mm}\\
20&. ~\frac{1}{32}\delta _{i_1i_2}\delta _{i_3i_4}\epsilon _{i_5i_6}\epsilon _{i_7i_8}\epsilon _{j_1j_2}\epsilon _{j_3j_4}\delta _{j_5j_6}\delta _{j_7j_8}\psi _2^{i_1j_1+}\psi _2^{i_2j_2+}\psi _2^{i_3j_3+}\psi _2^{i_4j_4+}\psi _0^{i_5j_5+}\psi _0^{i_6j_6+}\psi _3^{i_7j_7+}\psi _3^{i_8j_8+}
\end{align*}

\subsection{Group $(p_3,p_3)$}

\subsubsection{Type-I}
\begin{align*}
21&. ~\frac{1}{16}\delta _{i_1i_2}\delta _{i_3i_4}\delta _{i_5i_6}\delta _{i_7i_8}\epsilon _{j_1j_2}\epsilon _{j_3j_4}\epsilon _{j_5j_6}\epsilon _{j_7j_8}\psi _0^{i_1j_1+}\psi _0^{i_2j_2+}\psi _1^{i_3j_3+}\psi _1^{i_4j_4+}\psi _2^{i_5j_5+}\psi _2^{i_6j_6+}\psi _3^{i_7j_7+}\psi _3^{i_8j_8+}\\
22&. ~\frac{1}{16}\delta _{i_1i_2}\delta _{i_3i_4}\delta _{i_5i_6}\epsilon _{i_7i_8}\epsilon _{j_1j_2}\epsilon _{j_3j_4}\epsilon _{j_5j_6}\delta _{j_7j_8}\psi _0^{i_1j_1+}\psi _0^{i_2j_2+}\psi _1^{i_3j_3+}\psi _1^{i_4j_4+}\psi _2^{i_5j_5+}\psi _2^{i_6j_6+}\psi _3^{i_7j_7+}\psi _3^{i_8j_8+}\\
23&. ~\frac{1}{16}\delta _{i_1i_2}\delta _{i_3i_4}\epsilon _{i_5i_6}\delta _{i_7i_8}\epsilon _{j_1j_2}\epsilon _{j_3j_4}\delta _{j_5j_6}\epsilon _{j_7j_8}\psi _0^{i_1j_1+}\psi _0^{i_2j_2+}\psi _1^{i_3j_3+}\psi _1^{i_4j_4+}\psi _2^{i_5j_5+}\psi _2^{i_6j_6+}\psi _3^{i_7j_7+}\psi _3^{i_8j_8+}\\
24&. ~\frac{1}{16}\delta _{i_1i_2}\epsilon _{i_3i_4}\delta _{i_5i_6}\delta _{i_7i_8}\epsilon _{j_1j_2}\delta _{j_3j_4}\epsilon _{j_5j_6}\epsilon _{j_7j_8}\psi _0^{i_1j_1+}\psi _0^{i_2j_2+}\psi _1^{i_3j_3+}\psi _1^{i_4j_4+}\psi _2^{i_5j_5+}\psi _2^{i_6j_6+}\psi _3^{i_7j_7+}\psi _3^{i_8j_8+}\\
25&. ~\frac{1}{16}\epsilon _{i_1i_2}\delta _{i_3i_4}\delta _{i_5i_6}\delta _{i_7i_8}\delta _{j_1j_2}\epsilon _{j_3j_4}\epsilon _{j_5j_6}\epsilon _{j_7j_8}\psi _0^{i_1j_1+}\psi _0^{i_2j_2+}\psi _1^{i_3j_3+}\psi _1^{i_4j_4+}\psi _2^{i_5j_5+}\psi _2^{i_6j_6+}\psi _3^{i_7j_7+}\psi _3^{i_8j_8+}\\
26&. ~\frac{1}{16}\delta _{i_1i_2}\delta _{i_3i_4}\epsilon _{i_5i_6}\epsilon _{i_7i_8}\epsilon _{j_1j_2}\epsilon _{j_3j_4}\delta _{j_5j_6}\delta _{j_7j_8}\psi _0^{i_1j_1+}\psi _0^{i_2j_2+}\psi _1^{i_3j_3+}\psi _1^{i_4j_4+}\psi _2^{i_5j_5+}\psi _2^{i_6j_6+}\psi _3^{i_7j_7+}\psi _3^{i_8j_8+}\\
27&. ~\frac{1}{16}\delta _{i_1i_2}\epsilon _{i_3i_4}\delta _{i_5i_6}\epsilon _{i_7i_8}\epsilon _{j_1j_2}\delta _{j_3j_4}\epsilon _{j_5j_6}\delta _{j_7j_8}\psi _0^{i_1j_1+}\psi _0^{i_2j_2+}\psi _1^{i_3j_3+}\psi _1^{i_4j_4+}\psi _2^{i_5j_5+}\psi _2^{i_6j_6+}\psi _3^{i_7j_7+}\psi _3^{i_8j_8+}\\
28&. ~\frac{1}{16}\epsilon _{i_1i_2}\delta _{i_3i_4}\delta _{i_5i_6}\epsilon _{i_7i_8}\delta _{j_1j_2}\epsilon _{j_3j_4}\epsilon _{j_5j_6}\delta _{j_7j_8}\psi _0^{i_1j_1+}\psi _0^{i_2j_2+}\psi _1^{i_3j_3+}\psi _1^{i_4j_4+}\psi _2^{i_5j_5+}\psi _2^{i_6j_6+}\psi _3^{i_7j_7+}\psi _3^{i_8j_8+}\\
29&. ~\frac{1}{16}\delta _{i_1i_2}\epsilon _{i_3i_4}\epsilon _{i_5i_6}\delta _{i_7i_8}\epsilon _{j_1j_2}\delta _{j_3j_4}\delta _{j_5j_6}\epsilon _{j_7j_8}\psi _0^{i_1j_1+}\psi _0^{i_2j_2+}\psi _1^{i_3j_3+}\psi _1^{i_4j_4+}\psi _2^{i_5j_5+}\psi _2^{i_6j_6+}\psi _3^{i_7j_7+}\psi _3^{i_8j_8+}\\
30&. ~\frac{1}{16}\epsilon _{i_1i_2}\delta _{i_3i_4}\epsilon _{i_5i_6}\delta _{i_7i_8}\delta _{j_1j_2}\epsilon _{j_3j_4}\delta _{j_5j_6}\epsilon _{j_7j_8}\psi _0^{i_1j_1+}\psi _0^{i_2j_2+}\psi _1^{i_3j_3+}\psi _1^{i_4j_4+}\psi _2^{i_5j_5+}\psi _2^{i_6j_6+}\psi _3^{i_7j_7+}\psi _3^{i_8j_8+}\\
31&. ~\frac{1}{16}\epsilon _{i_1i_2}\epsilon _{i_3i_4}\delta _{i_5i_6}\delta _{i_7i_8}\delta _{j_1j_2}\delta _{j_3j_4}\epsilon _{j_5j_6}\epsilon _{j_7j_8}\psi _0^{i_1j_1+}\psi _0^{i_2j_2+}\psi _1^{i_3j_3+}\psi _1^{i_4j_4+}\psi _2^{i_5j_5+}\psi _2^{i_6j_6+}\psi _3^{i_7j_7+}\psi _3^{i_8j_8+}\\
32&. ~\frac{1}{16}\delta _{i_1i_2}\epsilon _{i_3i_4}\epsilon _{i_5i_6}\epsilon _{i_7i_8}\epsilon _{j_1j_2}\delta _{j_3j_4}\delta _{j_5j_6}\delta _{j_7j_8}\psi _0^{i_1j_1+}\psi _0^{i_2j_2+}\psi _1^{i_3j_3+}\psi _1^{i_4j_4+}\psi _2^{i_5j_5+}\psi _2^{i_6j_6+}\psi _3^{i_7j_7+}\psi _3^{i_8j_8+}\\
33&. ~\frac{1}{16}\epsilon _{i_1i_2}\delta _{i_3i_4}\epsilon _{i_5i_6}\epsilon _{i_7i_8}\delta _{j_1j_2}\epsilon _{j_3j_4}\delta _{j_5j_6}\delta _{j_7j_8}\psi _0^{i_1j_1+}\psi _0^{i_2j_2+}\psi _1^{i_3j_3+}\psi _1^{i_4j_4+}\psi _2^{i_5j_5+}\psi _2^{i_6j_6+}\psi _3^{i_7j_7+}\psi _3^{i_8j_8+}\\
34&. ~\frac{1}{16}\epsilon _{i_1i_2}\epsilon _{i_3i_4}\delta _{i_5i_6}\epsilon _{i_7i_8}\delta _{j_1j_2}\delta _{j_3j_4}\epsilon _{j_5j_6}\delta _{j_7j_8}\psi _0^{i_1j_1+}\psi _0^{i_2j_2+}\psi _1^{i_3j_3+}\psi _1^{i_4j_4+}\psi _2^{i_5j_5+}\psi _2^{i_6j_6+}\psi _3^{i_7j_7+}\psi _3^{i_8j_8+}\\
35&. ~\frac{1}{16}\epsilon _{i_1i_2}\epsilon _{i_3i_4}\epsilon _{i_5i_6}\delta _{i_7i_8}\delta _{j_1j_2}\delta _{j_3j_4}\delta _{j_5j_6}\epsilon _{j_7j_8}\psi _0^{i_1j_1+}\psi _0^{i_2j_2+}\psi _1^{i_3j_3+}\psi _1^{i_4j_4+}\psi _2^{i_5j_5+}\psi _2^{i_6j_6+}\psi _3^{i_7j_7+}\psi _3^{i_8j_8+}\\
36&. ~\frac{1}{16}\epsilon _{i_1i_2}\epsilon _{i_3i_4}\epsilon _{i_5i_6}\epsilon _{i_7i_8}\delta _{j_1j_2}\delta _{j_3j_4}\delta _{j_5j_6}\delta _{j_7j_8}\psi _0^{i_1j_1+}\psi _0^{i_2j_2+}\psi _1^{i_3j_3+}\psi _1^{i_4j_4+}\psi _2^{i_5j_5+}\psi _2^{i_6j_6+}\psi _3^{i_7j_7+}\psi _3^{i_8j_8+}
\end{align*}

\subsubsection{Type-II}

\begin{align*}
37&. ~\frac{1}{8}~\epsilon _{i_1i_2}\epsilon _{i_5i_6}\delta _{i_3i_4}\delta _{i_7i_8}\left(\delta _{j_1j_5}\delta _{j_2j_6}-\frac{1}{2}\delta _{j_1j_2}\delta _{j_5j_6}\right)\epsilon _{j_3j_4}\epsilon _{j_7j_8}\times\\
&\hspace{60 mm}\times\psi _0^{i_1j_1+}\psi _0^{i_2j_2+}\psi _1^{i_3j_3+}\psi _1^{i_4j_4+}\psi _2^{i_5j_5+}\psi _2^{i_6j_6+}\psi _3^{i_7j_7+}\psi _3^{i_8j_8+}\\
38&. ~\frac{1}{8}~\epsilon _{i_1i_2}\epsilon _{i_5i_6}\delta _{i_3i_4}\epsilon _{i_7i_8}\left(\delta _{j_1j_5}\delta _{j_2j_6}-\frac{1}{2}\delta _{j_1j_2}\delta _{j_5j_6}\right)\epsilon _{j_3j_4}\delta _{j_7j_8}\times\\
&\hspace{60 mm}\times\psi _0^{i_1j_1+}\psi _0^{i_2j_2+}\psi _1^{i_3j_3+}\psi _1^{i_4j_4+}\psi _2^{i_5j_5+}\psi _2^{i_6j_6+}\psi _3^{i_7j_7+}\psi _3^{i_8j_8+}\\
39&. ~\frac{1}{8}~\epsilon _{i_1i_2}\epsilon _{i_5i_6}\epsilon _{i_3i_4}\delta _{i_7i_8}\left(\delta _{j_1j_5}\delta _{j_2j_6}-\frac{1}{2}\delta _{j_1j_2}\delta _{j_5j_6}\right)\delta _{j_3j_4}\epsilon _{j_7j_8}\times\\
&\hspace{60 mm}\times\psi _0^{i_1j_1+}\psi _0^{i_2j_2+}\psi _1^{i_3j_3+}\psi _1^{i_4j_4+}\psi _2^{i_5j_5+}\psi _2^{i_6j_6+}\psi _3^{i_7j_7+}\psi _3^{i_8j_8+}\\
40&. ~\frac{1}{8}~\epsilon _{i_1i_2}\epsilon _{i_5i_6}\epsilon _{i_3i_4}\epsilon _{i_7i_8}\left(\delta _{j_1j_5}\delta _{j_2j_6}-\frac{1}{2}\delta _{j_1j_2}\delta _{j_5j_6}\right)\delta _{j_3j_4}\delta _{j_7j_8}\times\\
&\hspace{60 mm}\times\psi _0^{i_1j_1+}\psi _0^{i_2j_2+}\psi _1^{i_3j_3+}\psi _1^{i_4j_4+}\psi _2^{i_5j_5+}\psi _2^{i_6j_6+}\psi _3^{i_7j_7+}\psi _3^{i_8j_8+}\\
41&. ~\frac{1}{8}~\epsilon _{i_1i_2}\epsilon _{i_5i_6}\delta _{i_3i_4}\delta _{i_7i_8}\delta _{j_1j_5}\epsilon _{j_2j_6}\epsilon _{j_3j_4}\epsilon _{j_7j_8}\psi _0^{i_1j_1+}\psi _0^{i_2j_2+}\psi _1^{i_3j_3+}\psi _1^{i_4j_4+}\psi _2^{i_5j_5+}\psi _2^{i_6j_6+}\psi _3^{i_7j_7+}\psi _3^{i_8j_8+}\\
42&. ~\frac{1}{8}~\epsilon _{i_1i_2}\epsilon _{i_5i_6}\delta _{i_3i_4}\epsilon _{i_7i_8}\delta _{j_1j_5}\epsilon _{j_2j_6}\epsilon _{j_3j_4}\delta _{j_7j_8}\psi _0^{i_1j_1+}\psi _0^{i_2j_2+}\psi _1^{i_3j_3+}\psi _1^{i_4j_4+}\psi _2^{i_5j_5+}\psi _2^{i_6j_6+}\psi _3^{i_7j_7+}\psi _3^{i_8j_8+}\\
43&. ~\frac{1}{8}~\epsilon _{i_1i_2}\epsilon _{i_5i_6}\epsilon _{i_3i_4}\delta _{i_7i_8}\delta _{j_1j_5}\epsilon _{j_2j_6}\delta _{j_3j_4}\epsilon _{j_7j_8}\psi _0^{i_1j_1+}\psi _0^{i_2j_2+}\psi _1^{i_3j_3+}\psi _1^{i_4j_4+}\psi _2^{i_5j_5+}\psi _2^{i_6j_6+}\psi _3^{i_7j_7+}\psi _3^{i_8j_8+}\\
44&. ~\frac{1}{8}~\epsilon _{i_1i_2}\epsilon _{i_5i_6}\epsilon _{i_3i_4}\epsilon _{i_7i_8}\delta _{j_1j_5}\epsilon _{j_2j_6}\delta _{j_3j_4}\delta _{j_7j_8}\psi _0^{i_1j_1+}\psi _0^{i_2j_2+}\psi _1^{i_3j_3+}\psi _1^{i_4j_4+}\psi _2^{i_5j_5+}\psi _2^{i_6j_6+}\psi _3^{i_7j_7+}\psi _3^{i_8j_8+}
\end{align*}

\subsubsection{Type-III}

\begin{align*}
45&. ~\frac{1}{8}~\delta _{i_1i_2}\delta _{i_5i_6}\epsilon _{i_3i_4}\epsilon _{i_7i_8}\epsilon _{j_1j_2}\epsilon _{j_5j_6}\left(\delta _{j_3j_7}\delta _{j_4j_8}-\frac{1}{2}\delta _{j_3j_4}\delta _{j_7j_8}\right)\times\\
&\hspace{60 mm}\times\psi _0^{i_1j_1+}\psi _0^{i_2j_2+}\psi _1^{i_3j_3+}\psi _1^{i_4j_4+}\psi _2^{i_5j_5+}\psi _2^{i_6j_6+}\psi _3^{i_7j_7+}\psi _3^{i_8j_8+}\\
46&. ~\frac{1}{8}~\delta _{i_1i_2}\epsilon _{i_5i_6}\epsilon _{i_3i_4}\epsilon _{i_7i_8}\epsilon _{j_1j_2}\delta _{j_5j_6}\left(\delta _{j_3j_7}\delta _{j_4j_8}-\frac{1}{2}\delta _{j_3j_4}\delta _{j_7j_8}\right)\times\\
&\hspace{60 mm}\times\psi _0^{i_1j_1+}\psi _0^{i_2j_2+}\psi _1^{i_3j_3+}\psi _1^{i_4j_4+}\psi _2^{i_5j_5+}\psi _2^{i_6j_6+}\psi _3^{i_7j_7+}\psi _3^{i_8j_8+}\\
47&. ~\frac{1}{8}~\epsilon _{i_1i_2}\delta _{i_5i_6}\epsilon _{i_3i_4}\epsilon _{i_7i_8}\delta _{j_1j_2}\epsilon _{j_5j_6}\left(\delta _{j_3j_7}\delta _{j_4j_8}-\frac{1}{2}\delta _{j_3j_4}\delta _{j_7j_8}\right)\times\\
&\hspace{60 mm}\times\psi _0^{i_1j_1+}\psi _0^{i_2j_2+}\psi _1^{i_3j_3+}\psi _1^{i_4j_4+}\psi _2^{i_5j_5+}\psi _2^{i_6j_6+}\psi _3^{i_7j_7+}\psi _3^{i_8j_8+}\\
48&. ~\frac{1}{8}~\epsilon _{i_1i_2}\epsilon _{i_5i_6}\epsilon _{i_3i_4}\epsilon _{i_7i_8}\delta _{j_1j_2}\delta _{j_5j_6}\left(\delta _{j_3j_7}\delta _{j_4j_8}-\frac{1}{2}\delta _{j_3j_4}\delta _{j_7j_8}\right)\times\\
&\hspace{60 mm}\times\psi _0^{i_1j_1+}\psi _0^{i_2j_2+}\psi _1^{i_3j_3+}\psi _1^{i_4j_4+}\psi _2^{i_5j_5+}\psi _2^{i_6j_6+}\psi _3^{i_7j_7+}\psi _3^{i_8j_8+}\\
49&.  ~\frac{1}{8}~\delta _{i_1i_2}\delta _{i_5i_6}\epsilon _{i_3i_4}\epsilon _{i_7i_8}\epsilon _{j_1j_2}\epsilon _{j_5j_6}\delta _{j_3j_7}\epsilon _{j_4j_8}\psi _0^{i_1j_1+}\psi _0^{i_2j_2+}\psi _1^{i_3j_3+}\psi _1^{i_4j_4+}\psi _2^{i_5j_5+}\psi _2^{i_6j_6+}\psi _3^{i_7j_7+}\psi _3^{i_8j_8+}\\
50&.  ~\frac{1}{8}~\delta _{i_1i_2}\epsilon _{i_5i_6}\epsilon _{i_3i_4}\epsilon _{i_7i_8}\epsilon _{j_1j_2}\delta _{j_5j_6}\delta _{j_3j_7}\epsilon _{j_4j_8}\psi _0^{i_1j_1+}\psi _0^{i_2j_2+}\psi _1^{i_3j_3+}\psi _1^{i_4j_4+}\psi _2^{i_5j_5+}\psi _2^{i_6j_6+}\psi _3^{i_7j_7+}\psi _3^{i_8j_8+}\\
51&.  ~\frac{1}{8}~\epsilon _{i_1i_2}\delta _{i_5i_6}\epsilon _{i_3i_4}\epsilon _{i_7i_8}\delta _{j_1j_2}\epsilon _{j_5j_6}\delta _{j_3j_7}\epsilon _{j_4j_8}\psi _0^{i_1j_1+}\psi _0^{i_2j_2+}\psi _1^{i_3j_3+}\psi _1^{i_4j_4+}\psi _2^{i_5j_5+}\psi _2^{i_6j_6+}\psi _3^{i_7j_7+}\psi _3^{i_8j_8+}\\
52&.  ~\frac{1}{8}~\epsilon _{i_1i_2}\epsilon _{i_5i_6}\epsilon _{i_3i_4}\epsilon _{i_7i_8}\delta _{j_1j_2}\delta _{j_5j_6}\delta _{j_3j_7}\epsilon _{j_4j_8}\psi _0^{i_1j_1+}\psi _0^{i_2j_2+}\psi _1^{i_3j_3+}\psi _1^{i_4j_4+}\psi _2^{i_5j_5+}\psi _2^{i_6j_6+}\psi _3^{i_7j_7+}\psi _3^{i_8j_8+}
\end{align*}
\subsubsection{Type-IV}

\begin{align*}
53&. ~\frac{1}{8}~\left(\delta _{i_1i_3}\delta _{i_2i_4}-\frac{1}{2}\delta _{i_1i_2}\delta _{i_3i_4}\right)\delta _{i_5i_6}\delta _{i_7i_8}\epsilon _{j_1j_2}\epsilon _{j_3j_4}\epsilon _{j_5j_6}\epsilon _{j_7j_8}\times\\
&\hspace{60 mm}\times\psi _0^{i_1j_1+}\psi _0^{i_2j_2+}\psi _1^{i_3j_3+}\psi _1^{i_4j_4+}\psi _2^{i_5j_5+}\psi _2^{i_6j_6+}\psi _3^{i_7j_7+}\psi _3^{i_8j_8+}\\
54&. ~\frac{1}{8}~\left(\delta _{i_1i_3}\delta _{i_2i_4}-\frac{1}{2}\delta _{i_1i_2}\delta _{i_3i_4}\right)\delta _{i_5i_6}\epsilon _{i_7i_8}\epsilon _{j_1j_2}\epsilon _{j_3j_4}\epsilon _{j_5j_6}\delta _{j_7j_8}\times\\
&\hspace{60 mm}\times\psi _0^{i_1j_1+}\psi _0^{i_2j_2+}\psi _1^{i_3j_3+}\psi _1^{i_4j_4+}\psi _2^{i_5j_5+}\psi _2^{i_6j_6+}\psi _3^{i_7j_7+}\psi _3^{i_8j_8+}\\
55&. ~\frac{1}{8}~\left(\delta _{i_1i_3}\delta _{i_2i_4}-\frac{1}{2}\delta _{i_1i_2}\delta _{i_3i_4}\right)\epsilon _{i_5i_6}\delta _{i_7i_8}\epsilon _{j_1j_2}\epsilon _{j_3j_4}\delta _{j_5j_6}\epsilon _{j_7j_8}\times\\
&\hspace{60 mm}\times\psi _0^{i_1j_1+}\psi _0^{i_2j_2+}\psi _1^{i_3j_3+}\psi _1^{i_4j_4+}\psi _2^{i_5j_5+}\psi _2^{i_6j_6+}\psi _3^{i_7j_7+}\psi _3^{i_8j_8+}\\
56&. ~\frac{1}{8}~\left(\delta _{i_1i_3}\delta _{i_2i_4}-\frac{1}{2}\delta _{i_1i_2}\delta _{i_3i_4}\right)\epsilon _{i_5i_6}\epsilon _{i_7i_8}\epsilon _{j_1j_2}\epsilon _{j_3j_4}\delta _{j_5j_6}\delta _{j_7j_8}\times\\
&\hspace{60 mm}\times\psi _0^{i_1j_1+}\psi _0^{i_2j_2+}\psi _1^{i_3j_3+}\psi _1^{i_4j_4+}\psi _2^{i_5j_5+}\psi _2^{i_6j_6+}\psi _3^{i_7j_7+}\psi _3^{i_8j_8+}\\
57&. ~\frac{1}{8}~\delta _{i_1i_3}\epsilon _{i_2i_4}\delta _{i_5i_6}\delta _{i_7i_8}\epsilon _{j_1j_2}\epsilon _{j_3j_4}\epsilon _{j_5j_6}\epsilon _{j_7j_8}\psi _0^{i_1j_1+}\psi _0^{i_2j_2+}\psi _1^{i_3j_3+}\psi _1^{i_4j_4+}\psi _2^{i_5j_5+}\psi _2^{i_6j_6+}\psi _3^{i_7j_7+}\psi _3^{i_8j_8+}\\
58&. ~\frac{1}{8}~\delta _{i_1i_3}\epsilon _{i_2i_4}\delta _{i_5i_6}\epsilon _{i_7i_8}\epsilon _{j_1j_2}\epsilon _{j_3j_4}\epsilon _{j_5j_6}\delta _{j_7j_8}\psi _0^{i_1j_1+}\psi _0^{i_2j_2+}\psi _1^{i_3j_3+}\psi _1^{i_4j_4+}\psi _2^{i_5j_5+}\psi _2^{i_6j_6+}\psi _3^{i_7j_7+}\psi _3^{i_8j_8+}\\
59&. ~\frac{1}{8}~\delta _{i_1i_3}\epsilon _{i_2i_4}\epsilon _{i_5i_6}\delta _{i_7i_8}\epsilon _{j_1j_2}\epsilon _{j_3j_4}\delta _{j_5j_6}\epsilon _{j_7j_8}\psi _0^{i_1j_1+}\psi _0^{i_2j_2+}\psi _1^{i_3j_3+}\psi _1^{i_4j_4+}\psi _2^{i_5j_5+}\psi _2^{i_6j_6+}\psi _3^{i_7j_7+}\psi _3^{i_8j_8+}\\
60&. ~\frac{1}{8}~\delta _{i_1i_3}\epsilon _{i_2i_4}\epsilon _{i_5i_6}\epsilon _{i_7i_8}\epsilon _{j_1j_2}\epsilon _{j_3j_4}\delta _{j_5j_6}\delta _{j_7j_8}\psi _0^{i_1j_1+}\psi _0^{i_2j_2+}\psi _1^{i_3j_3+}\psi _1^{i_4j_4+}\psi _2^{i_5j_5+}\psi _2^{i_6j_6+}\psi _3^{i_7j_7+}\psi _3^{i_8j_8+}
\end{align*}

\subsubsection{Type-V}

\begin{align*}
61&. ~\frac{1}{8}~\delta _{i_1i_2}\delta _{i_3i_4}\left(\delta _{i_5i_7}\delta _{i_6i_8}-\frac{1}{2}\delta _{i_5i_6}\delta _{i_7i_8}\right)\epsilon _{j_1j_2}\epsilon _{j_3j_4}\epsilon _{j_5j_6}\epsilon _{j_7j_8}\times\\
&\hspace{60 mm}\times\psi _0^{i_1j_1+}\psi _0^{i_2j_2+}\psi _1^{i_3j_3+}\psi _1^{i_4j_4+}\psi _2^{i_5j_5+}\psi _2^{i_6j_6+}\psi _3^{i_7j_7+}\psi _3^{i_8j_8+}\\
62&. ~\frac{1}{8}~\delta _{i_1i_2}\epsilon _{i_3i_4}\left(\delta _{i_5i_7}\delta _{i_6i_8}-\frac{1}{2}\delta _{i_5i_6}\delta _{i_7i_8}\right)\epsilon _{j_1j_2}\delta _{j_3j_4}\epsilon _{j_5j_6}\epsilon _{j_7j_8}\times\\
&\hspace{60 mm}\times\psi _0^{i_1j_1+}\psi _0^{i_2j_2+}\psi _1^{i_3j_3+}\psi _1^{i_4j_4+}\psi _2^{i_5j_5+}\psi _2^{i_6j_6+}\psi _3^{i_7j_7+}\psi _3^{i_8j_8+}\\
63&. ~\frac{1}{8}~\epsilon _{i_1i_2}\delta _{i_3i_4}\left(\delta _{i_5i_7}\delta _{i_6i_8}-\frac{1}{2}\delta _{i_5i_6}\delta _{i_7i_8}\right)\delta _{j_1j_2}\epsilon _{j_3j_4}\epsilon _{j_5j_6}\epsilon _{j_7j_8}\times\\
&\hspace{60 mm}\times\psi _0^{i_1j_1+}\psi _0^{i_2j_2+}\psi _1^{i_3j_3+}\psi _1^{i_4j_4+}\psi _2^{i_5j_5+}\psi _2^{i_6j_6+}\psi _3^{i_7j_7+}\psi _3^{i_8j_8+}\\
64&. ~\frac{1}{8}~\epsilon _{i_1i_2}\epsilon _{i_3i_4}\left(\delta _{i_5i_7}\delta _{i_6i_8}-\frac{1}{2}\delta _{i_5i_6}\delta _{i_7i_8}\right)\delta _{j_1j_2}\delta _{j_3j_4}\epsilon _{j_5j_6}\epsilon _{j_7j_8}\times\\
&\hspace{60 mm}\times\psi _0^{i_1j_1+}\psi _0^{i_2j_2+}\psi _1^{i_3j_3+}\psi _1^{i_4j_4+}\psi _2^{i_5j_5+}\psi _2^{i_6j_6+}\psi _3^{i_7j_7+}\psi _3^{i_8j_8+}\\
65&. ~\frac{1}{8}~\delta _{i_1i_2}\delta _{i_3i_4}\delta _{i_5i_7}\epsilon _{i_6i_8}\epsilon _{j_1j_2}\epsilon _{j_3j_4}\epsilon _{j_5j_6}\epsilon _{j_7j_8}\psi _0^{i_1j_1+}\psi _0^{i_2j_2+}\psi _1^{i_3j_3+}\psi _1^{i_4j_4+}\psi _2^{i_5j_5+}\psi _2^{i_6j_6+}\psi _3^{i_7j_7+}\psi _3^{i_8j_8+}\\
66&. ~\frac{1}{8}~\delta _{i_1i_2}\epsilon _{i_3i_4}\delta _{i_5i_7}\epsilon _{i_6i_8}\epsilon _{j_1j_2}\delta _{j_3j_4}\epsilon _{j_5j_6}\epsilon _{j_7j_8}\psi _0^{i_1j_1+}\psi _0^{i_2j_2+}\psi _1^{i_3j_3+}\psi _1^{i_4j_4+}\psi _2^{i_5j_5+}\psi _2^{i_6j_6+}\psi _3^{i_7j_7+}\psi _3^{i_8j_8+}\\
67&. ~\frac{1}{8}~\epsilon _{i_1i_2}\delta _{i_3i_4}\delta _{i_5i_7}\epsilon _{i_6i_8}\delta _{j_1j_2}\epsilon _{j_3j_4}\epsilon _{j_5j_6}\epsilon _{j_7j_8}\psi _0^{i_1j_1+}\psi _0^{i_2j_2+}\psi _1^{i_3j_3+}\psi _1^{i_4j_4+}\psi _2^{i_5j_5+}\psi _2^{i_6j_6+}\psi _3^{i_7j_7+}\psi _3^{i_8j_8+}\\
68&. ~\frac{1}{8}~\epsilon _{i_1i_2}\epsilon _{i_3i_4}\delta _{i_5i_7}\epsilon _{i_6i_8}\delta _{j_1j_2}\delta _{j_3j_4}\epsilon _{j_5j_6}\epsilon _{j_7j_8}\psi _0^{i_1j_1+}\psi _0^{i_2j_2+}\psi _1^{i_3j_3+}\psi _1^{i_4j_4+}\psi _2^{i_5j_5+}\psi _2^{i_6j_6+}\psi _3^{i_7j_7+}\psi _3^{i_8j_8+}
\end{align*}

\subsubsection{Type-VI}

\begin{align*}
69&. ~\frac{1}{4}~\epsilon _{i_1i_2}\epsilon _{i_5i_6}\epsilon _{i_3i_4}\epsilon _{i_7i_8}\left(\delta _{j_1j_5}\delta _{j_2j_6}-\frac{1}{2}\delta _{j_1j_2}\delta _{j_5j_6}\right)\left(\delta _{j_3j_7}\delta _{j_4j_8}-\frac{1}{2}\delta _{j_3j_4}\delta _{j_7j_8}\right)\times\\
&\hspace{60 mm}\times\psi _0^{i_1j_1+}\psi _0^{i_2j_2+}\psi _1^{i_3j_3+}\psi _1^{i_4j_4+}\psi _2^{i_5j_5+}\psi _2^{i_6j_6+}\psi _3^{i_7j_7+}\psi _3^{i_8j_8+}\\
70&.~\frac{1}{4}~\epsilon _{i_1i_2}\epsilon _{i_5i_6}\epsilon _{i_3i_4}\epsilon _{i_7i_8}\delta _{j_1j_5}\epsilon _{j_2j_6} \left(\delta _{j_3j_7}\delta _{j_4j_8}-\frac{1}{2}\delta _{j_3j_4}\delta _{j_7j_8}\right)\times\\
&\hspace{60 mm}\times\psi _0^{i_1j_1+}\psi _0^{i_2j_2+}\psi _1^{i_3j_3+}\psi _1^{i_4j_4+}\psi _2^{i_5j_5+}\psi _2^{i_6j_6+}\psi _3^{i_7j_7+}\psi _3^{i_8j_8+}\\
71&. ~\frac{1}{4}~\epsilon _{i_1i_2}\epsilon _{i_5i_6}\epsilon _{i_3i_4}\epsilon _{i_7i_8}\left(\delta _{j_1j_5}\delta _{j_2j_6}-\frac{1}{2}\delta _{j_1j_2}\delta _{j_5j_6}\right)\delta _{j_3j_7}\epsilon _{j_4j_8}\times\\
&\hspace{60 mm}\times\psi _0^{i_1j_1+}\psi _0^{i_2j_2+}\psi _1^{i_3j_3+}\psi _1^{i_4j_4+}\psi _2^{i_5j_5+}\psi _2^{i_6j_6+}\psi _3^{i_7j_7+}\psi _3^{i_8j_8+}\\
72&. ~\frac{1}{4}~\epsilon _{i_1i_2}\epsilon _{i_5i_6}\epsilon _{i_3i_4}\epsilon _{i_7i_8}\delta _{j_1j_5}\epsilon _{j_2j_6}\delta _{j_3j_7}\epsilon _{j_4j_8}\psi _0^{i_1j_1+}\psi _0^{i_2j_2+}\psi _1^{i_3j_3+}\psi _1^{i_4j_4+}\psi _2^{i_5j_5+}\psi _2^{i_6j_6+}\psi _3^{i_7j_7+}\psi _3^{i_8j_8+}\\
\end{align*}

\subsubsection{Type- VII}

\begin{align*}
73&.  ~\frac{1}{8}~\left(\delta _{i_1i_3}\delta _{i_2i_4}-\frac{1}{2}\delta _{i_1i_2}\delta _{i_3i_4}\right)\left(\delta _{i_5i_7}\delta _{i_6i_8}-\frac{1}{2}\delta _{i_5i_6}\delta _{i_7i_8}\right)\epsilon _{j_1j_2}\epsilon _{j_3j_4}\epsilon _{j_5j_6}\epsilon _{j_7j_8}\times\\
&\hspace{60 mm}\times\psi _0^{i_1j_1+}\psi _0^{i_2j_2+}\psi _1^{i_3j_3+}\psi _1^{i_4j_4+}\psi _2^{i_5j_5+}\psi _2^{i_6j_6+}\psi _3^{i_7j_7+}\psi _3^{i_8j_8+}\\
74&.  ~\frac{1}{8}~\left(\delta _{i_1i_3}\delta _{i_2i_4}-\frac{1}{2}\delta _{i_1i_2}\delta _{i_3i_4}\right)\delta _{i_5i_7}\epsilon _{i_6i_8}\epsilon _{j_1j_2}\epsilon _{j_3j_4}\epsilon _{j_5j_6}\epsilon _{j_7j_8}\times\\
&\hspace{60 mm}\times\psi _0^{i_1j_1+}\psi _0^{i_2j_2+}\psi _1^{i_3j_3+}\psi _1^{i_4j_4+}\psi _2^{i_5j_5+}\psi _2^{i_6j_6+}\psi _3^{i_7j_7+}\psi _3^{i_8j_8+}\\
75&.  ~\frac{1}{8}~\delta _{i_1i_3}\epsilon _{i_2i_4}\left(\delta _{i_5i_7}\delta _{i_6i_8}-\frac{1}{2}\delta _{i_5i_6}\delta _{i_7i_8}\right)\epsilon _{j_1j_2}\epsilon _{j_3j_4}\epsilon _{j_5j_6}\epsilon _{j_7j_8}\times\\
&\hspace{60 mm}\times\psi _0^{i_1j_1+}\psi _0^{i_2j_2+}\psi _1^{i_3j_3+}\psi _1^{i_4j_4+}\psi _2^{i_5j_5+}\psi _2^{i_6j_6+}\psi _3^{i_7j_7+}\psi _3^{i_8j_8+}\\
76&.  ~\frac{1}{8}~\delta _{i_1i_3}\epsilon _{i_2i_4}\delta _{i_5i_7}\epsilon _{i_6i_8}\epsilon _{j_1j_2}\epsilon _{j_3j_4}\epsilon _{j_5j_6}\epsilon _{j_7j_8}\psi _0^{i_1j_1+}\psi _0^{i_2j_2+}\psi _1^{i_3j_3+}\psi _1^{i_4j_4+}\psi _2^{i_5j_5+}\psi _2^{i_6j_6+}\psi _3^{i_7j_7+}\psi _3^{i_8j_8+}
\end{align*}

\subsection{Group $(p_2,p_2)$}

\begin{align*}
77&. -\delta _{i_1i_2}\delta _{i_3i_4}\delta _{i_5i_6}\delta _{i_7i_8}\delta _{j_1j_7}\delta _{j_2j_3}\delta _{j_4j_5}\delta _{j_6j_8}~\psi _0^{i_1j_1+}\psi _0^{i_2j_2+}\psi _0^{i_3j_3+}\psi _1^{i_4j_4+}\psi _1^{i_5j_5+}\psi _1^{i_6j_6+}\psi _2^{i_7j_7+}\psi _3^{i_8j_8+}\\
78&. +\epsilon _{i_1i_2}\delta _{i_3i_4}\delta _{i_5i_6}\epsilon _{i_7i_8}\epsilon _{j_1j_7}\delta _{j_2j_3}\delta _{j_4j_5}\epsilon _{j_6j_8}~\psi _0^{i_1j_1+}\psi _0^{i_2j_2+}\psi _0^{i_3j_3+}\psi _1^{i_4j_4+}\psi _1^{i_5j_5+}\psi _1^{i_6j_6+}\psi _2^{i_7j_7+}\psi _3^{i_8j_8+}\\
79&. +\delta _{i_1i_2}\delta _{i_3i_4}\delta _{i_5i_6}\delta _{i_7i_8}\delta _{j_1j_7}\epsilon _{j_2j_3}\delta _{j_4j_5}\delta _{j_6j_8}~\psi _0^{i_1j_1+}\psi _0^{i_2j_2+}\psi _0^{i_3j_3+}\psi _1^{i_4j_4+}\psi _1^{i_5j_5+}\psi _1^{i_6j_6+}\psi _2^{i_7j_7+}\psi _3^{i_8j_8+}\\
80&.+\epsilon _{i_1i_2}\delta _{i_3i_4}\delta _{i_5i_6}\delta _{i_7i_8}\delta _{j_1j_7}\delta _{j_2j_3}\delta _{j_4j_5}\delta _{j_6j_8}~\psi _0^{i_1j_1+}\psi _0^{i_2j_2+}\psi _0^{i_3j_3+}\psi _1^{i_4j_4+}\psi _1^{i_5j_5+}\psi _1^{i_6j_6+}\psi _2^{i_7j_7+}\psi _3^{i_8j_8+}\\
81&.  -\delta _{i_1i_2}\delta _{i_3i_4}\delta _{i_5i_6}\delta _{i_7i_8}\delta _{j_1j_7}\delta _{j_2j_3}\delta _{j_4j_5}\epsilon _{j_6j_8}~\psi _0^{i_1j_1+}\psi _0^{i_2j_2+}\psi _0^{i_3j_3+}\psi _1^{i_4j_4+}\psi _1^{i_5j_5+}\psi _1^{i_6j_6+}\psi _2^{i_7j_7+}\psi _3^{i_8j_8+}\\
82&. -\delta _{i_1i_2}\delta _{i_3i_4}\delta _{i_5i_6}\epsilon _{i_7i_8}\delta _{j_1j_7}\delta _{j_2j_3}\delta _{j_4j_5}\delta _{j_6j_8}~\psi _0^{i_1j_1+}\psi _0^{i_2j_2+}\psi _0^{i_3j_3+}\psi _1^{i_4j_4+}\psi _1^{i_5j_5+}\psi _1^{i_6j_6+}\psi _2^{i_7j_7+}\psi _3^{i_8j_8+}\\
83&. +\epsilon _{i_1i_2}\delta _{i_3i_4}\delta _{i_5i_6}\epsilon _{i_7i_8}\delta _{j_1j_7}\delta _{j_2j_3}\delta _{j_4j_5}\epsilon _{j_6j_8}~\psi _0^{i_1j_1+}\psi _0^{i_2j_2+}\psi _0^{i_3j_3+}\psi _1^{i_4j_4+}\psi _1^{i_5j_5+}\psi _1^{i_6j_6+}\psi _2^{i_7j_7+}\psi _3^{i_8j_8+}\\
84&. +\delta _{i_1i_2}\delta _{i_3i_4}\delta _{i_5i_6}\epsilon _{i_7i_8}\delta _{j_1j_7}\epsilon _{j_2j_3}\delta _{j_4j_5}\epsilon _{j_6j_8}~\psi _0^{i_1j_1+}\psi _0^{i_2j_2+}\psi _0^{i_3j_3+}\psi _1^{i_4j_4+}\psi _1^{i_5j_5+}\psi _1^{i_6j_6+}\psi _2^{i_7j_7+}\psi _3^{i_8j_8+}\\
85&. +\epsilon _{i_1i_2}\delta _{i_3i_4}\delta _{i_5i_6}\delta _{i_7i_8}\epsilon _{j_1j_7}\delta _{j_2j_3}\delta _{j_4j_5}\epsilon _{j_6j_8}~\psi _0^{i_1j_1+}\psi _0^{i_2j_2+}\psi _0^{i_3j_3+}\psi _1^{i_4j_4+}\psi _1^{i_5j_5+}\psi _1^{i_6j_6+}\psi _2^{i_7j_7+}\psi _3^{i_8j_8+}\\
86&. +\epsilon _{i_1i_2}\delta _{i_3i_4}\delta _{i_5i_6}\epsilon _{i_7i_8}\epsilon _{j_1j_7}\delta _{j_2j_3}\delta _{j_4j_5}\delta _{j_6j_8}~\psi _0^{i_1j_1+}\psi _0^{i_2j_2+}\psi _0^{i_3j_3+}\psi _1^{i_4j_4+}\psi _1^{i_5j_5+}\psi _1^{i_6j_6+}\psi _2^{i_7j_7+}\psi _3^{i_8j_8+}\\
87&. -\delta _{i_1i_2}\delta _{i_3i_4}\delta _{i_5i_6}\epsilon _{i_7i_8}\delta _{j_1j_7}\delta _{j_2j_3}\delta _{j_4j_5}\epsilon _{j_6j_8}~\psi _0^{i_1j_1+}\psi _0^{i_2j_2+}\psi _0^{i_3j_3+}\psi _1^{i_4j_4+}\psi _1^{i_5j_5+}\psi _1^{i_6j_6+}\psi _2^{i_7j_7+}\psi _3^{i_8j_8+}\\
88&. +\epsilon _{i_1i_2}\delta _{i_3i_4}\delta _{i_5i_6}\delta _{i_7i_8}\epsilon _{j_1j_7}\delta _{j_2j_3}\delta _{j_4j_5}\delta _{j_6j_8}~\psi _0^{i_1j_1+}\psi _0^{i_2j_2+}\psi _0^{i_3j_3+}\psi _1^{i_4j_4+}\psi _1^{i_5j_5+}\psi _1^{i_6j_6+}\psi _2^{i_7j_7+}\psi _3^{i_8j_8+}\\
89&. +\epsilon _{i_1i_2}\delta _{i_3i_4}\delta _{i_5i_6}\delta _{i_7i_8}\delta _{j_1j_7}\delta _{j_2j_3}\delta _{j_4j_5}\epsilon _{j_6j_8}~\psi _0^{i_1j_1+}\psi _0^{i_2j_2+}\psi _0^{i_3j_3+}\psi _1^{i_4j_4+}\psi _1^{i_5j_5+}\psi _1^{i_6j_6+}\psi _2^{i_7j_7+}\psi _3^{i_8j_8+}\\
90&. -\delta _{i_1i_2}\delta _{i_3i_4}\delta _{i_5i_6}\epsilon _{i_7i_8}\delta _{j_1j_7}\epsilon _{j_2j_3}\delta _{j_4j_5}\delta _{j_6j_8}~\psi _0^{i_1j_1+}\psi _0^{i_2j_2+}\psi _0^{i_3j_3+}\psi _1^{i_4j_4+}\psi _1^{i_5j_5+}\psi _1^{i_6j_6+}\psi _2^{i_7j_7+}\psi _3^{i_8j_8+}\\
91&. +\epsilon _{i_1i_2}\delta _{i_3i_4}\delta _{i_5i_6}\epsilon _{i_7i_8}\delta _{j_1j_7}\delta _{j_2j_3}\delta _{j_4j_5}\delta _{j_6j_8}~\psi _0^{i_1j_1+}\psi _0^{i_2j_2+}\psi _0^{i_3j_3+}\psi _1^{i_4j_4+}\psi _1^{i_5j_5+}\psi _1^{i_6j_6+}\psi _2^{i_7j_7+}\psi _3^{i_8j_8+}\\
92&. +\delta _{i_1i_2}\delta _{i_3i_4}\delta _{i_5i_6}\delta _{i_7i_8}\delta _{j_1j_7}\epsilon _{j_2j_3}\delta _{j_4j_5}\epsilon _{j_6j_8}~\psi _0^{i_1j_1+}\psi _0^{i_2j_2+}\psi _0^{i_3j_3+}\psi _1^{i_4j_4+}\psi _1^{i_5j_5+}\psi _1^{i_6j_6+}\psi _2^{i_7j_7+}\psi _3^{i_8j_8+}
\end{align*}

\subsection{Group $(p_2,p_4)$}
\begin{align*}
93&. +\delta _{i_1i_2}\delta _{i_3i_4}\delta _{i_5i_6}\delta _{i_7i_8}\delta _{j_1j_7}\delta _{j_2j_3}\delta _{j_4j_5}\delta _{j_6j_8}~\psi _0^{j_1i_1+}\psi _0^{j_2i_2+}\psi _0^{j_3i_3+}\psi _2^{j_4i_4+}\psi _2^{j_5i_5+}\psi _2^{j_6i_6+}\psi _1^{j_7i_7+}\psi _3^{j_8i_8+}\\
94&. -\epsilon _{i_1i_2}\delta _{i_3i_4}\delta _{i_5i_6}\epsilon _{i_7i_8}\epsilon _{j_1j_7}\delta _{j_2j_3}\delta _{j_4j_5}\epsilon _{j_6j_8}~\psi _0^{j_1i_1+}\psi _0^{j_2i_2+}\psi _0^{j_3i_3+}\psi _2^{j_4i_4+}\psi _2^{j_5i_5+}\psi _2^{j_6i_6+}\psi _1^{j_7i_7+}\psi _3^{j_8i_8+}\\
95&. +\epsilon _{i_1i_2}\delta _{i_3i_4}\delta _{i_5i_6}\delta _{i_7i_8}\delta _{j_1j_7}\delta _{j_2j_3}\delta _{j_4j_5}\delta _{j_6j_8}~\psi _0^{j_1i_1+}\psi _0^{j_2i_2+}\psi _0^{j_3i_3+}\psi _2^{j_4i_4+}\psi _2^{j_5i_5+}\psi _2^{j_6i_6+}\psi _1^{j_7i_7+}\psi _3^{j_8i_8+}\\
96&.+\delta _{i_1i_2}\delta _{i_3i_4}\delta _{i_5i_6}\delta _{i_7i_8}\delta _{j_1j_7}\epsilon _{j_2j_3}\delta _{j_4j_5}\delta _{j_6j_8}~\psi _0^{j_1i_1+}\psi _0^{j_2i_2+}\psi _0^{j_3i_3+}\psi _2^{j_4i_4+}\psi _2^{j_5i_5+}\psi _2^{j_6i_6+}\psi _1^{j_7i_7+}\psi _3^{j_8i_8+}\\
97&.  -\delta _{i_1i_2}\delta _{i_3i_4}\delta _{i_5i_6}\epsilon _{i_7i_8}\delta _{j_1j_7}\delta _{j_2j_3}\delta _{j_4j_5}\delta _{j_6j_8}~\psi _0^{j_1i_1+}\psi _0^{j_2i_2+}\psi _0^{j_3i_3+}\psi _2^{j_4i_4+}\psi _2^{j_5i_5+}\psi _2^{j_6i_6+}\psi _1^{j_7i_7+}\psi _3^{j_8i_8+}\\
98&. +\delta _{i_1i_2}\delta _{i_3i_4}\delta _{i_5i_6}\delta _{i_7i_8}\delta _{j_1j_7}\delta _{j_2j_3}\delta _{j_4j_5}\epsilon _{j_6j_8}~\psi _0^{j_1i_1+}\psi _0^{j_2i_2+}\psi _0^{j_3i_3+}\psi _2^{j_4i_4+}\psi _2^{j_5i_5+}\psi _2^{j_6i_6+}\psi _1^{j_7i_7+}\psi _3^{j_8i_8+}\\
99&. +\delta _{i_1i_2}\delta _{i_3i_4}\delta _{i_5i_6}\epsilon _{i_7i_8}\delta _{j_1j_7}\epsilon _{j_2j_3}\delta _{j_4j_5}\epsilon _{j_6j_8}~\psi _0^{j_1i_1+}\psi _0^{j_2i_2+}\psi _0^{j_3i_3+}\psi _2^{j_4i_4+}\psi _2^{j_5i_5+}\psi _2^{j_6i_6+}\psi _1^{j_7i_7+}\psi _3^{j_8i_8+}\\
100&. +\epsilon _{i_1i_2}\delta _{i_3i_4}\delta _{i_5i_6}\epsilon _{i_7i_8}\delta _{j_1j_7}\delta _{j_2j_3}\delta _{j_4j_5}\epsilon _{j_6j_8}~\psi _0^{j_1i_1+}\psi _0^{j_2i_2+}\psi _0^{j_3i_3+}\psi _2^{j_4i_4+}\psi _2^{j_5i_5+}\psi _2^{j_6i_6+}\psi _1^{j_7i_7+}\psi _3^{j_8i_8+}\\
101&. -\epsilon _{i_1i_2}\delta _{i_3i_4}\delta _{i_5i_6}\epsilon _{i_7i_8}\epsilon _{j_1j_7}\delta _{j_2j_3}\delta _{j_4j_5}\delta _{j_6j_8}~\psi _0^{j_1i_1+}\psi _0^{j_2i_2+}\psi _0^{j_3i_3+}\psi _2^{j_4i_4+}\psi _2^{j_5i_5+}\psi _2^{j_6i_6+}\psi _1^{j_7i_7+}\psi _3^{j_8i_8+}\\
102&. +\epsilon _{i_1i_2}\delta _{i_3i_4}\delta _{i_5i_6}\delta _{i_7i_8}\epsilon _{j_1j_7}\delta _{j_2j_3}\delta _{j_4j_5}\epsilon _{j_6j_8}~\psi _0^{j_1i_1+}\psi _0^{j_2i_2+}\psi _0^{j_3i_3+}\psi _2^{j_4i_4+}\psi _2^{j_5i_5+}\psi _2^{j_6i_6+}\psi _1^{j_7i_7+}\psi _3^{j_8i_8+}\\
103&. +\delta _{i_1i_2}\delta _{i_3i_4}\delta _{i_5i_6}\epsilon _{i_7i_8}\delta _{j_1j_7}\delta _{j_2j_3}\delta _{j_4j_5}\epsilon _{j_6j_8}~\psi _0^{j_1i_1+}\psi _0^{j_2i_2+}\psi _0^{j_3i_3+}\psi _2^{j_4i_4+}\psi _2^{j_5i_5+}\psi _2^{j_6i_6+}\psi _1^{j_7i_7+}\psi _3^{j_8i_8+}\\
104&. -\epsilon _{i_1i_2}\delta _{i_3i_4}\delta _{i_5i_6}\delta _{i_7i_8}\epsilon _{j_1j_7}\delta _{j_2j_3}\delta _{j_4j_5}\delta _{j_6j_8}~\psi _0^{j_1i_1+}\psi _0^{j_2i_2+}\psi _0^{j_3i_3+}\psi _2^{j_4i_4+}\psi _2^{j_5i_5+}\psi _2^{j_6i_6+}\psi _1^{j_7i_7+}\psi _3^{j_8i_8+}\\
105&. +\delta _{i_1i_2}\delta _{i_3i_4}\delta _{i_5i_6}\epsilon _{i_7i_8}\delta _{j_1j_7}\epsilon _{j_2j_3}\delta _{j_4j_5}\delta _{j_6j_8}~\psi _0^{j_1i_1+}\psi _0^{j_2i_2+}\psi _0^{j_3i_3+}\psi _2^{j_4i_4+}\psi _2^{j_5i_5+}\psi _2^{j_6i_6+}\psi _1^{j_7i_7+}\psi _3^{j_8i_8+}\\
106&. +\epsilon _{i_1i_2}\delta _{i_3i_4}\delta _{i_5i_6}\delta _{i_7i_8}\delta _{j_1j_7}\delta _{j_2j_3}\delta _{j_4j_5}\epsilon _{j_6j_8}~\psi _0^{j_1i_1+}\psi _0^{j_2i_2+}\psi _0^{j_3i_3+}\psi _2^{j_4i_4+}\psi _2^{j_5i_5+}\psi _2^{j_6i_6+}\psi _1^{j_7i_7+}\psi _3^{j_8i_8+}\\
107&. +\delta _{i_1i_2}\delta _{i_3i_4}\delta _{i_5i_6}\delta _{i_7i_8}\delta _{j_1j_7}\epsilon _{j_2j_3}\delta _{j_4j_5}\epsilon _{j_6j_8}~\psi _0^{j_1i_1+}\psi _0^{j_2i_2+}\psi _0^{j_3i_3+}\psi _2^{j_4i_4+}\psi _2^{j_5i_5+}\psi _2^{j_6i_6+}\psi _1^{j_7i_7+}\psi _3^{j_8i_8+}\\
108&. +\epsilon _{i_1i_2}\delta _{i_3i_4}\delta _{i_5i_6}\epsilon _{i_7i_8}\delta _{j_1j_7}\delta _{j_2j_3}\delta _{j_4j_5}\delta _{j_6j_8}~\psi _0^{j_1i_1+}\psi _0^{j_2i_2+}\psi _0^{j_3i_3+}\psi _2^{j_4i_4+}\psi _2^{j_5i_5+}\psi _2^{j_6i_6+}\psi _1^{j_7i_7+}\psi _3^{j_8i_8+}
\end{align*}

\subsection{Group $(p_4,p_2)$}

\begin{align*}
109&. +\delta _{i_1i_2}\delta _{i_3i_4}\delta _{i_5i_6}\delta _{i_7i_8}\delta _{j_1j_7}\delta _{j_2j_3}\delta _{j_4j_5}\delta _{j_6j_8}~\psi _3^{j_1i_1+}\psi _3^{j_2i_2+}\psi _3^{j_3i_3+}\psi _1^{j_4i_4+}\psi _1^{j_5i_5+}\psi _1^{j_6i_6+}\psi _2^{j_7i_7+}\psi _0^{j_8i_8+}\\
110&. -\epsilon _{i_1i_2}\delta _{i_3i_4}\delta _{i_5i_6}\epsilon _{i_7i_8}\epsilon _{j_1j_7}\delta _{j_2j_3}\delta _{j_4j_5}\epsilon _{j_6j_8}~\psi _3^{j_1i_1+}\psi _3^{j_2i_2+}\psi _3^{j_3i_3+}\psi _1^{j_4i_4+}\psi _1^{j_5i_5+}\psi _1^{j_6i_6+}\psi _2^{j_7i_7+}\psi _0^{j_8i_8+}\\
111&.  +\delta _{i_1i_2}\delta _{i_3i_4}\delta _{i_5i_6}\epsilon _{i_7i_8}\delta _{j_1j_7}\delta _{j_2j_3}\delta _{j_4j_5}\delta _{j_6j_8}~\psi _3^{j_1i_1+}\psi _3^{j_2i_2+}\psi _3^{j_3i_3+}\psi _1^{j_4i_4+}\psi _1^{j_5i_5+}\psi _1^{j_6i_6+}\psi _2^{j_7i_7+}\psi _0^{j_8i_8+}\\
112&. +\delta _{i_1i_2}\delta _{i_3i_4}\delta _{i_5i_6}\delta _{i_7i_8}\delta _{j_1j_7}\delta _{j_2j_3}\delta _{j_4j_5}\epsilon _{j_6j_8}~\psi _3^{j_1i_1+}\psi _3^{j_2i_2+}\psi _3^{j_3i_3+}\psi _1^{j_4i_4+}\psi _1^{j_5i_5+}\psi _1^{j_6i_6+}\psi _2^{j_7i_7+}\psi _0^{j_8i_8+}\\
113&. -\epsilon _{i_1i_2}\delta _{i_3i_4}\delta _{i_5i_6}\delta _{i_7i_8}\delta _{j_1j_7}\delta _{j_2j_3}\delta _{j_4j_5}\delta _{j_6j_8}~\psi _3^{j_1i_1+}\psi _3^{j_2i_2+}\psi _3^{j_3i_3+}\psi _1^{j_4i_4+}\psi _1^{j_5i_5+}\psi _1^{j_6i_6+}\psi _2^{j_7i_7+}\psi _0^{j_8i_8+}\\
114&. +\delta _{i_1i_2}\delta _{i_3i_4}\delta _{i_5i_6}\delta _{i_7i_8}\delta _{j_1j_7}\epsilon _{j_2j_3}\delta _{j_4j_5}\delta _{j_6j_8}~\psi _3^{j_1i_1+}\psi _3^{j_2i_2+}\psi _3^{j_3i_3+}\psi _1^{j_4i_4+}\psi _1^{j_5i_5+}\psi _1^{j_6i_6+}\psi _2^{j_7i_7+}\psi _0^{j_8i_8+}\\
115&. -\epsilon _{i_1i_2}\delta _{i_3i_4}\delta _{i_5i_6}\delta _{i_7i_8}\epsilon _{j_1j_7}\delta _{j_2j_3}\delta _{j_4j_5}\epsilon _{j_6j_8}~\psi _3^{j_1i_1+}\psi _3^{j_2i_2+}\psi _3^{j_3i_3+}\psi _1^{j_4i_4+}\psi _1^{j_5i_5+}\psi _1^{j_6i_6+}\psi _2^{j_7i_7+}\psi _0^{j_8i_8+}\\
116&. +\epsilon _{i_1i_2}\delta _{i_3i_4}\delta _{i_5i_6}\epsilon _{i_7i_8}\epsilon _{j_1j_7}\delta _{j_2j_3}\delta _{j_4j_5}\delta _{j_6j_8}~\psi _3^{j_1i_1+}\psi _3^{j_2i_2+}\psi _3^{j_3i_3+}\psi _1^{j_4i_4+}\psi _1^{j_5i_5+}\psi _1^{j_6i_6+}\psi _2^{j_7i_7+}\psi _0^{j_8i_8+}\\
117&. -\epsilon _{i_1i_2}\delta _{i_3i_4}\delta _{i_5i_6}\epsilon _{i_7i_8}\delta _{j_1j_7}\delta _{j_2j_3}\delta _{j_4j_5}\epsilon _{j_6j_8}~\psi _3^{j_1i_1+}\psi _3^{j_2i_2+}\psi _3^{j_3i_3+}\psi _1^{j_4i_4+}\psi _1^{j_5i_5+}\psi _1^{j_6i_6+}\psi _2^{j_7i_7+}\psi _0^{j_8i_8+}\\
118&. +\delta _{i_1i_2}\delta _{i_3i_4}\delta _{i_5i_6}\epsilon _{i_7i_8}\delta _{j_1j_7}\epsilon _{j_2j_3}\delta _{j_4j_5}\epsilon _{j_6j_8}~\psi _3^{j_1i_1+}\psi _3^{j_2i_2+}\psi _3^{j_3i_3+}\psi _1^{j_4i_4+}\psi _1^{j_5i_5+}\psi _1^{j_6i_6+}\psi _2^{j_7i_7+}\psi _0^{j_8i_8+}\\
119&.  -\epsilon _{i_1i_2}\delta _{i_3i_4}\delta _{i_5i_6}\delta _{i_7i_8}\epsilon _{j_1j_7}\delta _{j_2j_3}\delta _{j_4j_5}\delta _{j_6j_8}~\psi _3^{j_1i_1+}\psi _3^{j_2i_2+}\psi _3^{j_3i_3+}\psi _1^{j_4i_4+}\psi _1^{j_5i_5+}\psi _1^{j_6i_6+}\psi _2^{j_7i_7+}\psi _0^{j_8i_8+}\\
120&. +\delta _{i_1i_2}\delta _{i_3i_4}\delta _{i_5i_6}\epsilon _{i_7i_8}\delta _{j_1j_7}\delta _{j_2j_3}\delta _{j_4j_5}\epsilon _{j_6j_8}~\psi _3^{j_1i_1+}\psi _3^{j_2i_2+}\psi _3^{j_3i_3+}\psi _1^{j_4i_4+}\psi _1^{j_5i_5+}\psi _1^{j_6i_6+}\psi _2^{j_7i_7+}\psi _0^{j_8i_8+}\\
121&. +\epsilon _{i_1i_2}\delta _{i_3i_4}\delta _{i_5i_6}\delta _{i_7i_8}\delta _{j_1j_7}\delta _{j_2j_3}\delta _{j_4j_5}\epsilon _{j_6j_8}~\psi _3^{j_1i_1+}\psi _3^{j_2i_2+}\psi _3^{j_3i_3+}\psi _1^{j_4i_4+}\psi _1^{j_5i_5+}\psi _1^{j_6i_6+}\psi _2^{j_7i_7+}\psi _0^{j_8i_8+}\\
122&. +\delta _{i_1i_2}\delta _{i_3i_4}\delta _{i_5i_6}\epsilon _{i_7i_8}\delta _{j_1j_7}\epsilon _{j_2j_3}\delta _{j_4j_5}\delta _{j_6j_8}~\psi _3^{j_1i_1+}\psi _3^{j_2i_2+}\psi _3^{j_3i_3+}\psi _1^{j_4i_4+}\psi _1^{j_5i_5+}\psi _1^{j_6i_6+}\psi _2^{j_7i_7+}\psi _0^{j_8i_8+}\\
123&. +\delta _{i_1i_2}\delta _{i_3i_4}\delta _{i_5i_6}\delta _{i_7i_8}\delta _{j_1j_7}\epsilon _{j_2j_3}\delta _{j_4j_5}\epsilon _{j_6j_8}~\psi _3^{j_1i_1+}\psi _3^{j_2i_2+}\psi _3^{j_3i_3+}\psi _1^{j_4i_4+}\psi _1^{j_5i_5+}\psi _1^{j_6i_6+}\psi _2^{j_7i_7+}\psi _0^{j_8i_8+}\\
124&. +\epsilon _{i_1i_2}\delta _{i_3i_4}\delta _{i_5i_6}\epsilon _{i_7i_8}\delta _{j_1j_7}\delta _{j_2j_3}\delta _{j_4j_5}\delta _{j_6j_8}~\psi _3^{j_1i_1+}\psi _3^{j_2i_2+}\psi _3^{j_3i_3+}\psi _1^{j_4i_4+}\psi _1^{j_5i_5+}\psi _1^{j_6i_6+}\psi _2^{j_7i_7+}\psi _0^{j_8i_8+}
\end{align*}

\subsection{Group $(p_4,p_4)$}

\begin{align*}
125&. -\delta _{i_1i_2}\delta _{i_3i_4}\delta _{i_5i_6}\delta _{i_7i_8}\delta _{j_1j_7}\delta _{j_2j_3}\delta _{j_4j_5}\delta _{j_6j_8}~\psi _3^{i_1j_1+}\psi _3^{i_2j_2+}\psi _3^{i_3j_3+}\psi _2^{i_4j_4+}\psi _2^{i_5j_5+}\psi _2^{i_6j_6+}\psi _1^{i_7j_7+}\psi _0^{i_8j_8+}\\
126&. +\epsilon _{i_1i_2}\delta _{i_3i_4}\delta _{i_5i_6}\epsilon _{i_7i_8}\epsilon _{j_1j_7}\delta _{j_2j_3}\delta _{j_4j_5}\epsilon _{j_6j_8}~\psi _3^{i_1j_1+}\psi _3^{i_2j_2+}\psi _3^{i_3j_3+}\psi _2^{i_4j_4+}\psi _2^{i_5j_5+}\psi _2^{i_6j_6+}\psi _1^{i_7j_7+}\psi _0^{i_8j_8+}\\
127&. -\delta _{i_1i_2}\delta _{i_3i_4}\delta _{i_5i_6}\delta _{i_7i_8}\delta _{j_1j_7}\delta _{j_2j_3}\delta _{j_4j_5}\epsilon _{j_6j_8}~\psi _3^{i_1j_1+}\psi _3^{i_2j_2+}\psi _3^{i_3j_3+}\psi _2^{i_4j_4+}\psi _2^{i_5j_5+}\psi _2^{i_6j_6+}\psi _1^{i_7j_7+}\psi _0^{i_8j_8+}\\
128&. -\delta _{i_1i_2}\delta _{i_3i_4}\delta _{i_5i_6}\epsilon _{i_7i_8}\delta _{j_1j_7}\delta _{j_2j_3}\delta _{j_4j_5}\delta _{j_6j_8}~\psi _3^{i_1j_1+}\psi _3^{i_2j_2+}\psi _3^{i_3j_3+}\psi _2^{i_4j_4+}\psi _2^{i_5j_5+}\psi _2^{i_6j_6+}\psi _1^{i_7j_7+}\psi _0^{i_8j_8+}\\
129&. +\delta _{i_1i_2}\delta _{i_3i_4}\delta _{i_5i_6}\delta _{i_7i_8}\delta _{j_1j_7}\epsilon _{j_2j_3}\delta _{j_4j_5}\delta _{j_6j_8}~\psi _3^{i_1j_1+}\psi _3^{i_2j_2+}\psi _3^{i_3j_3+}\psi _2^{i_4j_4+}\psi _2^{i_5j_5+}\psi _2^{i_6j_6+}\psi _1^{i_7j_7+}\psi _0^{i_8j_8+}\\
130&. +\epsilon _{i_1i_2}\delta _{i_3i_4}\delta _{i_5i_6}\delta _{i_7i_8}\delta _{j_1j_7}\delta _{j_2j_3}\delta _{j_4j_5}\delta _{j_6j_8}~\psi _3^{i_1j_1+}\psi _3^{i_2j_2+}\psi _3^{i_3j_3+}\psi _2^{i_4j_4+}\psi _2^{i_5j_5+}\psi _2^{i_6j_6+}\psi _1^{i_7j_7+}\psi _0^{i_8j_8+}\\
131&. +\epsilon _{i_1i_2}\delta _{i_3i_4}\delta _{i_5i_6}\epsilon _{i_7i_8}\epsilon _{j_1j_7}\delta _{j_2j_3}\delta _{j_4j_5}\delta _{j_6j_8}~\psi _3^{i_1j_1+}\psi _3^{i_2j_2+}\psi _3^{i_3j_3+}\psi _2^{i_4j_4+}\psi _2^{i_5j_5+}\psi _2^{i_6j_6+}\psi _1^{i_7j_7+}\psi _0^{i_8j_8+}\\
132&. -\epsilon _{i_1i_2}\delta _{i_3i_4}\delta _{i_5i_6}\delta _{i_7i_8}\epsilon _{j_1j_7}\delta _{j_2j_3}\delta _{j_4j_5}\epsilon _{j_6j_8}~\psi _3^{i_1j_1+}\psi _3^{i_2j_2+}\psi _3^{i_3j_3+}\psi _2^{i_4j_4+}\psi _2^{i_5j_5+}\psi _2^{i_6j_6+}\psi _1^{i_7j_7+}\psi _0^{i_8j_8+}\\
133&. +\delta _{i_1i_2}\delta _{i_3i_4}\delta _{i_5i_6}\epsilon _{i_7i_8}\delta _{j_1j_7}\epsilon _{j_2j_3}\delta _{j_4j_5}\epsilon _{j_6j_8}~\psi _3^{i_1j_1+}\psi _3^{i_2j_2+}\psi _3^{i_3j_3+}\psi _2^{i_4j_4+}\psi _2^{i_5j_5+}\psi _2^{i_6j_6+}\psi _1^{i_7j_7+}\psi _0^{i_8j_8+}\\
134&. +\epsilon _{i_1i_2}\delta _{i_3i_4}\delta _{i_5i_6}\epsilon _{i_7i_8}\delta _{j_1j_7}\delta _{j_2j_3}\delta _{j_4j_5}\epsilon _{j_6j_8}~\psi _3^{i_1j_1+}\psi _3^{i_2j_2+}\psi _3^{i_3j_3+}\psi _2^{i_4j_4+}\psi _2^{i_5j_5+}\psi _2^{i_6j_6+}\psi _1^{i_7j_7+}\psi _0^{i_8j_8+}\\
135&. +\epsilon _{i_1i_2}\delta _{i_3i_4}\delta _{i_5i_6}\delta _{i_7i_8}\epsilon _{j_1j_7}\delta _{j_2j_3}\delta _{j_4j_5}\delta _{j_6j_8}~\psi _3^{i_1j_1+}\psi _3^{i_2j_2+}\psi _3^{i_3j_3+}\psi _2^{i_4j_4+}\psi _2^{i_5j_5+}\psi _2^{i_6j_6+}\psi _1^{i_7j_7+}\psi _0^{i_8j_8+}\\
136&. -\delta _{i_1i_2}\delta _{i_3i_4}\delta _{i_5i_6}\epsilon _{i_7i_8}\delta _{j_1j_7}\delta _{j_2j_3}\delta _{j_4j_5}\epsilon _{j_6j_8}~\psi _3^{i_1j_1+}\psi _3^{i_2j_2+}\psi _3^{i_3j_3+}\psi _2^{i_4j_4+}\psi _2^{i_5j_5+}\psi _2^{i_6j_6+}\psi _1^{i_7j_7+}\psi _0^{i_8j_8+}\\
137&. -\delta _{i_1i_2}\delta _{i_3i_4}\delta _{i_5i_6}\epsilon _{i_7i_8}\delta _{j_1j_7}\epsilon _{j_2j_3}\delta _{j_4j_5}\delta _{j_6j_8}~\psi _3^{i_1j_1+}\psi _3^{i_2j_2+}\psi _3^{i_3j_3+}\psi _2^{i_4j_4+}\psi _2^{i_5j_5+}\psi _2^{i_6j_6+}\psi _1^{i_7j_7+}\psi _0^{i_8j_8+}\\
138&. -\epsilon _{i_1i_2}\delta _{i_3i_4}\delta _{i_5i_6}\delta _{i_7i_8}\delta _{j_1j_7}\delta _{j_2j_3}\delta _{j_4j_5}\epsilon _{j_6j_8}~\psi _3^{i_1j_1+}\psi _3^{i_2j_2+}\psi _3^{i_3j_3+}\psi _2^{i_4j_4+}\psi _2^{i_5j_5+}\psi _2^{i_6j_6+}\psi _1^{i_7j_7+}\psi _0^{i_8j_8+}\\
139&.  +\epsilon _{i_1i_2}\delta _{i_3i_4}\delta _{i_5i_6}\epsilon _{i_7i_8}\delta _{j_1j_7}\delta _{j_2j_3}\delta _{j_4j_5}\delta _{j_6j_8}~\psi _3^{i_1j_1+}\psi _3^{i_2j_2+}\psi _3^{i_3j_3+}\psi _2^{i_4j_4+}\psi _2^{i_5j_5+}\psi _2^{i_6j_6+}\psi _1^{i_7j_7+}\psi _0^{i_8j_8+}\\
140&. +\delta _{i_1i_2}\delta _{i_3i_4}\delta _{i_5i_6}\delta _{i_7i_8}\delta _{j_1j_7}\epsilon _{j_2j_3}\delta _{j_4j_5}\epsilon _{j_6j_8}~\psi _3^{i_1j_1+}\psi _3^{i_2j_2+}\psi _3^{i_3j_3+}\psi _2^{i_4j_4+}\psi _2^{i_5j_5+}\psi _2^{i_6j_6+}\psi _1^{i_7j_7+}\psi _0^{i_8j_8+}
\end{align*}

\section{List of singlets- Method-II}\label{method-II}

In this appendix, we explicitly list all the 140 singlets in the $n=2$ Gurau-Witten model obtained via method-II of section \ref{singlets of SO(n)} by classifying them based on the groups they belong to.

\subsection{Groups $(p_1,p_1)$, $(p_1,p_5)$, $(p_5,p_1)$ and $(p_5,p_5)$}
\begin{align}
1&. ~\psi _0^{11+}\psi _0^{12+}\psi _0^{21+}\psi _0^{22+}\psi _1^{11+}\psi _1^{12+}\psi _1^{21+}\psi _1^{22+}|~\rangle \hspace{75 mm}\\
2&. ~\psi _0^{11+}\psi _0^{12+}\psi _0^{21+}\psi _0^{22+}\psi _2^{11+}\psi _2^{12+}\psi _2^{21+}\psi _2^{22+}|~\rangle\\
3&. ~\psi _3^{11+}\psi _3^{12+}\psi _3^{21+}\psi _3^{22+}\psi _1^{11+}\psi _1^{12+}\psi _1^{21+}\psi _1^{22+}|~\rangle\\
4&. ~\psi _3^{11+}\psi _3^{12+}\psi _3^{21+}\psi _3^{22+}\psi _2^{11+}\psi _2^{12+}\psi _2^{21+}\psi _2^{22+}|~\rangle
\end{align}

\subsection{Groups $(p_1,p_3)$, $(p_5,p_3)$, $(p_3,p_1)$ and $(p_3,p_5)$}
\begin{align}
5&. ~\psi _0^{11+}\psi _0^{12+}\psi _0^{21+}\psi _0^{22+}\left(\psi _1^{11+}\psi _1^{12+}+\psi _1^{21+}\psi _1^{22+}\right)\left(\psi _2^{11+}\psi _2^{12+}+\psi _2^{21+}\psi _2^{22+}\right)|~\rangle\hspace{15 mm}\\
6&. ~\psi _0^{11+}\psi _0^{12+}\psi _0^{21+}\psi _0^{22+}\left(\psi _1^{11+}\psi _1^{12+}+\psi _1^{21+}\psi _1^{22+}\right)\left(\psi _2^{11+}\psi _2^{21+}+\psi _2^{12+}\psi _2^{22+}\right)|~\rangle\\
7&. ~\psi _0^{11+}\psi _0^{12+}\psi _0^{21+}\psi _0^{22+}\left(\psi _1^{11+}\psi _1^{21+}+\psi _1^{12+}\psi _1^{22+}\right)\left(\psi _2^{11+}\psi _2^{12+}+\psi _2^{21+}\psi _2^{22+}\right)|~\rangle\\
8&. ~\psi _0^{11+}\psi _0^{12+}\psi _0^{21+}\psi _0^{22+}\left(\psi _1^{11+}\psi _1^{21+}+\psi _1^{12+}\psi _1^{22+}\right)\left(\psi _2^{11+}\psi _2^{21+}+\psi _2^{12+}\psi _2^{22+}\right)|~\rangle\\
&\hspace{48 mm} \diamondsuit \hspace{10 mm}\diamondsuit \hspace{10 mm}\diamondsuit \nonumber \\
9&. ~\psi _3^{11+}\psi _3^{12+}\psi _3^{21+}\psi _3^{22+}\left(\psi _1^{11+}\psi _1^{12+}+\psi _1^{21+}\psi _1^{22+}\right)\left(\psi _2^{11+}\psi _2^{12+}+\psi _2^{21+}\psi _2^{22+}\right)|~\rangle\hspace{15 mm}\\
10&. ~\psi _3^{11+}\psi _3^{12+}\psi _3^{21+}\psi _3^{22+}\left(\psi _1^{11+}\psi _1^{12+}+\psi _1^{21+}\psi _1^{22+}\right)\left(\psi _2^{11+}\psi _2^{21+}+\psi _2^{12+}\psi _2^{22+}\right)|~\rangle\\
11&. ~\psi _3^{11+}\psi _3^{12+}\psi _3^{21+}\psi _3^{22+}\left(\psi _1^{11+}\psi _1^{21+}+\psi _1^{12+}\psi _1^{22+}\right)\left(\psi _2^{11+}\psi _2^{12+}+\psi _2^{21+}\psi _2^{22+}\right)|~\rangle\\
12&. ~\psi _3^{11+}\psi _3^{12+}\psi _3^{21+}\psi _3^{22+}\left(\psi _1^{11+}\psi _1^{21+}+\psi _1^{12+}\psi _1^{22+}\right)\left(\psi _2^{11+}\psi _2^{21+}+\psi _2^{12+}\psi _2^{22+}\right)|~\rangle\\
&\hspace{48 mm} \diamondsuit \hspace{10 mm}\diamondsuit \hspace{10 mm}\diamondsuit \nonumber \\
13&. ~\psi _1^{11+}\psi _1^{12+}\psi _1^{21+}\psi _1^{22+}\left(\psi _0^{11+}\psi _0^{12+}+\psi _0^{21+}\psi _0^{22+}\right)\left(\psi _3^{11+}\psi _3^{12+}+\psi _3^{21+}\psi _3^{22+}\right)|~\rangle\hspace{15 mm}\\
14&.~\psi _1^{11+}\psi _1^{12+}\psi _1^{21+}\psi _1^{22+}\left(\psi _0^{11+}\psi _0^{12+}+\psi _0^{21+}\psi _0^{22+}\right)\left(\psi _3^{11+}\psi _3^{21+}+\psi _3^{12+}\psi _3^{22+}\right)|~\rangle\\
15&. ~\psi _1^{11+}\psi _1^{12+}\psi _1^{21+}\psi _1^{22+}\left(\psi _0^{11+}\psi _0^{21+}+\psi _0^{12+}\psi _0^{22+}\right)\left(\psi _3^{11+}\psi _3^{12+}+\psi _3^{21+}\psi _3^{22+}\right)|~\rangle\\
16&. ~\psi _1^{11+}\psi _1^{12+}\psi _1^{21+}\psi _1^{22+}\left(\psi _0^{11+}\psi _0^{21+}+\psi _0^{12+}\psi _0^{22+}\right)\left(\psi _3^{11+}\psi _3^{21+}+\psi _3^{12+}\psi _3^{22+}\right)|~\rangle\\
&\hspace{48 mm} \diamondsuit \hspace{10 mm}\diamondsuit \hspace{10 mm}\diamondsuit \nonumber \\
17&. ~\psi _2^{11+}\psi _2^{12+}\psi _2^{21+}\psi _2^{22+}\left(\psi _0^{11+}\psi _0^{12+}+\psi _0^{21+}\psi _0^{22+}\right)\left(\psi _3^{11+}\psi _3^{12+}+\psi _3^{21+}\psi _3^{22+}\right)|~\rangle\hspace{15 mm}\\
18&.~\psi _2^{11+}\psi _2^{12+}\psi _2^{21+}\psi _2^{22+}\left(\psi _0^{11+}\psi _0^{12+}+\psi _0^{21+}\psi _0^{22+}\right)\left(\psi _3^{11+}\psi _3^{21+}+\psi _3^{12+}\psi _3^{22+}\right)|~\rangle\\
19&. ~\psi _2^{11+}\psi _2^{12+}\psi _2^{21+}\psi _2^{22+}\left(\psi _0^{11+}\psi _0^{21+}+\psi _0^{12+}\psi _0^{22+}\right)\left(\psi _3^{11+}\psi _3^{12+}+\psi _3^{21+}\psi _3^{22+}\right)|~\rangle\\
20&. ~\psi _2^{11+}\psi _2^{12+}\psi _2^{21+}\psi _2^{22+}\left(\psi _0^{11+}\psi _0^{21+}+\psi _0^{12+}\psi _0^{22+}\right)\left(\psi _3^{11+}\psi _3^{21+}+\psi _3^{12+}\psi _3^{22+}\right)|~\rangle
\end{align}

\subsection{Group $(p_3,p_3)$}

\subsubsection{Type-I}
\begin{align}
21&. ~\left(\psi _0^{11+}\psi _0^{12+}+\psi _0^{21+}\psi _0^{22+}\right)\left(\psi _1^{11+}\psi _1^{12+}+\psi _1^{21+}\psi _1^{22+}\right)\times \nonumber \\
&\hspace{41 mm}\times \left(\psi _2^{11+}\psi _2^{12+}+\psi _2^{21+}\psi _2^{22+}\right)\left(\psi _3^{11+}\psi _3^{12+}+\psi _3^{21+}\psi _3^{22+}\right)|~\rangle \\
22&. ~\left(\psi _0^{11+}\psi _0^{12+}+\psi _0^{21+}\psi _0^{22+}\right)\left(\psi _1^{11+}\psi _1^{12+}+\psi _1^{21+}\psi _1^{22+}\right)\times \nonumber \\
&\hspace{41 mm}\times \left(\psi _2^{11+}\psi _2^{12+}+\psi _2^{21+}\psi _2^{22+}\right)\left(\psi _3^{11+}\psi _3^{21+}+\psi _3^{12+}\psi _3^{22+}\right)|~\rangle\\
23&. ~\left(\psi _0^{11+}\psi _0^{12+}+\psi _0^{21+}\psi _0^{22+}\right)\left(\psi _1^{11+}\psi _1^{12+}+\psi _1^{21+}\psi _1^{22+}\right)\times\nonumber \\
&\hspace{41 mm}\times \left(\psi _2^{11+}\psi _2^{21+}+\psi _2^{12+}\psi _2^{22+}\right)\left(\psi _3^{11+}\psi _3^{12+}+\psi _3^{21+}\psi _3^{22+}\right)|~\rangle\\
24&. ~\left(\psi _0^{11+}\psi _0^{12+}+\psi _0^{21+}\psi _0^{22+}\right)\left(\psi _1^{11+}\psi _1^{21+}+\psi _1^{12+}\psi _1^{22+}\right)\times\nonumber \\
&\hspace{41 mm}\times \left(\psi _2^{11+}\psi _2^{12+}+\psi _2^{21+}\psi _2^{22+}\right)\left(\psi _3^{11+}\psi _3^{12+}+\psi _3^{21+}\psi _3^{22+}\right)|~\rangle\\
25&. ~\left(\psi _0^{11+}\psi _0^{21+}+\psi _0^{12+}\psi _0^{22+}\right)\left(\psi _1^{11+}\psi _1^{12+}+\psi _1^{21+}\psi _1^{22+}\right)\times\nonumber \\
&\hspace{41 mm}\times \left(\psi _2^{11+}\psi _2^{12+}+\psi _2^{21+}\psi _2^{22+}\right)\left(\psi _3^{11+}\psi _3^{12+}+\psi _3^{21+}\psi _3^{22+}\right)|~\rangle\\
26&. ~\left(\psi _0^{11+}\psi _0^{12+}+\psi _0^{21+}\psi _0^{22+}\right)\left(\psi _1^{11+}\psi _1^{12+}+\psi _1^{21+}\psi _1^{22+}\right)\times\nonumber \\
&\hspace{41 mm}\times \left(\psi _2^{11+}\psi _2^{21+}+\psi _2^{12+}\psi _2^{22+}\right)\left(\psi _3^{11+}\psi _3^{21+}+\psi _3^{12+}\psi _3^{22+}\right)|~\rangle\\
27&. ~\left(\psi _0^{11+}\psi _0^{12+}+\psi _0^{21+}\psi _0^{22+}\right)\left(\psi _1^{11+}\psi _1^{21+}+\psi _1^{12+}\psi _1^{22+}\right)\times\nonumber \\
&\hspace{41 mm}\times \left(\psi _2^{11+}\psi _2^{12+}+\psi _2^{21+}\psi _2^{22+}\right)\left(\psi _3^{11+}\psi _3^{21+}+\psi _3^{12+}\psi _3^{22+}\right)|~\rangle\\
28&. ~\left(\psi _0^{11+}\psi _0^{21+}+\psi _0^{12+}\psi _0^{22+}\right)\left(\psi _1^{11+}\psi _1^{12+}+\psi _1^{21+}\psi _1^{22+}\right)\times\nonumber \\
&\hspace{41 mm}\times \left(\psi _2^{11+}\psi _2^{12+}+\psi _2^{21+}\psi _2^{22+}\right)\left(\psi _3^{11+}\psi _3^{21+}+\psi _3^{12+}\psi _3^{22+}\right)|~\rangle\\
29&. ~\left(\psi _0^{11+}\psi _0^{12+}+\psi _0^{21+}\psi _0^{22+}\right)\left(\psi _1^{11+}\psi _1^{21+}+\psi _1^{12+}\psi _1^{22+}\right)\times\nonumber \\
&\hspace{41 mm}\times \left(\psi _2^{11+}\psi _2^{21+}+\psi _2^{12+}\psi _2^{22+}\right)\left(\psi _3^{11+}\psi _3^{12+}+\psi _3^{21+}\psi _3^{22+}\right)|~\rangle\\
30&. ~\left(\psi _0^{11+}\psi _0^{21+}+\psi _0^{12+}\psi _0^{22+}\right)\left(\psi _1^{11+}\psi _1^{12+}+\psi _1^{21+}\psi _1^{22+}\right)\times\nonumber \\
&\hspace{41 mm}\times \left(\psi _2^{11+}\psi _2^{21+}+\psi _2^{12+}\psi _2^{22+}\right)\left(\psi _3^{11+}\psi _3^{12+}+\psi _3^{21+}\psi _3^{22+}\right)|~\rangle\\
31&. ~\left(\psi _0^{11+}\psi _0^{21+}+\psi _0^{12+}\psi _0^{22+}\right)\left(\psi _1^{11+}\psi _1^{21+}+\psi _1^{12+}\psi _1^{22+}\right)\times\nonumber \\
&\hspace{41 mm}\times \left(\psi _2^{11+}\psi _2^{12+}+\psi _2^{21+}\psi _2^{22+}\right)\left(\psi _3^{11+}\psi _3^{12+}+\psi _3^{21+}\psi _3^{22+}\right)|~\rangle\\
32&. ~\left(\psi _0^{11+}\psi _0^{12+}+\psi _0^{21+}\psi _0^{22+}\right)\left(\psi _1^{11+}\psi _1^{21+}+\psi _1^{12+}\psi _1^{22+}\right)\times\nonumber \\
&\hspace{41 mm}\times \left(\psi _2^{11+}\psi _2^{21+}+\psi _2^{12+}\psi _2^{22+}\right)\left(\psi _3^{11+}\psi _3^{21+}+\psi _3^{12+}\psi _3^{22+}\right)|~\rangle\\
33&. ~\left(\psi _0^{11+}\psi _0^{21+}+\psi _0^{12+}\psi _0^{22+}\right)\left(\psi _1^{11+}\psi _1^{12+}+\psi _1^{21+}\psi _1^{22+}\right)\times\nonumber \\
&\hspace{41 mm}\times \left(\psi _2^{11+}\psi _2^{21+}+\psi _2^{12+}\psi _2^{22+}\right)\left(\psi _3^{11+}\psi _3^{21+}+\psi _3^{12+}\psi _3^{22+}\right)|~\rangle\\
34&. ~\left(\psi _0^{11+}\psi _0^{21+}+\psi _0^{12+}\psi _0^{22+}\right)\left(\psi _1^{11+}\psi _1^{21+}+\psi _1^{12+}\psi _1^{22+}\right)\times\nonumber \\
&\hspace{41 mm}\times \left(\psi _2^{11+}\psi _2^{12+}+\psi _2^{21+}\psi _2^{22+}\right)\left(\psi _3^{11+}\psi _3^{21+}+\psi _3^{12+}\psi _3^{22+}\right)|~\rangle\\
35&. ~\left(\psi _0^{11+}\psi _0^{21+}+\psi _0^{12+}\psi _0^{22+}\right)\left(\psi _1^{11+}\psi _1^{21+}+\psi _1^{12+}\psi _1^{22+}\right)\times\nonumber \\
&\hspace{41 mm}\times \left(\psi _2^{11+}\psi _2^{21+}+\psi _2^{12+}\psi _2^{22+}\right)\left(\psi _3^{11+}\psi _3^{12+}+\psi _3^{21+}\psi _3^{22+}\right)|~\rangle\\
36&. ~\left(\psi _0^{11+}\psi _0^{21+}+\psi _0^{12+}\psi _0^{22+}\right)\left(\psi _1^{11+}\psi _1^{21+}+\psi _1^{12+}\psi _1^{22+}\right)\times\nonumber \\
&\hspace{41 mm}\times \left(\psi _2^{11+}\psi _2^{21+}+\psi _2^{12+}\psi _2^{22+}\right)\left(\psi _3^{11+}\psi _3^{21+}+\psi _3^{12+}\psi _3^{22+}\right)|~\rangle
\end{align}

\subsubsection{Type-II}

\begin{align}
37&. \left[\left(\psi _0^{11+}\psi _0^{21+}-\psi _0^{12+}\psi _0^{22+}\right)\left(0\leftrightarrow 2\right)+\left(\psi _0^{11+}\psi _0^{22+}+\psi _0^{12+}\psi _0^{21+}\right)\left(0\leftrightarrow 2\right)\right]\times \nonumber \\
&\hspace{41 mm}\times \left(\psi _1^{11+}\psi _1^{12+}+\psi _1^{21+}\psi _1^{22+}\right)\left(\psi _3^{11+}\psi _3^{12+}+\psi _3^{21+}\psi _3^{22+}\right)|~\rangle\\
38&. \left[\left(\psi _0^{11+}\psi _0^{21+}-\psi _0^{12+}\psi _0^{22+}\right)\left(0\leftrightarrow 2\right)+\left(\psi _0^{11+}\psi _0^{22+}+\psi _0^{12+}\psi _0^{21+}\right)\left(0\leftrightarrow 2\right)\right]\times \nonumber \\
&\hspace{41 mm}\times\left(\psi _1^{11+}\psi _1^{12+}+\psi _1^{21+}\psi _1^{22+}\right)\left(\psi _3^{11+}\psi _3^{21+}+\psi _3^{12+}\psi _3^{22+}\right)|~\rangle\\
39&. \left[\left(\psi _0^{11+}\psi _0^{21+}-\psi _0^{12+}\psi _0^{22+}\right)\left(0\leftrightarrow 2\right)+\left(\psi _0^{11+}\psi _0^{22+}+\psi _0^{12+}\psi _0^{21+}\right)\left(0\leftrightarrow 2\right)\right]\times \nonumber \\
&\hspace{41 mm}\times\left(\psi _1^{11+}\psi _1^{21+}+\psi _1^{12+}\psi _1^{22+}\right)\left(\psi _3^{11+}\psi _3^{12+}+\psi _3^{21+}\psi _3^{22+}\right)|~\rangle\\
40&. \left[\left(\psi _0^{11+}\psi _0^{21+}-\psi _0^{12+}\psi _0^{22+}\right)\left(0\leftrightarrow 2\right)+\left(\psi _0^{11+}\psi _0^{22+}+\psi _0^{12+}\psi _0^{21+}\right)\left(0\leftrightarrow 2\right)\right]\times \nonumber \\
&\hspace{41 mm}\times\left(\psi _1^{11+}\psi _1^{21+}+\psi _1^{12+}\psi _1	^{22+}\right)\left(\psi _3^{11+}\psi _3^{21+}+\psi _3^{12+}\psi _3^{22+}\right)|~\rangle\\
41&. \left[\left(\psi _0^{11+}\psi _0^{21+}-\psi _0^{12+}\psi _0^{22+}\right)\left(\psi _2^{11+}\psi _2^{22+}+\psi _2^{12+}\psi _2^{21+}\right)-\left(0\leftrightarrow 2\right)\right]\times \nonumber \\
&\hspace{41 mm}\times\left(\psi _1^{11+}\psi _1^{12+}+\psi _1^{21+}\psi _1^{22+}\right)\left(\psi _3^{11+}\psi _3^{12+}+\psi _3^{21+}\psi _3^{22+}\right)|~\rangle\\
42&. \left[\left(\psi _0^{11+}\psi _0^{21+}-\psi _0^{12+}\psi _0^{22+}\right)\left(\psi _2^{11+}\psi _2^{22+}+\psi _2^{12+}\psi _2^{21+}\right)-\left(0\leftrightarrow 2\right)\right]\times \nonumber \\
&\hspace{41 mm}\times\left(\psi _1^{11+}\psi _1^{12+}+\psi _1^{21+}\psi _1^{22+}\right)\left(\psi _3^{11+}\psi _3^{21+}+\psi _3^{12+}\psi _3^{22+}\right)|~\rangle\\
43&. \left[\left(\psi _0^{11+}\psi _0^{21+}-\psi _0^{12+}\psi _0^{22+}\right)\left(\psi _2^{11+}\psi _2^{22+}+\psi _2^{12+}\psi _2^{21+}\right)-\left(0\leftrightarrow 2\right)\right]\times \nonumber \\
&\hspace{41 mm}\times\left(\psi _1^{11+}\psi _1^{21+}+\psi _1^{12+}\psi _1^{22+}\right)\left(\psi _3^{11+}\psi _3^{12+}+\psi _3^{21+}\psi _3^{22+}\right)|~\rangle\\
44&. \left[\left(\psi _0^{11+}\psi _0^{21+}-\psi _0^{12+}\psi _0^{22+}\right)\left(\psi _2^{11+}\psi _2^{22+}+\psi _2^{12+}\psi _2^{21+}\right)-\left(0\leftrightarrow 2\right)\right]\times \nonumber \\
&\hspace{41 mm}\times\left(\psi _1^{11+}\psi _1^{21+}+\psi _1^{12+}\psi _1^{22+}\right)\left(\psi _3^{11+}\psi _3^{21+}+\psi _3^{12+}\psi _3^{22+}\right)|~\rangle
\end{align}

\subsubsection{Type-III}

\begin{align}
45&. \left[\left(\psi _1^{11+}\psi _1^{21+}-\psi _1^{12+}\psi _1^{22+}\right)\left(1\leftrightarrow 3\right)+\left(\psi _1^{11+}\psi _1^{22+}+\psi _1^{12+}\psi _1^{21+}\right)\left(1\leftrightarrow 3\right)\right]\times \nonumber \\
&\hspace{41 mm}\times\left(\psi _0^{11+}\psi _0^{12+}+\psi _0^{21+}\psi _0^{22+}\right)\left(\psi _2^{11+}\psi _2^{12+}+\psi _2^{21+}\psi _2^{22+}\right)|~\rangle\\
46&. \left[\left(\psi _1^{11+}\psi _1^{21+}-\psi _1^{12+}\psi _1^{22+}\right)\left(1\leftrightarrow 3\right)+\left(\psi _1^{11+}\psi _1^{22+}+\psi _1^{12+}\psi _1^{21+}\right)\left(1\leftrightarrow 3\right)\right]\times \nonumber \\
&\hspace{41 mm}\times\left(\psi _0^{11+}\psi _0^{12+}+\psi _0^{21+}\psi _0^{22+}\right)\left(\psi _2^{11+}\psi _2^{21+}+\psi _2^{12+}\psi _2^{22+}\right)|~\rangle\\
47&. \left[\left(\psi _1^{11+}\psi _1^{21+}-\psi _1^{12+}\psi _1^{22+}\right)\left(1\leftrightarrow 3\right)+\left(\psi _1^{11+}\psi _1^{22+}+\psi _1^{12+}\psi _1^{21+}\right)\left(1\leftrightarrow 3\right)\right]\times \nonumber \\
&\hspace{41 mm}\times\left(\psi _0^{11+}\psi _0^{21+}+\psi _0^{12+}\psi _0^{22+}\right)\left(\psi _2^{11+}\psi _2^{12+}+\psi _2^{21+}\psi _2^{22+}\right)|~\rangle\\
48&. \left[\left(\psi _1^{11+}\psi _1^{21+}-\psi _1^{12+}\psi _1^{22+}\right)\left(1\leftrightarrow 3\right)+\left(\psi _1^{11+}\psi _1^{22+}+\psi _1^{12+}\psi _1^{21+}\right)\left(1\leftrightarrow 3\right)\right]\times \nonumber \\
&\hspace{41 mm}\times\left(\psi _0^{11+}\psi _0^{21+}+\psi _0^{12+}\psi _0^{22+}\right)\left(\psi _2^{11+}\psi _2^{21+}+\psi _2^{12+}\psi _2^{22+}\right)|~\rangle\\
49&. \left[\left(\psi _1^{11+}\psi _1^{21+}-\psi _1^{12+}\psi _1^{22+}\right)\left(\psi _3^{11+}\psi _3^{22+}+\psi _3^{12+}\psi _3^{21+}\right)-\left(1\leftrightarrow 3\right)\right]\times \nonumber \\
&\hspace{41 mm}\times\left(\psi _0^{11+}\psi _0^{12+}+\psi _0^{21+}\psi _0^{22+}\right)\left(\psi _2^{11+}\psi _2^{12+}+\psi _2^{21+}\psi _2^{22+}\right)|~\rangle\\
50&. \left[\left(\psi _1^{11+}\psi _1^{21+}-\psi _1^{12+}\psi _1^{22+}\right)\left(\psi _3^{11+}\psi _3^{22+}+\psi _3^{12+}\psi _3^{21+}\right)-\left(1\leftrightarrow 3\right)\right]\times \nonumber \\
&\hspace{41 mm}\times\left(\psi _0^{11+}\psi _0^{12+}+\psi _0^{21+}\psi _0^{22+}\right)\left(\psi _2^{11+}\psi _2^{21+}+\psi _2^{12+}\psi _2^{22+}\right)|~\rangle\\
51&. \left[\left(\psi _1^{11+}\psi _1^{21+}-\psi _1^{12+}\psi _1^{22+}\right)\left(\psi _3^{11+}\psi _3^{22+}+\psi _3^{12+}\psi _3^{21+}\right)-\left(1\leftrightarrow 3\right)\right]\times \nonumber \\
&\hspace{41 mm}\times\left(\psi _0^{11+}\psi _0^{21+}+\psi _0^{12+}\psi _0^{22+}\right)\left(\psi _2^{11+}\psi _2^{12+}+\psi _2^{21+}\psi _2^{22+}\right)|~\rangle\\
52&. \left[\left(\psi _1^{11+}\psi _1^{21+}-\psi _1^{12+}\psi _1^{22+}\right)\left(\psi _3^{11+}\psi _3^{22+}+\psi _3^{12+}\psi _3^{21+}\right)-\left(1\leftrightarrow 3\right)\right]\times \nonumber \\
&\hspace{41 mm}\times\left(\psi _0^{11+}\psi _0^{21+}+\psi _0^{12+}\psi _0^{22+}\right)\left(\psi _2^{11+}\psi _2^{21+}+\psi _2^{12+}\psi _2^{22+}\right)|~\rangle
\end{align}

\subsubsection{Type-IV}

\begin{align}
53&. \left[\left(\psi _0^{11+}\psi _0^{12+}-\psi _0^{21+}\psi _0^{22+}\right)\left(0\leftrightarrow 1\right)+\left(\psi _0^{11+}\psi _0^{22+}-\psi _0^{12+}\psi _0^{21+}\right)\left(0\leftrightarrow 1\right)\right]\times \nonumber \\
&\hspace{41 mm}\times\left(\psi _2^{11+}\psi _2^{12+}+\psi _2^{21+}\psi _2^{22+}\right)\left(\psi _3^{11+}\psi _3^{12+}+\psi _3^{21+}\psi _3^{22+}\right)|~\rangle\\
54&. \left[\left(\psi _0^{11+}\psi _0^{12+}-\psi _0^{21+}\psi _0^{22+}\right)\left(0\leftrightarrow 1\right)+\left(\psi _0^{11+}\psi _0^{22+}-\psi _0^{12+}\psi _0^{21+}\right)\left(0\leftrightarrow 1\right)\right]\times \nonumber \\
&\hspace{41 mm}\times\left(\psi _2^{11+}\psi _2^{12+}+\psi _2^{21+}\psi _2^{22+}\right)\left(\psi _3^{11+}\psi _3^{21+}+\psi _3^{12+}\psi _3^{22+}\right)|~\rangle\\
55&. \left[\left(\psi _0^{11+}\psi _0^{12+}-\psi _0^{21+}\psi _0^{22+}\right)\left(0\leftrightarrow 1\right)+\left(\psi _0^{11+}\psi _0^{22+}-\psi _0^{12+}\psi _0^{21+}\right)\left(0\leftrightarrow 1\right)\right]\times \nonumber \\
&\hspace{41 mm}\times\left(\psi _2^{11+}\psi _2^{21+}+\psi _2^{12+}\psi _2^{22+}\right)\left(\psi _3^{11+}\psi _3^{12+}+\psi _3^{21+}\psi _3^{22+}\right)|~\rangle\\
56&. \left[\left(\psi _0^{11+}\psi _0^{12+}-\psi _0^{21+}\psi _0^{22+}\right)\left(0\leftrightarrow 1\right)+\left(\psi _0^{11+}\psi _0^{22+}-\psi _0^{12+}\psi _0^{21+}\right)\left(0\leftrightarrow 1\right)\right]\times \nonumber \\
&\hspace{41 mm}\times\left(\psi _2^{11+}\psi _2^{21+}+\psi _2^{12+}\psi _2^{22+}\right)\left(\psi _3^{11+}\psi _3^{21+}+\psi _3^{12+}\psi _3^{22+}\right)|~\rangle\\
57&. \left[\left(\psi _0^{11+}\psi _0^{12+}-\psi _0^{21+}\psi _0^{22+}\right)\left(\psi _1^{11+}\psi _1^{22+}-\psi _1^{12+}\psi _1^{21+}\right)-\left(0\leftrightarrow 1\right)\right]\times \nonumber \\
&\hspace{41 mm}\times\left(\psi _2^{11+}\psi _2^{12+}+\psi _2^{21+}\psi _2^{22+}\right)\left(\psi _3^{11+}\psi _3^{12+}+\psi _3^{21+}\psi _3^{22+}\right)|~\rangle\\
58&. \left[\left(\psi _0^{11+}\psi _0^{12+}-\psi _0^{21+}\psi _0^{22+}\right)\left(\psi _1^{11+}\psi _1^{22+}-\psi _1^{12+}\psi _1^{21+}\right)-\left(0\leftrightarrow 1\right)\right]\times \nonumber \\
&\hspace{41 mm}\times\left(\psi _2^{11+}\psi _2^{12+}+\psi _2^{21+}\psi _2^{22+}\right)\left(\psi _3^{11+}\psi _3^{21+}+\psi _3^{12+}\psi _3^{22+}\right)|~\rangle\\
59&. \left[\left(\psi _0^{11+}\psi _0^{12+}-\psi _0^{21+}\psi _0^{22+}\right)\left(\psi _1^{11+}\psi _1^{22+}-\psi _1^{12+}\psi _1^{21+}\right)-\left(0\leftrightarrow 1\right)\right]\times \nonumber \\
&\hspace{41 mm}\times\left(\psi _2^{11+}\psi _2^{21+}+\psi _2^{12+}\psi _2^{22+}\right)\left(\psi _3^{11+}\psi _3^{12+}+\psi _3^{21+}\psi _3^{22+}\right)|~\rangle\\
60&. \left[\left(\psi _0^{11+}\psi _0^{12+}-\psi _0^{21+}\psi _0^{22+}\right)\left(\psi _1^{11+}\psi _1^{22+}-\psi _1^{12+}\psi _1^{21+}\right)-\left(0\leftrightarrow 1\right)\right]\times \nonumber \\
&\hspace{41 mm}\times\left(\psi _2^{11+}\psi _2^{21+}+\psi _2^{12+}\psi _2^{22+}\right)\left(\psi _3^{11+}\psi _3^{21+}+\psi _3^{12+}\psi _3^{22+}\right)|~\rangle
\end{align}

\subsubsection{Type-V}

\begin{align}
61&. \left[\left(\psi _2^{11+}\psi _2^{12+}-\psi _2^{21+}\psi _2^{22+}\right)\left(2\leftrightarrow 3\right)+\left(\psi _2^{11+}\psi _2^{22+}-\psi _2^{12+}\psi _2^{21+}\right)\left(2\leftrightarrow 3\right)\right]\times \nonumber \\
&\hspace{41 mm}\times\left(\psi _0^{11+}\psi _0^{12+}+\psi _0^{21+}\psi _0^{22+}\right)\left(\psi _1^{11+}\psi _1^{12+}+\psi _1^{21+}\psi _1^{22+}\right)|~\rangle\\
62&. \left[\left(\psi _2^{11+}\psi _2^{12+}-\psi _2^{21+}\psi _2^{22+}\right)\left(2\leftrightarrow 3\right)+\left(\psi _2^{11+}\psi _2^{22+}-\psi _2^{12+}\psi _2^{21+}\right)\left(2\leftrightarrow 3\right)\right]\times \nonumber \\
&\hspace{41 mm}\times\left(\psi _0^{11+}\psi _0^{12+}+\psi _0^{21+}\psi _0^{22+}\right)\left(\psi _1^{11+}\psi _1^{21+}+\psi _1^{12+}\psi _1^{22+}\right)|~\rangle\\
63&. \left[\left(\psi _2^{11+}\psi _2^{12+}-\psi _2^{21+}\psi _2^{22+}\right)\left(2\leftrightarrow 3\right)+\left(\psi _2^{11+}\psi _2^{22+}-\psi _2^{12+}\psi _2^{21+}\right)\left(2\leftrightarrow 3\right)\right]\times \nonumber \\
&\hspace{41 mm}\times\left(\psi _0^{11+}\psi _0^{21+}+\psi _0^{12+}\psi _0^{22+}\right)\left(\psi _1^{11+}\psi _1^{12+}+\psi _1^{21+}\psi _1^{22+}\right)|~\rangle\\
64&. \left[\left(\psi _2^{11+}\psi _2^{12+}-\psi _2^{21+}\psi _2^{22+}\right)\left(2\leftrightarrow 3\right)+\left(\psi _2^{11+}\psi _2^{22+}-\psi _2^{12+}\psi _2^{21+}\right)\left(2\leftrightarrow 3\right)\right]\times \nonumber \\
&\hspace{41 mm}\times\left(\psi _0^{11+}\psi _0^{21+}+\psi _0^{12+}\psi _0^{22+}\right)\left(\psi _1^{11+}\psi _1^{21+}+\psi _1^{12+}\psi _1^{22+}\right)|~\rangle\\
65&. \left[\left(\psi _2^{11+}\psi _2^{12+}-\psi _2^{21+}\psi _2^{22+}\right)\left(\psi _3^{11+}\psi _3^{22+}-\psi _3^{12+}\psi _3^{21+}\right)-\left(2\leftrightarrow 3\right)\right]\times \nonumber \\
&\hspace{41 mm}\times\left(\psi _0^{11+}\psi _0^{12+}+\psi _0^{21+}\psi _0^{22+}\right)\left(\psi _1^{11+}\psi _1^{12+}+\psi _1^{21+}\psi _1^{22+}\right)|~\rangle\\
66&. \left[\left(\psi _2^{11+}\psi _2^{12+}-\psi _2^{21+}\psi _2^{22+}\right)\left(\psi _3^{11+}\psi _3^{22+}-\psi _3^{12+}\psi _3^{21+}\right)-\left(2\leftrightarrow 3\right)\right]\times \nonumber \\
&\hspace{41 mm}\times\left(\psi _0^{11+}\psi _0^{12+}+\psi _0^{21+}\psi _0^{22+}\right)\left(\psi _1^{11+}\psi _1^{21+}+\psi _1^{12+}\psi _1^{22+}\right)|~\rangle\\
67&. \left[\left(\psi _2^{11+}\psi _2^{12+}-\psi _2^{21+}\psi _2^{22+}\right)\left(\psi _3^{11+}\psi _3^{22+}-\psi _3^{12+}\psi _3^{21+}\right)-\left(2\leftrightarrow 3\right)\right]\times \nonumber \\
&\hspace{41 mm}\times\left(\psi _0^{11+}\psi _0^{21+}+\psi _0^{12+}\psi _0^{22+}\right)\left(\psi _1^{11+}\psi _1^{12+}+\psi _1^{21+}\psi _1^{22+}\right)|~\rangle\\
68&. \left[\left(\psi _2^{11+}\psi _2^{12+}-\psi _2^{21+}\psi _2^{22+}\right)\left(\psi _3^{11+}\psi _3^{22+}-\psi _3^{12+}\psi _3^{21+}\right)-\left(2\leftrightarrow 3\right)\right]\times \nonumber \\
&\hspace{41 mm}\times\left(\psi _0^{11+}\psi _0^{21+}+\psi _0^{12+}\psi _0^{22+}\right)\left(\psi _1^{11+}\psi _1^{21+}+\psi _1^{12+}\psi _1^{22+}\right)|~\rangle
\end{align}

\subsubsection{Type-VI}

\begin{align}
69&. ~~\left[\left(\psi _0^{11+}\psi _0^{21+}-\psi _0^{12+}\psi _0^{22+}\right)\left(0\leftrightarrow 2\right)+\left(\psi _0^{11+}\psi _0^{22+}+\psi _0^{12+}\psi _0^{21+}\right)\left(0\leftrightarrow 2\right)\right]\times\nonumber \\
&\times\left[\left(\psi _1^{11+}\psi _1^{21+}-\psi _1^{12+}\psi _1^{22+}\right)\left(1\leftrightarrow 3\right)+\left(\psi _1^{11+}\psi _1^{22+}+\psi _1^{12+}\psi _1^{21+}\right)\left(1\leftrightarrow 3\right)\right]|~\rangle\\
70&. ~~\left[\left(\psi _0^{11+}\psi _0^{21+}-\psi _0^{12+}\psi _0^{22+}\right)\left(\psi _2^{11+}\psi _2^{22+}+\psi _2^{12+}\psi _2^{21+}\right)-\left(0\leftrightarrow 2\right)\right]\times\nonumber \\
&\times\left[\left(\psi _1^{11+}\psi _1^{21+}-\psi _1^{12+}\psi _1^{22+}\right)\left(1\leftrightarrow 3\right)+\left(\psi _1^{11+}\psi _1^{22+}+\psi _1^{12+}\psi _1^{21+}\right)\left(1\leftrightarrow 3\right)\right]|~\rangle\\
71&. ~~\left[\left(\psi _0^{11+}\psi _0^{21+}-\psi _0^{12+}\psi _0^{22+}\right)\left(0\leftrightarrow 2\right)+\left(\psi _0^{11+}\psi _0^{22+}+\psi _0^{12+}\psi _0^{21+}\right)\left(0\leftrightarrow 2\right)\right]\times\nonumber \\
&\times\left[\left(\psi _1^{11+}\psi _1^{21+}-\psi _1^{12+}\psi _1^{22+}\right)\left(\psi _3^{11+}\psi _3^{22+}+\psi _3^{12+}\psi _3^{21+}\right)-\left(1\leftrightarrow 3\right)\right]|~\rangle\\
72&. ~~\left[\left(\psi _0^{11+}\psi _0^{21+}-\psi _0^{12+}\psi _0^{22+}\right)\left(\psi _2^{11+}\psi _2^{22+}+\psi _2^{12+}\psi _2^{21+}\right)-\left(0\leftrightarrow 2\right)\right]\times\nonumber \\
&\times\left[\left(\psi _1^{11+}\psi _1^{21+}-\psi _1^{12+}\psi _1^{22+}\right)\left(\psi _3^{11+}\psi _3^{22+}+\psi _3^{12+}\psi _3^{21+}\right)-\left(1\leftrightarrow 3\right)\right]|~\rangle
\end{align}

\subsubsection{Type-VII}

\begin{align}
73&. ~~\left[\left(\psi _0^{11+}\psi _0^{12+}-\psi _0^{21+}\psi _0^{22+}\right)\left(0\leftrightarrow 1\right)+\left(\psi _0^{11+}\psi _0^{22+}-\psi _0^{12+}\psi _0^{21+}\right)\left(0\leftrightarrow 1\right)\right]\times\nonumber \\
&\times\left[\left(\psi _2^{11+}\psi _2^{12+}-\psi _2^{21+}\psi _2^{22+}\right)\left(2\leftrightarrow 3\right)+\left(\psi _2^{11+}\psi _2^{22+}-\psi _2^{12+}\psi _2^{21+}\right)\left(2\leftrightarrow 3\right)\right]|~\rangle\\
74&. ~~\left[\left(\psi _0^{11+}\psi _0^{12+}-\psi _0^{21+}\psi _0^{22+}\right)\left(0\leftrightarrow 1\right)+\left(\psi _0^{11+}\psi _0^{22+}-\psi _0^{12+}\psi _0^{21+}\right)\left(0\leftrightarrow 1\right)\right]\times\nonumber \\
&\times\left[\left(\psi _2^{11+}\psi _2^{12+}-\psi _2^{21+}\psi _2^{22+}\right)\left(\psi _3^{11+}\psi _3^{22+}-\psi _3^{12+}\psi _3^{21+}\right)-\left(2\leftrightarrow 3\right)\right]|~\rangle\\
75&. ~~\left[\left(\psi _0^{11+}\psi _0^{12+}-\psi _0^{21+}\psi _0^{22+}\right)\left(\psi _1^{11+}\psi _1^{22+}-\psi _1^{12+}\psi _1^{21+}\right)-\left(0\leftrightarrow 1\right)\right]\times\nonumber \\
&\times\left[\left(\psi _2^{11+}\psi _2^{12+}-\psi _2^{21+}\psi _2^{22+}\right)\left(2\leftrightarrow 3\right)+\left(\psi _2^{11+}\psi _2^{22+}-\psi _2^{12+}\psi _2^{21+}\right)\left(2\leftrightarrow 3\right)\right]|~\rangle\\
76&. ~~\left[\left(\psi _0^{11+}\psi _0^{12+}-\psi _0^{21+}\psi _0^{22+}\right)\left(\psi _1^{11+}\psi _1^{22+}-\psi _1^{12+}\psi _1^{21+}\right)-\left(0\leftrightarrow 1\right)\right]\times\nonumber \\
&\times \left[\left(\psi _2^{11+}\psi _2^{12+}-\psi _2^{21+}\psi _2^{22+}\right)\left(\psi _3^{11+}\psi _3^{22+}-\psi _3^{12+}\psi _3^{21+}\right)-\left(2\leftrightarrow 3\right)\right]|~\rangle
\end{align}

\subsection{Group $(p_2,p_2)$}

Thanks to the structure of the singlet \eqref{general p2,p2 singlet} in this group, we can construct the entire singlet uniquely starting from any one of the sixteen terms. We use this fact to represent these singlet states using one of the terms in them. For instance, consider the following singlet state:
\begin{align}
|a\rangle \equiv&\left(\psi _2^{11+}\psi _3^{11+}+\psi _2^{21+}\psi _3^{21+}\right)\left(\psi _0^{12+}\psi _0^{21+}\psi _0^{22+}\psi _1^{12+}\psi _1^{21+}\psi _1^{22+}+\psi _0^{11+}\psi _0^{12+}\psi _0^{22+}\psi _1^{11+}\psi _1^{12+}\psi _1^{22+}\right)\nonumber \\
-&\left(\psi _2^{11+}\psi _3^{12+}+\psi _2^{21+}\psi _3^{22+}\right)\left(\psi _0^{12+}\psi _0^{21+}\psi _0^{22+}\psi _1^{11+}\psi _1^{21+}\psi _1^{22+}+\psi _0^{11+}\psi _0^{12+}\psi _0^{22+}\psi _1^{11+}\psi _1^{12+}\psi _1^{21+}\right)\nonumber \\
-&\left(\psi _2^{12+}\psi _3^{11+}+\psi _2^{22+}\psi _3^{21+}\right)\left(\psi _0^{11+}\psi _0^{21+}\psi _0^{22+}\psi _1^{12+}\psi _1^{21+}\psi _1^{22+}+\psi _0^{11+}\psi _0^{12+}\psi _0^{21+}\psi _1^{11+}\psi _1^{12+}\psi _1^{22+}\right)\nonumber \\
+&\left(\psi _2^{12+}\psi _3^{12+}+\psi _2^{22+}\psi _3^{22+}\right)\left(\psi _0^{11+}\psi _0^{21+}\psi _0^{22+}\psi _1^{11+}\psi _1^{21+}\psi _1^{22+}+\psi _0^{11+}\psi _0^{12+}\psi _0^{21+}\psi _1^{11+}\psi _1^{12+}\psi _1^{21+}\right)
\end{align}
This singlet state can be represented using any one of the terms as follows: 
\begin{align}
|a\rangle &=|(12,21,22),(12,21,22),(11),(11)\rangle =|(12,21,22),(12,21,22),(21),(21)\rangle \nonumber \\
&=-|(11,21,22),(12,21,22),(12),(11)\rangle =|(11,12,21),(11,12,21),(22),(22)\rangle =\ldots	 
\end{align}
Using this notation, we can list the singlet states in this group as follows:\\$\left(\psi _0^{i_1j_1+}\psi _0^{i_2j_2+}\psi _0^{i_3j_3+}\psi _1^{i_4j_4+}\psi _1^{i_5j_5+}\psi _1^{i_6j_6+}\psi _2^{i_7j_7+}\psi _3^{i_8j_8+}\right)$
\begin{align}
77&. ~|(12,21,22),(12,21,22),(11),(11)\rangle \hspace{80 mm} \\
78&. ~|(11,12,21),(12,21,22),(11),(22)\rangle  \\
79&. ~|(11,21,22),(12,21,22),(11),(11)\rangle \\
80&. ~|(11,12,22),(12,21,22),(11),(11)\rangle \\
81&. ~|(12,21,22),(11,21,22),(11),(11)\rangle  \\
82&. ~|(12,21,22),(12,21,22),(11),(21)\rangle  \\
83&. ~ |(12,21,22),(11,12,21),(21),(11)\rangle \\
84&. ~|(11,21,22),(12,21,22),(11),(22)\rangle \\
85&. ~|(11,12,21),(12,21,22),(11),(12)\rangle  \\
86&. ~|(12,21,22),(11,12,22),(22),(11)\rangle \\
87&. ~|(12,21,22),(12,21,22),(11),(22)\rangle \\
88&. ~|(11,12,21),(12,21,22),(11),(11)\rangle \\
89&. ~|(12,21,22),(11,12,21),(11),(11)\rangle \\
90&. ~|(12,21,22),(12,21,22),(22),(11)\rangle \\
91&. ~|(12,21,22),(11,12,22),(21),(11)\rangle \\
92&. ~|(12,21,22),(11,21,22),(12),(11)\rangle 
\end{align}

\subsection{Group $(p_2,p_4)$}
\begin{align}
93&. ~|(12,21,22),(11),(12,21,22),(11)\rangle \hspace{77 mm}\\
94&. ~|(11,12,21),(11),(12,21,22),(22)\rangle \\
95&. ~|(11,21,22),(11),(12,21,22),(11)\rangle \hspace{5 mm} \\
96&. ~|(11,12,22),(11),(12,21,22),(11)\rangle \\
97&. ~|(12,21,22),(12),(12,21,22),(11)\rangle \hspace{5 mm} \\
98&. ~|(12,21,22),(11),(12,21,22),(21)\rangle \\
99&. ~|(11,12,22),(11),(12,21,22),(22)\rangle \hspace{5 mm} \\
100&. ~|(11,21,22),(11),(12,21,22),(22)\rangle \\
101&. ~|(11,12,21),(11),(12,21,22),(12)\rangle \hspace{5 mm} \\
102&. ~|(12,21,22),(22),(11,12,21),(12)\rangle \\
103&. ~|(12,21,22),(11),(12,21,22),(22)\rangle \hspace{5 mm} \\
104&. ~|(11,12,21),(11),(12,21,22),(11)\rangle \\
105&. ~|(12,21,22),(22),(12,21,22),(11)\rangle \hspace{5 mm} \\
106&. ~|(12,21,22),(11),(11,12,21),(11)\rangle \\
107&. ~|(12,21,22),(21),(11,12,22),(11)\rangle \hspace{5 mm} \\
108&. ~|(12,21,22),(12),(11,21,22),(11)\rangle
\end{align}

\subsection{Group $(p_4,p_2)$}
\begin{align}
109&. ~|(11),(12,21,22),(11),(12,21,22)\rangle \hspace{75 mm} \\
110&. ~|(22),(12,21,22),(11),(11,12,21)\rangle \\
111&. ~|(12),(12,21,22),(11),(12,21,22)\rangle \hspace{5 mm} \\
112&. ~|(21),(12,21,22),(11),(12,21,22)\rangle \\
113&. ~|(11),(11,21,22),(11),(12,21,22)\rangle \hspace{5 mm} \\
114&. ~|(11),(12,21,22),(11),(11,12,22)\rangle \\
115&. ~|(21),(12,21,22),(11),(11,12,21)\rangle \hspace{5 mm} \\
116&. ~|(11),(11,21,22),(22),(12,21,22)\rangle \\
117&. ~|(11),(11,12,21),(12),(12,21,22)\rangle \hspace{5 mm} \\
118&. ~|(22),(12,21,22),(11),(11,12,22)\rangle \\
119&. ~|(11),(12,21,22),(11),(11,12,21)\rangle \hspace{5 mm} \\
120&. ~|(22),(12,21,22),(11),(12,21,22)\rangle \\
121&. ~|(11),(11,12,21),(11),(12,21,22)\rangle \hspace{5 mm} \\
122&. ~|(11),(12,21,22),(22),(12,21,22)\rangle \\
123&. ~|(11),(11,12,22),(21),(12,21,22)\rangle \hspace{5 mm} \\
124&. ~|(12),(12,21,22),(11),(11,21,22)\rangle
\end{align}

\subsection{Group $(p_4,p_4)$}
\begin{align}
125&. ~|(11),(11),(12,21,22),(12,21,22)\rangle \hspace{75 mm} \\
126&. ~|(22),(11),(12,21,22),(11,12,21)\rangle \\
127&. ~|(12),(11),(12,21,22),(12,21,22)\rangle \hspace{5 mm} \\
128&. ~|(21),(11),(12,21,22),(12,21,22)\rangle \\
129&. ~|(11),(11),(12,21,22),(11,21,22)\rangle \hspace{5 mm} \\
130&. ~|(11),(11),(12,21,22),(11,12,22)\rangle \\
131&. ~|(21),(11),(12,21,22),(11,12,21)\rangle \hspace{5 mm} \\
132&. ~|(21),(22),(11,12,21),(12,21,22)\rangle \\
133&. ~|(22),(11),(12,21,22),(11,21,22)\rangle \hspace{5 mm} \\
134&. ~|(22),(11),(12,21,22),(11,12,22)\rangle \\
135&. ~|(11),(11),(12,21,22),(11,12,21)\rangle \hspace{5 mm} \\
136&. ~|(22),(11),(12,21,22),(12,21,22)\rangle \\
137&. ~|(11),(22),(12,21,22),(12,21,22)\rangle \hspace{5 mm} \\
138&. ~|(11),(11),(11,12,21),(12,21,22)\rangle \\
139&. ~|(21),(11),(12,21,22),(11,12,22)\rangle \hspace{5 mm} \\
140&. ~|(11),(12),(11,21,22),(12,21,22)\rangle
\end{align}

\section{Independent sets of eigenstates} \label{eigenstates}

\subsection{Set 1:16 states}
\begin{align*}
\textbf{1}&. ~|36\rangle -|21\rangle \\
\textbf{2}&. ~|53\rangle -|61\rangle \\
\textbf{3}&. ~|73\rangle -4 ~|21\rangle \\
\textbf{4}&. ~|48\rangle -|40\rangle \\
\textbf{5}&. ~|69\rangle -4~|36\rangle \\
\textbf{6}&. ~|36\rangle -|1\rangle -|2\rangle -|3\rangle -|4\rangle \\
\textbf{7}&. ~|40\rangle -2~ |36\rangle +4~(|1\rangle +|4\rangle)\\
\textbf{8}&. ~|53\rangle -2~ |21\rangle +4~(|2\rangle +|3\rangle)\\
\textbf{9,10}&. ~4~(|1\rangle -|4\rangle)\pm (|77\rangle -|125\rangle)\\
\textbf{11,12}&. ~4~(|2\rangle -|3\rangle)\pm (|93\rangle -|109\rangle)\\
\textbf{13,14}&. ~4~(|1\rangle -|2\rangle-|3\rangle -|4\rangle)-|40\rangle -|48\rangle +|53\rangle +|61\rangle \pm \sqrt{3} (|77\rangle -|93\rangle -|109\rangle +|125\rangle)\\
\textbf{15,16}&. ~4~(|1\rangle +|2\rangle +|3\rangle +|4\rangle)+|36\rangle +|21\rangle +|73\rangle +|69\rangle  \pm \sqrt{\frac{7}{2}} (|77\rangle +|93\rangle +|109\rangle +|125\rangle)
\end{align*}
The first eight states have a zero eigenvalue, the next two have an eigenvalue of $\pm 4$ whereas the next two states have eigenvalues of $\pm 4\sqrt{3}$ and $\pm 2\sqrt{14}$ respectively. 

\subsection{Set 2:6 states}

\begin{align*}
\textbf{17}&. ~|78\rangle -|94\rangle \\
\textbf{18}&. ~|78\rangle -|126\rangle \\
\textbf{19}&. ~|78\rangle -|110\rangle \\
\textbf{20}&. ~|28\rangle -|29\rangle \\
\textbf{21,22}&. ~|78\rangle +|94\rangle +|110\rangle +|126\rangle \pm \sqrt{2}(|28\rangle +|29\rangle )\hspace{60 mm}
\end{align*}
The first four states have zero energy and the last one has eigenvalue of $\pm 2\sqrt{2}$.

\subsection{Set 3: 6 states}

\begin{align*}
\textbf{23}&. ~|79\rangle +|127\rangle \\
\textbf{24}&. ~|95\rangle +|111\rangle \\
\textbf{25,26}&. ~|79\rangle -|95\rangle +|111\rangle -|127\rangle \mp \sqrt{2} |44\rangle \\
\textbf{27,28}&. ~|79\rangle +|95\rangle -|111\rangle -|127\rangle \pm  |70\rangle \hspace{75 mm}\\
\end{align*}
The first two eigenstates have zero energy and the last two have eigenvalues of $\pm 2\sqrt{2}$ and $\pm 4$ respectively.

\subsection{Set 4: 6 states}

\begin{align*}
\textbf{29}&. ~|96\rangle -|112\rangle \\
\textbf{30}&. ~|80\rangle -|128\rangle \\
\textbf{31,32}&. ~|80\rangle -|96\rangle -|112\rangle +|128\rangle \pm \sqrt{2} |57\rangle \\
\textbf{33,34}&. ~|80\rangle +|96\rangle +|112\rangle +|128\rangle \pm  |75\rangle \hspace{75 mm}
\end{align*}
The first two eigenstates have zero energy and the last two have eigenvalues of $\pm 2\sqrt{2}$ and $\pm 4$ respectively.

\subsection{Set 5: 6 states}

\begin{align*}
\textbf{35}&. ~|81\rangle +|129\rangle \\
\textbf{36}&. ~|97\rangle +|113\rangle \\
\textbf{37,38}&. ~|81\rangle +|97\rangle -|113\rangle -|129\rangle \mp \sqrt{2} |52\rangle \\
\textbf{39,40}&. ~|81\rangle -|97\rangle +|113\rangle -|129\rangle \mp  |71\rangle \hspace{75 mm}
\end{align*}
The first two eigenstates have zero energy and the last two have eigenvalues of $\pm 2\sqrt{2}$ and $\pm 4$ respectively.

\subsection{Set 6: 6 states}

\begin{align*}
\textbf{41}&. ~|82\rangle -|130\rangle \\
\textbf{42}&. ~|98\rangle -|114\rangle \\
\textbf{43,44}&. ~|82\rangle -|98\rangle -|114\rangle +|130\rangle \pm \sqrt{2} |65\rangle \\
\textbf{45,46}&. ~|82\rangle +|98\rangle +|114\rangle +|130\rangle \mp  |74\rangle \hspace{75 mm}
\end{align*}
The first two eigenstates have zero energy and the last two have eigenvalues of $\pm 2\sqrt{2}$ and $\pm 4$ respectively.

\subsection{Set 7: 7 states}

\begin{align*}
\textbf{47}&. ~|83\rangle -|99\rangle +|115\rangle -|131\rangle \\
\textbf{48,49}&. ~|83\rangle -|99\rangle -|115\rangle +|131\rangle \pm \sqrt{2}|58\rangle \\
\textbf{50,51}&. ~|83\rangle +|99\rangle +|115\rangle +|131\rangle \pm \sqrt{2}|66\rangle \\
\textbf{52,53}&. ~|83\rangle +|99\rangle -|115\rangle -|131\rangle \mp \sqrt{2}|49\rangle \hspace{75 mm}
\end{align*}
The first state has zero energy and the rest of them have $\pm 2\sqrt{2}$ eigenvalue.

\subsection{Set 8: 7 states}

\begin{align*}
\textbf{54}&. ~|84\rangle -|100\rangle +|116\rangle -|132\rangle \\
\textbf{55,56}&. ~|84\rangle +|100\rangle +|116\rangle +|132\rangle \mp \sqrt{2}|51\rangle \\
\textbf{57,58}&. ~|84\rangle -|100\rangle -|116\rangle +|132\rangle \pm \sqrt{2}|43\rangle \\
\textbf{59,60}&. ~|84\rangle +|100\rangle -|116\rangle -|132\rangle \pm \sqrt{2}|68\rangle \hspace{75 mm}
\end{align*}
The first state has zero energy and the rest of them have $\pm 2\sqrt{2}$ eigenvalue.

\subsection{Set 9: 7 states}

\begin{align*}
\textbf{61}&. ~|85\rangle -|101\rangle +|117\rangle +|133\rangle \\
\textbf{62,63}&. ~|85\rangle +|101\rangle +|117\rangle -|133\rangle \mp \sqrt{2}|42\rangle \\
\textbf{64,65}&. ~|85\rangle -|101\rangle -|117\rangle -|133\rangle \pm \sqrt{2}|50\rangle \\
\textbf{66,67}&. ~|85\rangle +|101\rangle -|117\rangle +|133\rangle \mp \sqrt{2}|60\rangle \hspace{75 mm}
\end{align*}
The first state has zero energy and the rest of them have $\pm 2\sqrt{2}$ eigenvalue.

\subsection{Set 10: 7 states}

\begin{align*}
\textbf{68}&. ~|86\rangle +|102\rangle +|118\rangle -|134\rangle \\
\textbf{69,70}&. ~|86\rangle +|102\rangle -|118\rangle +|134\rangle \mp \sqrt{2}|67\rangle \\
\textbf{71,72}&. ~|86\rangle -|102\rangle +|118\rangle +|134\rangle \mp \sqrt{2}|59\rangle \\
\textbf{73,74}&. ~|86\rangle -|102\rangle -|118\rangle -|134\rangle \mp \sqrt{2}|41\rangle \hspace{75 mm}
\end{align*}
The first state has zero energy and the rest of them have $\pm 2\sqrt{2}$ eigenvalue.

\subsection{Set 11: 10 states}

\begin{align*}
\textbf{75}&. ~|35\rangle -|22\rangle \\
\textbf{76}&. ~|54\rangle +|39\rangle \\
\textbf{77}&. ~|87\rangle +|103\rangle -|119\rangle -|135\rangle \\
\textbf{78}&. ~|39\rangle +2 (|6\rangle +|10\rangle )\\
\textbf{79,80}&. ~|87\rangle -|103\rangle +|119\rangle -|135\rangle \mp 2\sqrt{2} (|6\rangle -|10\rangle )\\ 
\textbf{81,82}&. ~|87\rangle +|103\rangle +|119\rangle +|135\rangle \pm \sqrt{2} (|35\rangle +|22\rangle )\\ 
\textbf{83,84}&. ~|87\rangle -|103\rangle -|119\rangle +|135\rangle \mp \sqrt{\frac{8}{3}} (|6\rangle +|10\rangle )\mp \sqrt{\frac{2}{3}} (|54\rangle -|39\rangle )\hspace{60 mm}
\end{align*}
The first four states have zero energy and the next two have eigenvalue of $\pm 2\sqrt{2}$ whereas the last one has eigenvalue of $\pm 2\sqrt{6}$.

\subsection{Set 12: 10 states}

\begin{align*}
\textbf{85}&. ~|25\rangle -|32\rangle \\
\textbf{86}&. ~|63\rangle +|46\rangle \\
\textbf{87}&. ~|88\rangle +|104\rangle -|120\rangle -|136\rangle \\
\textbf{88}&. ~|63\rangle -2 (|7\rangle +|11\rangle )\\
\textbf{89,90}&. ~|88\rangle -|104\rangle +|120\rangle -|136\rangle \mp 2\sqrt{2} (|7\rangle -|11\rangle )\\ 
\textbf{91,92}&. ~|88\rangle +|104\rangle +|120\rangle +|136\rangle \mp \sqrt{2} (|25\rangle +|32\rangle )\\ 
\textbf{93,94}&. ~|88\rangle -|104\rangle -|120\rangle +|136\rangle \mp \sqrt{\frac{8}{3}} (|7\rangle +|11\rangle )\mp \sqrt{\frac{2}{3}} (|63\rangle -|46\rangle ) \hspace{50 mm}
\end{align*}
The first four states have zero energy and the next two have eigenvalue of $\pm 2\sqrt{2}$ whereas the last one has eigenvalue of $\pm 2\sqrt{6}$.

\subsection{Set 13: 10 states}

\begin{align*}
\textbf{95}&. ~|24\rangle -|33\rangle \\
\textbf{96}&. ~|62\rangle +|38\rangle \\
\textbf{97}&. ~|89\rangle -|105\rangle +|121\rangle -|137\rangle \\
\textbf{98}&. ~|62\rangle -2 (|15\rangle +|19\rangle )\\
\textbf{99,100}&. ~|89\rangle +|105\rangle -|121\rangle -|137\rangle \mp 2\sqrt{2} (|15\rangle -|19\rangle )\\ 
\textbf{101,102}&. ~|89\rangle +|105\rangle +|121\rangle +|137\rangle \mp \sqrt{2} (|24\rangle +|33\rangle )\\ 
\textbf{103,104}&. ~|89\rangle -|105\rangle -|121\rangle +|137\rangle \mp \sqrt{\frac{8}{3}} (|15\rangle +|19\rangle )\mp \sqrt{\frac{2}{3}} (|62\rangle -|38\rangle ) \hspace{50 mm}
\end{align*}
The first four states have zero energy and the next two have eigenvalue of $\pm 2\sqrt{2}$ whereas the last one has eigenvalue of $\pm 2\sqrt{6}$.

\subsection{Set 14: 10 states}

\begin{align*}
\textbf{105}&. ~|23\rangle -|34\rangle \\
\textbf{106}&. ~|55\rangle +|47\rangle \\
\textbf{107}&. ~|90\rangle -|106\rangle +|122\rangle -|138\rangle \\
\textbf{108}&. ~|55\rangle -2 (|14\rangle +|18\rangle )\\
\textbf{109,110}&. ~|90\rangle +|106\rangle -|122\rangle -|138\rangle \mp 2\sqrt{2} (|14\rangle -|18\rangle )\\ 
\textbf{111,112}&. ~|90\rangle +|106\rangle +|122\rangle +|138\rangle \mp \sqrt{2} (|23\rangle +|34\rangle )\\ 
\textbf{113,114}&. ~|90\rangle -|106\rangle -|122\rangle +|138\rangle \mp \sqrt{\frac{8}{3}} (|14\rangle +|18\rangle )\mp \sqrt{\frac{2}{3}} (|55\rangle -|47\rangle ) \hspace{50 mm}
\end{align*}
The first four states have zero energy and the next two have eigenvalue of $\pm 2\sqrt{2}$ whereas the last one has eigenvalue of $\pm 2\sqrt{6}$.

\subsection{Set 15: 13 states}

\begin{align*}
\textbf{115}&. ~|8\rangle +|12\rangle -|16\rangle -|20\rangle \\
\textbf{116}&. ~|27\rangle -|30\rangle \\
\textbf{117}&. ~|37\rangle -|45\rangle \\
\textbf{118}&. ~|37\rangle +|20\rangle +|16\rangle +|8\rangle +|12\rangle \\
\textbf{119}&. ~|76\rangle -2 |45\rangle \\
\textbf{120,121}&. ~|91\rangle -|139\rangle \pm \sqrt{2} (|8\rangle -|12\rangle +|16\rangle -|20\rangle )\\
\textbf{122,123}&. ~|107\rangle -|123\rangle \pm \sqrt{2} (|8\rangle -|12\rangle -|16\rangle +|20\rangle )\\
\textbf{124,125}&. ~|91\rangle -|107\rangle -|123\rangle +|139\rangle \pm \sqrt{2} (|30\rangle +|27\rangle )\\
\textbf{126,127}&. ~|91\rangle +|107\rangle +|123\rangle +|139\rangle \mp \sqrt{\frac{1}{3}} (|37\rangle +|45\rangle +|76\rangle -2|8\rangle -2|12\rangle -2|16\rangle -2|20\rangle )\hspace{20 mm}
\end{align*}
The first five states have zero eigenvalue and the next four have eigenvalues $\pm 2\sqrt{2}$ whereas the last one has eigenvalues $\pm 4\sqrt{3}$.

\subsection{Set 16: 13 states}

\begin{align*}
\textbf{128}&. ~|13\rangle +|17\rangle -|5\rangle -|9\rangle \\
\textbf{129}&. ~|31\rangle -|26\rangle \\
\textbf{130}&. ~|56\rangle -|64\rangle \\
\textbf{131}&. ~|56\rangle -|13\rangle -|17\rangle -|5\rangle -|9\rangle \\
\textbf{132}&. ~|72\rangle -2 |64\rangle \\
\textbf{133,134}&. ~|108\rangle -|124\rangle \pm \sqrt{2} (|5\rangle -|9\rangle -|13\rangle +|17\rangle )\\
\textbf{135,136}&. ~|92\rangle -|140\rangle \pm \sqrt{2} (|5\rangle -|9\rangle +|13\rangle -|17\rangle )\\
\textbf{137,138}&. ~|92\rangle -|108\rangle -|124\rangle +|140\rangle \pm \sqrt{2} (|26\rangle +|31\rangle )\\
\textbf{139,140}&. ~|92\rangle +|108\rangle +|124\rangle +|140\rangle \pm \sqrt{\frac{1}{3}} (|56\rangle +|64\rangle +|72\rangle +2|5\rangle +2|9\rangle +2|13\rangle +2|17\rangle )\hspace{20 mm}
\end{align*}
The first five states have zero eigenvalue and the next four have eigenvalues $\pm 2\sqrt{2}$ whereas the last one has eigenvalues $\pm 4\sqrt{3}$.

\section{$n=2$ uncolored model}

In this appendix, we give relevant details of the $n=2$ uncolored model. These are useful especially for finding singlets via method-II. For more details on the $n=2$ uncolored model, see \cite{dario}.

 The Noether charges Of $n=2$ uncolored model are given by:
\begin{align}
Q_1^{12}&=i \left(\psi ^{111^+}\psi ^{211^-}+\psi ^{111^-}\psi ^{211^+}+\psi ^{121^+}\psi ^{221^-}+\psi ^{121^-}\psi ^{221^+}\right)\nonumber \\
Q_2^{12}&=i \left(\psi ^{111^+}\psi ^{121^-}+\psi ^{111^-}\psi ^{121^+}+\psi ^{211^+}\psi ^{221^-}+\psi ^{211^-}\psi ^{221^+}\right)\nonumber \\
Q_3^{12}&=2-\psi ^{111^+}\psi ^{111^-}-\psi ^{121^+}\psi ^{121^-}-\psi ^{211^+}\psi ^{211^-}-\psi ^{221^+}\psi ^{221^-}
\end{align}
The action of the first two charges on level 0 and level 4 states is given by:
\begin{align}
Q_{1,2}^{12}|~\rangle &=0 \nonumber \\
Q_{1,2}^{12}\left(\psi ^{111^+}\psi ^{121^+}\psi ^{211^+}\psi ^{221^+}\right)|~\rangle &=0 \nonumber \\
\end{align}
At level 1, we have:
\begin{align}
Q_1^{12}~\psi ^{111^+}|~\rangle &=-\psi ^{211^+}\nonumber \\
Q_1^{12}~\psi ^{211^+}|~\rangle &=+\psi ^{111^+}\nonumber \\
Q_1^{12}~\psi ^{121^+}|~\rangle &=-\psi ^{221^+}\nonumber \\
Q_1^{12}~\psi ^{221^+}|~\rangle &=+\psi ^{121^+}
\end{align}
At level 3, we have:
\begin{align}
Q_{1}^{12}\left(\psi ^{111^+}\psi ^{121^+}\psi ^{211^+}\right)|~\rangle &=+\psi ^{111^+}\psi ^{211^+}\psi ^{221^+}|~\rangle \nonumber \\
Q_{1}^{12}\left(\psi ^{111^+}\psi ^{211^+}\psi ^{221^+}\right)|~\rangle &=-\psi ^{111^+}\psi ^{121^+}\psi ^{211^+}|~\rangle \nonumber \\
Q_{1}^{12}\left(\psi ^{111^+}\psi ^{121^+}\psi ^{221^+}\right)|~\rangle &=+\psi ^{121^+}\psi ^{211^+}\psi ^{221^+}|~\rangle \nonumber \\
Q_{1}^{12}\left(\psi ^{121^+}\psi ^{211^+}\psi ^{221^+}\right)|~\rangle &=-\psi ^{111^+}\psi ^{121^+}\psi ^{221^+}|~\rangle
\end{align}
From these relations, we can see that:
\begin{align}
\left((Q_1)^2+1\right) |\text{Level} ~1/3 ~\text{state}~\rangle &=0  
\end{align}
At level 2:
\begin{align}
Q_{1}^{12}\left(\psi ^{111^+}\psi ^{211^+}\right) |~\rangle &=0 \nonumber \\
Q_{1}^{12}\left(\psi ^{121^+}\psi ^{221^+}\right) |~\rangle &=0 \nonumber \\
Q_{1}^{12}\left(\psi ^{111^+}\psi ^{221^+}+\psi ^{121^+}\psi ^{211^+}\right) |~\rangle &=0 \nonumber \\
Q_{1}^{12}\left(\psi ^{111^+}\psi ^{121^+}+\psi ^{211^+}\psi ^{221^+}\right) |~\rangle &=0 \nonumber \\
Q_{1}^{12}\left(\psi ^{111^+}\psi ^{121^+}-\psi ^{211^+}\psi ^{221^+}\right) |~\rangle &=2 \left(\psi ^{121^+}\psi ^{211^+}-\psi ^{111^+}\psi ^{221^+}\right) |~\rangle\nonumber \\
Q_{1}^{12}\left(\psi ^{111^+}\psi ^{221^+}-\psi ^{121^+}\psi ^{211^+}\right) |~\rangle &=2 \left(\psi ^{111^+}\psi ^{121^+}-\psi ^{211^+}\psi ^{221^+}\right) |~\rangle
\end{align}

Now, we consider the action of $Q_2$ charge. At level 1, we have the following relations:
\begin{align}
Q_2^{12}~\psi ^{111^+}|~\rangle &=-\psi ^{121^+}\nonumber \\
Q_2^{12}~\psi ^{121^+}|~\rangle &=+\psi ^{111^+}\nonumber \\
Q_2^{12}~\psi ^{211^+}|~\rangle &=-\psi ^{221^+}\nonumber \\
Q_2^{12}~\psi ^{221^+}|~\rangle &=+\psi ^{211^+}
\end{align}
At level 3:
\begin{align}
Q_{2}^{12}\left(\psi ^{111^+}\psi ^{121^+}\psi ^{211^+}\right)|~\rangle &=-\psi ^{111^+}\psi ^{121^+}\psi ^{221^+}|~\rangle \nonumber \\
Q_{2}^{12}\left(\psi ^{111^+}\psi ^{121^+}\psi ^{221^+}\right)|~\rangle &=+\psi ^{111^+}\psi ^{121^+}\psi ^{211^+}|~\rangle \nonumber \\
Q_{2}^{12}\left(\psi ^{111^+}\psi ^{211^+}\psi ^{221^+}\right)|~\rangle &=-\psi ^{121^+}\psi ^{211^+}\psi ^{221^+}|~\rangle \nonumber \\
Q_{2}^{12}\left(\psi ^{121^+}\psi ^{211^+}\psi ^{221^+}\right)|~\rangle &=+\psi ^{111^+}\psi ^{211^+}\psi ^{221^+}|~\rangle
\end{align}
As in the case of $Q_1$, we have:
\begin{align}
\left((Q_2)^2+1\right) |\text{Level} ~1/3 ~\text{state}~\rangle &=0  
\end{align}
At level 2, we have:
\begin{align}
Q_{2}^{12}\left(\psi ^{111^+}\psi ^{121^+}\right) |~\rangle &=0 \nonumber \\
Q_{2}^{12}\left(\psi ^{211^+}\psi ^{221^+}\right) |~\rangle &=0 \nonumber \\
Q_{2}^{12}\left(\psi ^{111^+}\psi ^{221^+}-\psi ^{121^+}\psi ^{211^+}\right) |~\rangle &=0 \nonumber \\
Q_{2}^{12}\left(\psi ^{111^+}\psi ^{211^+}+\psi ^{121^+}\psi ^{221^+}\right) |~\rangle &=0 \nonumber \\
Q_{2}^{12}\left(\psi ^{111^+}\psi ^{211^+}-\psi ^{121^+}\psi ^{221^+}\right) |~\rangle &=-2 \left(\psi ^{121^+}\psi ^{211^+}+\psi ^{111^+}\psi ^{221^+}\right) |~\rangle\nonumber \\
Q_{2}^{12}\left(\psi ^{111^+}\psi ^{221^+}+\psi ^{121^+}\psi ^{211^+}\right) |~\rangle &=2 \left(\psi ^{111^+}\psi ^{211^+}-\psi ^{121^+}\psi ^{221^+}\right) |~\rangle
\end{align}

We conclude by pointing out that the charges $Q_1$ and $Q_2$ commute and this fact is useful in determining the singlets of $(p_{2,4},p_{2,4})$ and also in uniquely fixing such singlets to be of the form \eqref{general p2,p2 singlet} .

\end{document}